\documentclass[showpacs,twocolumn,aps,floatfix,superscriptaddress,noshowpacs]{revtex4-1}
\usepackage[mathlines]{lineno}

\usepackage{amsthm,amsmath,amssymb,graphicx,bm,color,mathrsfs,verbatim,epstopdf,dcolumn,dsfont,cancel,subfigure}

\usepackage{algorithm}
\usepackage[noend]{algpseudocode}
\makeatletter
\def\BState{\State\hskip-\ALG@thistlm}
\makeatother

\usepackage{bbold} 
\usepackage{graphicx}

\usepackage{ulem}
\graphicspath{{./}{figs/}{figs/main_text_fig}{figs/appendix_fig}}

\usepackage{hyperref}
\hypersetup{ 
	colorlinks = true
}

\usepackage{color}

\begin{document}


\title{Thermalization and Prethermalization in \\
	Periodically Kicked Quantum Spin Chains}%

\author{Christoph Fleckenstein}
\email{christoph.fleckenstein@physik.uni-wuerzburg.de}
\affiliation{Universit\"at W\"urzburg, Am Hubland, 97074 W\"urzburg, Germany}
\affiliation{Department of Physics, KTH Royal Institute of Technology, SE-106 91 Stockholm, Sweden}
\author{Marin Bukov}
\email{mgbukov@phys.uni-sofia.bg}
\affiliation{Department of Physics, University of California, Berkeley, CA 94720, USA}
\affiliation{Department of Physics, St.~Kliment Ohridski University of Sofia, 5 James Bourchier Blvd, 1164 Sofia, Bulgaria}

\begin{abstract}
We study the dynamics of periodically-kicked many-body systems away from the high-frequency regime, and discuss a family of Floquet systems where the notion of prethermalization can be naturally extended to intermediate and low driving frequencies. 
We investigate numerically the dynamics of both integrable and nonintegrable systems, and report on the formation of a long-lived prethermal plateau, akin to the high-frequency limit, where the system thermalizes with respect to an effective Hamiltonian captured by the inverse-frequency expansion (IFE). Unlike the high-frequency regime, we find that the relevant heating times are model dependent:
we analyze the stability of the prethermal plateau to small perturbations in the drive period, and show that, in a spin chain whose IFE is intractable, the plateau duration is insensitive to the perturbation strength, in contrast to a chain where the IFE admits the resummation of an entire subseries.
Infinitesimal perturbations are enough to restore the ergodic properties of the system, and decrease residual finite-size effects.
Although the regime where the Floquet system leaves the prethermal plateau and starts heating up to infinite temperature is not captured by the IFE, we provide evidence that the evolved subsystem is described well by a thermal state w.r.t.~the IFE Hamiltonian, with a gradually changing temperature, in accord with the Eigenstate Thermalization Hypothesis. 

\end{abstract} 

\date{\today}
\maketitle

\section{\label{sec:intro}Introduction}

Periodic drives provide a versatile toolbox to investigate properties of quantum many-body systems~\cite{goldman2014periodically,goldman2015case,eckardt2017atomic,bukov2015universal,rodriguez2020moir}. 
Based on dynamical localization and stabilization, high-frequency modulations represent a state-of-the-art experimental technique to enhance magnetic correlations~\cite{gorg2018enhancement}, to emulate artificial gauge fields~\cite{struck_13,aidelsburger_13,miyake_13,jotzu_15,nascimbene2015dynamic,  price2017synthetic,tarnowski2019measuring,gorg2019realization}, to study phases of matter with no static analogues~\cite{wintersperger2020realization,quelle2017driving}, to simulate $\mathbb{Z}_2$-lattice gauge theories~\cite{schweizer2019floquet,barbiero2019coupling} and strongly-correlated systems~\cite{sandholzer2019quantum} in ultracold atomic gases, and to induce topological properties~\cite{rechtsman_13,hafezi_14,mittal_14}; more recently they have also found applications in quantum materials~\cite{topp2019topological, mciver2020light, nuske2020floquet}.

However, attempts to extend this Floquet engineering toolbox towards strongly-interacting many-body systems reveal a bottleneck set by detrimental heating processes~\cite{bukov2015prethermal,canovi2016stroboscopic,mweinberg_15,lellouch_17,reitter_17,lellouch_18,wintersperger2020paramteric,boulier2019paramteric}.
From the perspective of thermalizing dynamics, periodically-driven (Floquet) systems share striking similarities with their static counterparts. For this reason, they provide an important paradigm to understand thermalization in quantum many-body systems~\cite{moessner2017equilibration}.

The cornerstone of the theory of periodically-driven systems, modeled by a Hamiltonian $H(t) \!=\! H(t+T)$, is Floquet's theorem \cite{Shirley1965,Sambe1973}. It states that, at times integer-multiple of the drive period (i.e.,~stroboscopically), the evolution operator $U(\ell T,0)$ is generated by the time-independent Floquet Hamiltonian $H_F$:
\begin{equation}
\label{eq:Floquet_thm}
U(\ell T,0) = \left(U_F\right)^\ell,\quad
U_F = \mathrm e^{-i  T H_F},\qquad 
\ell \in \mathbb{N}.
\end{equation}
This comes in stark contrast to generic time-dependent Hamiltonians, where the evolution operator is given by a complicated time-ordered exponential with no obvious simplification. Yet, periodically driven systems do not conserve energy. 

Theoretically, Floquet systems are of conceptual importance, since they feature a nontrivial controllable equilibrium limit: 
at infinite drive frequencies, energy conservation is restored, and the Floquet Hamiltonian is a static local operator whose properties are indistinguishable from those of static many-body systems. 
Hence, Floquet systems provide a systematic approach to understand and analyze nonequilibrium behavior. 
Away from the infinite-frequency limit, energy absorption may occur, and generic local many-body Floquet systems are currently believed to heat up to infinite temperature at infinite times~\cite{dalessio_14,lazarides_14,bar2017absence,weidinger2017floquet} [but see also Refs.~\cite{prosen_98a, prosen_99, dalessio2013many,haldar2018onset,kai2018suppression}]. 

The existence of the infinite-frequency limit affects significantly the dynamics of Floquet systems: when the drive frequency is much larger than the typical single-particle energy scales of the non-driven Hamiltonian, following a quick constrained thermalization stage, fast-driven systems enter an exponentially long-lived prethermal plateau~\cite{Berges2004}, before unconstrained thermalization brings the system to a featureless infinite-temperature state~\cite{abanin_15,mori_15}.

The physics in the prethermal plateau is well captured by the inverse-frequency expansion (IFE)~\cite{bukov2015universal} for the effective approximate local Hamiltonian $H_\mathrm{eff}\!\approx\! H_F$~\cite{de2019very}.
This equilibrium-like regime facilitates significantly the analysis of Floquet systems. Moreover, it provides a playground for the ideas of Floquet engineering.
It opens up a long prethermal time window which, under suitable conditions, supports phases of matter inaccessible in static systems~\cite{else_16,khemani_16,yao_17,haldar2018onset,else2020long,wintersperger2020realization,machado2020Long}. 
Curiously, the physics of the prethermal plateau has been found to exist in isolated (semi-)classical Floquet systems, which suggests that it is \emph{not} caused by quantum mechanical processes~\cite{notarnicola2018from,rajak2018stability, howell2019asymptotic, mori2018floquet,rajak2019characterizations,huveneers2020prethermalization,torre2020statistical}. 
Recently, it was shown that a similar prethermal plateau exists for random dipolar driving~\cite{zhao2020random}, the periodically-driven SYK model~\cite{kuhlenkamp2020periodically}, Floquet models exhibiting quantum scars~\cite{Mukherjee2020a,Mukherjee2020b} and for quasi-periodically driven systems where its duration is controlled by a stretched exponential~\cite{dumitrescu2018logarithmically,else2020long,zhao2020random}; the latter have also been shown to exhibit topological phenomena~\cite{martin2019topological,crowley2019halfinteger}. In fact, prethermalization is a widely investigated phenomenon, observed also in non-periodically driven many-body systems~\cite{Moeckel2008,Eckstein2009,Moeckel_2010}.

At lower drive frequencies, the system starts absorbing increased amounts of energy via a proliferation of Floquet many-body resonances~\cite{bukov_15_res}. In order for a Floquet system to absorb energy from the periodic drive, two conditions must be met: (i) the existence of many-body eigenstates in the non-driven system whose energies differ by an integer multiple of the drive frequency (the so-called spectrum folding criterion), and (ii) a finite transition matrix element between these states when exposed to the periodic drive.
The prethermal plateau shrinks gradually with decreasing the drive frequency until it disappears completely when the drive frequency becomes of the order of the single-particle energy scales in the non-driven Hamiltonian. This is correlated with a progressively more nonlocal operator structure of the exact Floquet `Hamiltonian', whose inverse-frequency approximation breaks down as an asymptotic series with the onset of infinite-temperature heating~\cite{mori_15,bukov_15_res}. Recently, techniques have been developed to find approximations to the Floquet Hamiltonian in the intermediate and low-frequency regimes, based on the empty-lattice-type approximation~\cite{vogl2020effective}, the Flow equation approach~\cite{verdeny2013accurate,vogl2019flow}, Floquet perturbation theory~\cite{rodriguez2018floquet,sen2021analytic}, and the Replica expansion~\cite{vajna2018replica}. This has facilitated the study of prethermal transients in interacting topological models~\cite{Lindner2017,Gulden2020,gawatz2021prethermalization}.
 
\section{\label{sec:results}Summary of the Main Results} 

In this paper, we discuss in detail an extension of the notion of prethermalization to the intermediate and low-frequency regime, introduced in Ref.~\cite{fleckenstein_short}. We investigate three different step-driven integrable and nonintegrable drives: the mixed-field Ising model, the transverse field Ising model and the Ising model without quantum fluctuations, in the vicinity of commensurate driving periods $T^\ast_k$, for which energy conservation is restored exactly. By using three different types of drives, we aim to investigate how drive-induced integrability breaking colludes with prethermalization at intermediate and low frequencies.
Indeed, we find a rich thermalization behavior: unconstrained thermalization to infinite temperature is suppressed with drive-dependent heating rates following both powerlaw and non-powerlaw behavior as a function of the distance $\varepsilon$ to the commensurate point $T^\ast_k$, and the model under investigation: while the mixed-field and transverse-field Ising models show a pure powerlaw dependence following Fermi's Golden Rule, the Ising model without quantum fluctuations exhibits a more sophisticated heating beahviour, ranging in between an exponential and power law scaling with an anomalous power. This is surprising and interesting as it suggests a non-Markovian dynamics where the evolved state retains some information throughout the evolution.
We believe that this suppressed thermalization originates from the suppressed magnitude of matrix elements between resonant many-body states (the spectrum folding criterion being readily satisfied close to $T^\ast_k$).

The intermediate-to-low frequency regime enhances the ergodic properties of the dynamics, and allows us to obtain clean data already at moderate system sizes. Yet, we observe that periodically driven pure states are prevented from reaching a featureless infinite temperature state at very long times. Instead, in analogy to many-body dynamical localization~\cite{Rozenbaum2017,Rylands2020,Fava2020}, at finite system size, thermalization comes to a halt in a finite time, indicating a not entirely ergodic dynamics. Interestingly, we find that small perturbations in the driving protocol suffice to restore ergodicity also at finite system sizes. Yet, prethermalization is resilient against such perturbations in the driving period.  
We further consider the evolution of both pure states and thermal ensembles, and demonstrate that prethermalized Floquet systems evolve into a featureless infinite-temperature state by gradually changing their temperature with respect to a local effective Floquet Hamiltonian, although the latter is computed with the help of an asymptotic IFE. Thus, the IFE can provide a useful static description even outside the prethermal plateau.

The paper is structured as follows. In Sec.~\ref{sec:setup}, we introduce a class of Floquet systems which later on allows us to extend the notion of Floquet prethermalization to intermediate and low drive frequencies. 
In Sec.~\ref{sec:nonintegrable}, we analyze the thermalization dynamics generated by a generic nonintegrable Hamiltonian -- the driven mixed-field Ising model -- starting from a pure initial state (Secs.~\ref{subsec:pure_state} and~\ref{subsec:subsys}); we define and discuss the qualitative behavior of heating rates (Sec.~\ref{subsec:heating_rate}), as well as its robustness to perturbations in the driving protocol (Sec.~\ref{subsec:noise}). Next, in Sec.~\ref{subsec:thermal}, we investigate the prethermal properties, starting from a thermal initial ensemble. The section concludes with Sec.~\ref{subsec:cont_drive}, where we investigate continuous drives.
In the second part of the study (Sec.~\ref{sec:integrable}), we investigate Floquet dynamics generated by two integrable Hamiltonians: the transverse-field Ising model , and the Ising model without quantum fluctuations. 
Finally, in Sec.~\ref{sec:discussion} we conclude and summarize our results.
Additional data, including finite-size scaling, and the complete replica derivation of the effective Hamiltonian, are shown in the Appendix.

\tableofcontents

\section{\label{sec:setup}Realizing Prethermal Behavior away from the High-Frequency Limit}

Following Ref.~\cite{fleckenstein_short}, we consider the family of Floquet unitaries:
\begin{equation}
\label{eq:U_F}
U_F(T) = \mathrm e^{-i T H/4}\mathrm e^{-i T V/2}\mathrm e^{-i T H/4},
\end{equation}
where $T=2\pi/\Omega$ is the drive period with the associated frequency of switching (henceforth called the drive frequency). The operator $V$ is required to have a commensurate spectrum, where level spacings between adjacent levels are integer multiple of some fixed fundamental number $\gamma$; $H$ is an arbitrary local many-body Hamiltonian, such that the average Hamiltonian $H_\mathrm{ave}\!=\! H+V$ is nonintegrable (i.e., it does not possess an extensive number of local conserved integrals of motion). Since $H_\mathrm{ave}$ is the leading-order term in the IFE, we assume that the nonintegrability of $H_\mathrm{ave}$ implies the nonintegrability of $H_F$~\footnote{While we do not provide a proof for this statement, we strongly believe it to be plausible for generic enough systems.}. 

Note that the commensurability condition on $V$ is not excessively restrictive, since merely all short-range interaction terms of density-density type in bosonic, fermionic, and spin systems, readily satisfy it. Thus, this setup applies to a large family of systems. For simplicity, in this paper we consider spin-$1/2$ systems, and choose
\begin{equation}
\label{eq:kick}
V = \gamma \sum_{j=1}^L \sigma^x_j,
\end{equation} 
is a global magnetic field of strength $\gamma$  along the $x$-direction; the Pauli matrices obey $[\sigma^\alpha_i,\sigma^\beta_j]=2i\delta_{ij}\epsilon^{\alpha\beta\gamma}\sigma^\gamma_j$. We shall discuss both integrable and nonintegrable drive Hamiltonians $H$. 

The choice of a symmetric drive in Eq.~\eqref{eq:U_F}: $\{T/4,T/2,T/4\}$, as compared to $\{T/2,T/2\}$, results in a real-valued generator $H_F$ of stroboscopic dynamics. We verified that it does not affect our results and conclusions.
We mention in passing that, although Eq.~\eqref{eq:U_F} bears a formal resemblance with Floquet time crystals~\cite{else_16,khemani_16,yao_17,pizzi2020time,pizzi2020time}, investigating time-crystalline behavior is beyond the scope of the present study. 

Due to the commensurate structure in the spectrum of $V$, there exists a sequence of drive periods $T=2T^\ast_k \!=\!2 \pi k/\gamma$ with $k\in\mathbb{N}$ (for $L$ even and $V$ from Eq.~\eqref{eq:kick}), and associated frequencies $\Omega^\ast_k=2\pi/(2T_k^\ast)$, where $\exp(-i T^\ast_k V) \!=\! \mathbb{1}$. Thus, for the class of Floquet unitaries under consideration, we have
\begin{eqnarray}
\qquad U_F(2T^\ast_k)=\exp(-iT^\ast_k H),\quad T^\ast_k \!=\! \frac{\pi k}{\gamma}.
\end{eqnarray}
Hence, at the special points $T^\ast_k$, the dynamics of the kicked system in Eq.~\eqref{eq:U_F} reduces to a quench problem to the static local Hamiltonian $H$. Therefore, by construction, at $T\!=\!2 T^\ast_k$, energy is conserved and the system is prevented from heating up. In this work, we discuss the behavior of this class of systems in the vicinity of $T^\ast_k$.

Note that, for $k\!=\!0$, we recover the familiar infinite-frequency point, surrounded by an interval of large but finite frequencies for which the Floquet system exhibits a prethermal plateau~\cite{abanin_15, mori_15}. Thus, the setup in Eq.~\eqref{eq:U_F} provides a natural candidate to extend the Floquet prethermal physics to finite frequencies. Observe that, for $k\!>\!0$, the drive frequency $\Omega$ can also happen to be in the intermediate-to-low frequency regime [compared to a typical single-particle energy scale in $H$]. Therefore, the present construction allows to induce stable isolated points $\Omega^\ast_k$ on the frequency axis, by means of inhibiting Floquet resonances.  

\begin{figure}[t!]
	\includegraphics[width=1.0\columnwidth]{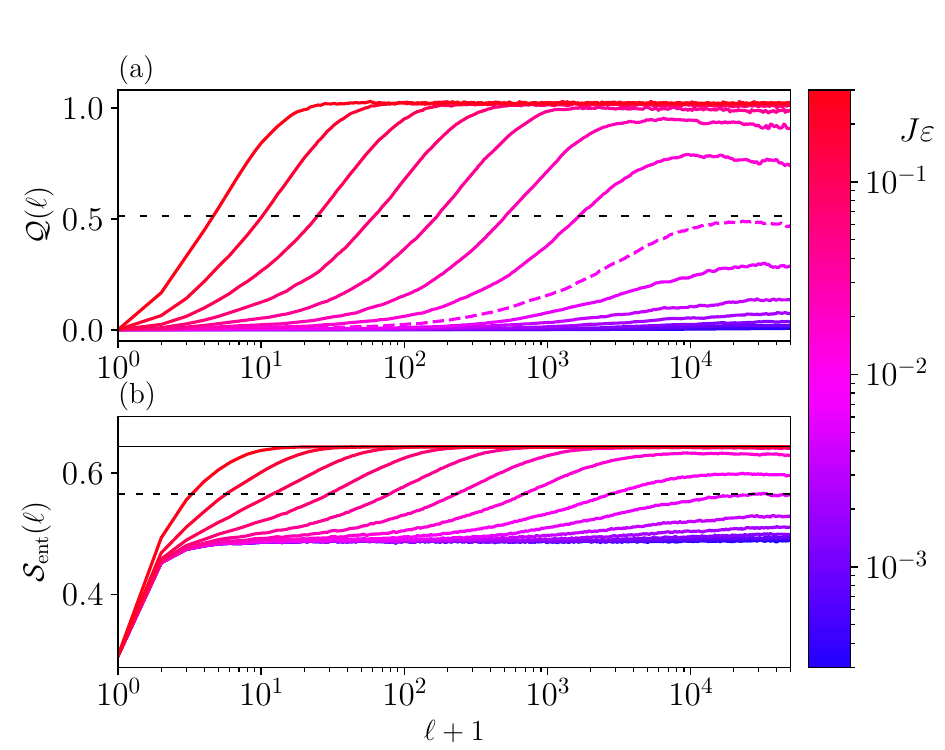}
	\caption{\label{fig:E_vs_ell_pure}
		Stroboscopic evolution using $H_1$.
		\textbf{(a)} Rescaled energy density $\mathcal{Q}(\ell)$.
		\textbf{(b)} Entanglement entropy density $\mathcal{S}_\mathrm{ent}$ of the half chain, with the Page-corrected value shown by the solid horizontal black line~\cite{Page1993}. 
		Both panels show the formation of a prethermal plateau over a few decades of driving cycles, whose duration increases parametrically out to infinity as $\varepsilon\to0$ at the commensurate point $T^\ast_k$.
		The two dashed horizontal black lines correspond to the right-hand-side in Eq.~\eqref{Eq:scaling_values} for $O=\mathcal{Q}$ and $O=\mathcal{S}_\mathrm{ent}$, respectively.
		The purple dashed curve in (a) highlights one curve to better compare it to its counterpart shown in Fig.~\ref{fig:noise}.  
		We choose 15 logarithmically spaced $\varepsilon$ values, $\varepsilon \in [3\times 10^{-4},3\times 10^{-1}]$ (the interval limits including). 
		The parameters are $h_z/J=0.809$, $h_x/J=0.9045$, $\gamma/J=1$, and $k=2$; the frequency of switching is $\Omega^\ast_2/J=1/2$. The system and subsystem sizes are $L=20$ and $L_A=10$, respectively. 
		We show both quantities against $\ell+1$ to make the initial value $\ell=0$ visible on the log scale. We display a logarithmically decreasing number of data points at large $\ell$. 	
	}
\end{figure}

In this study, we focus on the vicinity of the stable points $T\!=2(\!T^\ast_k+\varepsilon)$ ($\varepsilon\ll T^\ast_k$) where resonances are expected to be suppressed, and
\begin{equation}
\label{eq:UF_ast}
U_F\left(T\!=\!2(T^\ast_k + \varepsilon)\right) =  \mathrm e^{-i T H/4}\mathrm e^{-i \varepsilon V}\mathrm e^{-i T H/4}.
\end{equation}
The problem reduces to that of the system $H$ subject to small periodic kicks $V$ of strength $\varepsilon$~\cite{prosen_98a, prosen_99}. 

To get an intuition for why prethermal behavior can be expected even at low frequencies in this class of systems, consider the (oversimplied) model $H_0=2\gamma \sum_{j} \sigma^z_{j+1}\sigma^z_j + \sigma^z_j$. Notice that, in this case, the spectra of both $V$ and $H_0$ are commensurate with the same $T^\ast_k$; yet, $H_\mathrm{ave}\!=\!H_0\!+\!V$ is the mixed-field Ising model which is a nonintegrable Hamiltonian; hence, the Floquet system is expected to display thermalizing dynamics and heat up at intermediate to low frequencies. However, for this choice of $H_0$ and $V$, it is easy to see that $U_F \left( 2(T^\ast_k \! + \! \varepsilon ) \right)  = U_F(\varepsilon)$ for all $k$, and the period axis compactifies to a circle. Therefore, despite $T^\ast_k$ corresponding to a low drive frequency $\Omega$ at large $k$, the behavior of the system around higher-order commensurate points $T^\ast_k$ is exactly the same as around infinite frequency (i.e.~$k\!=\!0$). In particular, it follows that the dynamics features an exponentially long prethermal plateau for $T\approx 2 T^\ast_k$, while energy conservation is restored exactly (and thus the plateau lifetime becomes infinite) for $T=2 T^\ast_k$. This toy model showcases that, in order for a Floquet system to heat up to infinite temperature, it must possess finite matrix elements between the states of the non-driven Hamiltonian whose energies differ by integer multiples of $\Omega$; in other words, the criterion for folding the spectrum is a necessary but not a sufficient condition. 

\begin{figure*}[t!]
	\includegraphics[width=1\textwidth]{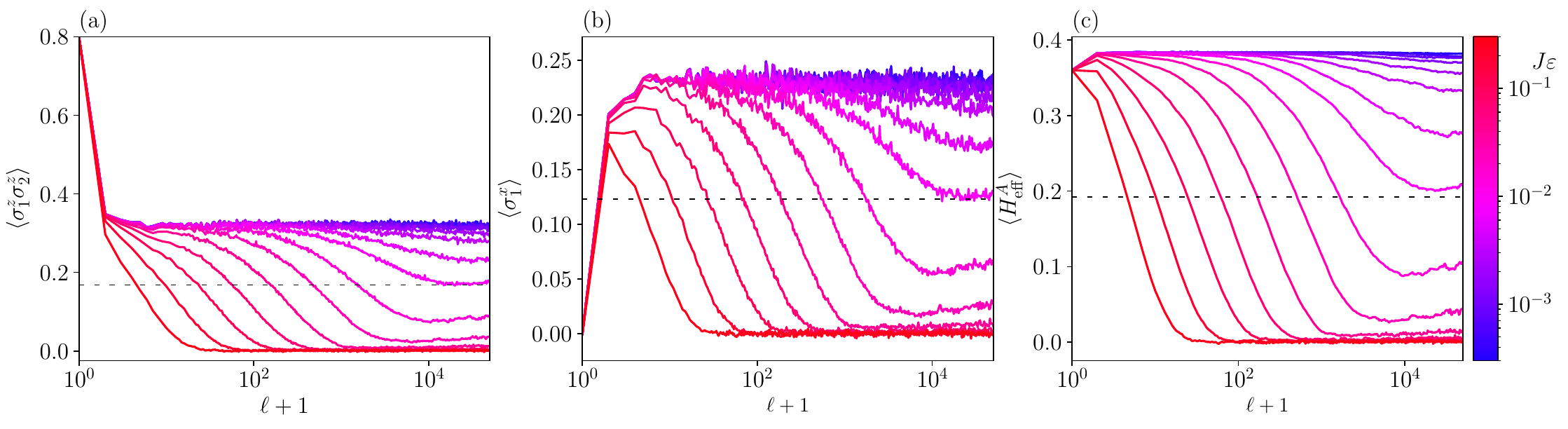}
	\caption{\label{fig:Obs_vs_ell_pure}
		Stroboscopic time evolution using $H_1$ of local observables display the four stages of thermalization dynamics in generic Floquet systems [see text] in the $\varepsilon$ vicinity of the commensurate point $T^\ast_k$, including a long-lived prethermal plateau. 
		\textbf{(a)}: local correlator $\langle \sigma^z_1\sigma^z_2\rangle$, \textbf{(b)} $x$- magnetization $\langle\sigma^x\rangle$, and \textbf{(c)} subsystem energy $\langle H_{\mathrm{eff}}^A\rangle$. Here, $H_{\mathrm{eff}}^A$ is the effective Floquet Hamiltonian restricted to subsystem $A$, as defined in Sec. \ref{subsec:subsys}.
		The parameters are the same as in Fig.~\ref{fig:E_vs_ell_pure}. 
	}
	
\end{figure*} 

Throughout this paper, we consider integrable and nonintegrable Hamiltonians $H_j$ with non-commensurate spectra, where the description of the behavior around $T^\ast_k$ is not immediately obvious. Specifically, we attempt to answer the following questions: 
(i) Under what conditions can there exist a prethermal plateau in the vicinity of $T^\ast_k$? Notice that for classical integrable systems, Nekhoroshev's estimate w.r.t.~breaking energy conservation in the vicinity of $T^\ast_k$ postulates that integrals of motion are conserved up to exponentially long times in $\varepsilon^{-1}$~\cite{nekhoroshev1971behavior,Kunihiko1989,Konishi_1990}. However, these estimates carry a system-size dependence and, to the best of our knowledge, there is no formal proof which holds in the thermodynamic limit, or in cases where the non-driven system already breaks integrability. Quantum mechanically, Fermi's Golden Rule (FGR) postulates that the system should start absorbing energy for infinitesimally small $\varepsilon$, but we do not have expressions for how the magnitude of the transition matrix elements depends on $\varepsilon$ [recent results indicate that the latter is captured by ETH for small drive amplitudes~\cite{mallaya2019heating}]. 
(ii) What are qualitative differences between the prethermal plateaus at $k=0$ (infinite frequency) and $k>0$ (moderate to low frequencies)? 
(iii) Is there an effective approximate analytical description for the dynamics of the system in the vicinity of $T^\ast_k$, similar to the IFE? 
(iv) Does the thermalization dynamics depend on whether the drive $H$ is integrable or nonintegrable [given that $H_\mathrm{ave}$ is assumed nonintegrable]? 
(v) Is the state of the Floquet system, after the initial transient is over, fully thermal, or are there any drive-induced synchronization effects~\cite{howell2019asymptotic}, such as many-body dynamical localization?

Along the way, we also investigate the following hypothesis: if the thermalization dynamics of a pure state subject to a Floquet drive exhibits a prethermal plateau, then the subsequent approach to the infinite temperature state, caused by unconstrained thermalization, is a quasi-static process; in particular, a subsystem goes through a series of (approximately) thermal states of gradually changing temperature. However, when the system heats up to infinite temperature without going through a prethermal plateau, equilibration is first reached at energy densities corresponding to infinite temperature.

\section{\label{sec:nonintegrable}Nonintegrable Drives}

Consider first the kicked system~\eqref{eq:U_F}, where $V$ is given by Eq.~\eqref{eq:kick}, for the nonintegrable spin-$1/2$ mixed-field Ising model
\begin{eqnarray}
\label{eq:Ising_mixed}
H_1= \sum_{j} J \sigma^z_{j+1}\sigma^z_j + h_z \sigma^z_j + h_x \sigma^x_j
\end{eqnarray}
with periodic boundary conditions on a lattice of $L$ sites [$L$ is chosen even for convenience]; we set $h_z/J=0.809$, $h_x/J=0.9045$, and $\gamma/J=1$ [cf.~Eq.~\eqref{eq:kick}]. We work in the zero momentum sector of positive parity, where $H_1$ has no local conservation laws other than energy itself. In this paper, parity refers to reflection with respect to the middle of the spin chain. The Hamiltonian $H_1$ and hence $H_\mathrm{ave}=H_1/2+\mathcal{O}(\varepsilon)$ [cf.~Eq.~\eqref{eq:UF_ast} for $k>0$] both exhibit Wigner-Dyson level-spacing statistics. 

The dynamics of the kicked system generated by Eqs.~\eqref{eq:UF_ast} and~\eqref{eq:Ising_mixed} violates energy conservation; thus, according to the Eigenstate Thermalization Hypothesis (ETH), we expect to observe thermalizing dynamics~\cite{dalessio2016quantum}.

\begin{figure*}[t!]
	\includegraphics[width=1\textwidth]{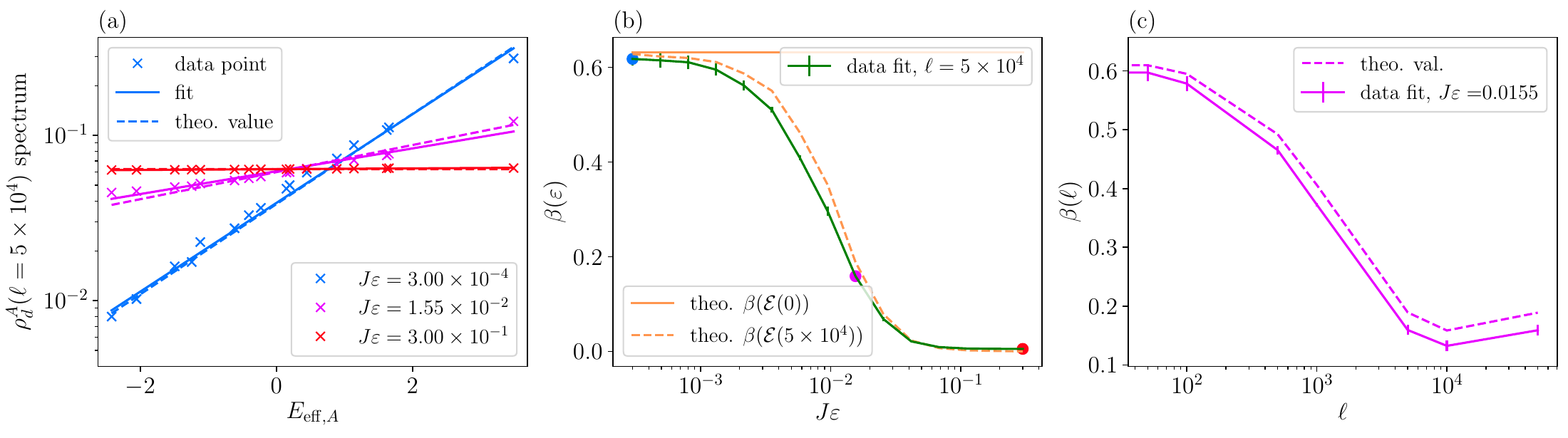}
	\caption{\label{fig:beta_pure}
		Verifying ETH in the vicinity of the commensurate points $T^\ast_k$ for $H_1$. 
		\textbf{(a)} spectrum of the reduced density matrix against the eigenvalues of $H_\mathrm{eff}^A$ for three fixed values of $\varepsilon$ [cf.~color scheme] and open boundary conditions (OBCs) on the subsystem. Crosses indicate the eigenvalues of $\rho_d^A(\ell)$ extracted from the dynamical simulation; solid lines show the best least-squares fit to the crosses. The dashed line indicates the ETH prediction with $\beta$ determined using Eq.~\eqref{eq:beta_of_E}. 
		\textbf{(b)} inverse temperature $\beta$ as a function of $\varepsilon$. The solid green line with error bars marks the values extracted from the fits in (a) at $\ell=5\times10^4$. The error bars display the uncertainty of the least square fit: assuming vanishing covariance (i.e.~uncorrelated samples), it is computed using $\Delta\beta= \sqrt{\mathrm{var}(\rho_d^A)/(2^{L_A} \mathrm{var}(E_A))}$. The solid orange line is the prediction for the prethermal plateau value according to $H_\mathrm{eff}$, while the dashed orange line is the solution to Eq.~\eqref{eq:beta_of_E} using the instantaneous value of the energy density $\mathcal{E}(\ell)$ at $\ell=5\times10^4$. Filled dots indicate the three values of $\varepsilon$ shown in (a) [color-marked]. 
		\textbf{(c)} time-dependence of the inverse temperature $\beta(\ell)$ for $\varepsilon=0.0095$. The solid line marks the least-square fit; the dashed line is the solution to Eq.~\eqref{eq:beta_of_E} using the instantaneous values of the energy density $\mathcal{E}(\ell)$. The subsystem size is $L_A=4$ and the rest of the simulation parameters are the same as in Fig.~\ref{fig:E_vs_ell_pure}.
	}
\end{figure*}

\subsection{\label{subsec:pure_state}Dynamics of a Pure Initial State}

The high-frequency ($k=0$) behavior in Floquet systems is distinguished by a long-lived prethermal plateau, and our first goal is to investigate the behavior of the  kicked Floquet system close to the commensurate point $T^\ast_k$ for $k>0$. To this end, we prepare the system in the domain wall state $\mathcal{P}|\uparrow\dots\uparrow\downarrow\dots\downarrow\rangle$, where $\mathcal{P}$ is the projector onto the zero-momentum sector of positive parity. We use symmetries in order to achieve larger Hilbert space sizes.
Ordered pure states are of particular importance in view of recent progress in Floquet engineering~\cite{goldman2014periodically,goldman2015case,eckardt2017atomic,bukov2015universal}. That said, we verified that the conclusions laid out below, do not depend on the choice of the initial pure state [although thermalization and equilibration timescales typically do].

We compute the exact evolution of the system numerically up to $5\times10^4$ driving cycles, and do measurements of the energy density $\mathcal{E}(\ell)=\langle\psi(\ell)| H_\mathrm{ave} |\psi(\ell)\rangle/L$ in the time-evolved state $|\psi(\ell)\rangle= U_F^\ell |\psi_i\rangle$ at stroboscopic times $\ell T$. Let us define the rescaled energy
\begin{equation}
\mathcal{Q}(\ell) = \frac{\mathcal{E}(\ell) - \mathcal{E}(0) }{ \langle H_\mathrm{ave} \rangle_{\beta=0} - \mathcal{E}(0) },
\end{equation}
where $\langle H_\mathrm{ave} \rangle_{\beta=0} \approx 0$ is the infinite-temperature expectation value of the average Hamiltonian. The quantity $\mathcal{Q}(\ell)$ measures energy absorption relative to the energy of the initial state. 

Figure~\ref{fig:E_vs_ell_pure}a shows that a qualitatively very similar behavior to the familiar infinite-frequency point ($k=0$), occurs in the vicinity of the commensurate points $T^\ast_k$ with $k>0$ [cf.~also App.~\ref{app:k-dep}]. Because, $k>0$ falls in the low-frequency driving regime, one can potentially make use of this parametrically long-lived stable regime to extend ideas from Floquet engineering to the low-frequency regime.

In the limit $\varepsilon\to 0$, $H_\mathrm{ave}$ is close to the exact Floquet Hamiltonian $H_F$ which is conserved.
Therefore, to guarantee that the observed prethermal behavior is not a property of the energy observable $H_\mathrm{ave}$, we also show the time evolution of the entanglement entropy density. Denoting a chain subsystem by $A$, and the corresponding reduced density matrix by $\rho^A=\mathrm{tr}_{\bar A}|\psi(\ell)\rangle\langle\psi(\ell)|$, we define
\begin{equation}
\mathcal{S}_\mathrm{ent}(\ell) = -\frac{1}{L_A}\mathrm{tr}_A\rho^A\log\rho^A.
\end{equation}
As anticipated, the prethermal plateau is also clearly visible in Fig.~\ref{fig:E_vs_ell_pure}b. In particular, we observe the same four stages of thermalization in the vicinity of the commensurate points, familiar from the high-frequency regime: (I) a transient of constrained thermalization precedes (II) a prethermal plateau, followed by (III) a second transient of unconstrained thermalization leading eventually to (IV) a featureless infinite-temperature state. Observing the prethermal physics in $\mathcal{S}_\mathrm{ent}$, we anticipate that this behavior is generic, i.e.~it applies to all local observables; we confirm this numerically in Fig.~\ref{fig:Obs_vs_ell_pure}.

\subsection{\label{subsec:subsys}Local Equilibration and Subsystem Thermalization}

Consider the quench problem of preparing a system in some initial state, and then evolving it under a generic Hamiltonian. 
A defining prediction of ETH is that a subsystem, evolving under a nonintegrable Hamiltonian, thermalizes at a temperature, corresponding to the energy density of the initial state~\cite{dalessio2016quantum,deutsch2018eigenstate}. In short, the reduced density matrix $\rho^A$ is expected to evolve into the thermal state $\rho_\mathrm{th}^A$~\cite{garrison2018does, dymarsky2018subsystem}.

In Floquet systems, it has been established that the prethermal plateau around $k=0$ is well described by an effective Hamiltonian $H_\mathrm{eff}$ obtained using the IFE~\cite{dalessio2013many,abanin_15,mori_15,bukov_15_res,howell2019asymptotic}. We now study numerically the applicability of ETH in the vicinity of commensurate points $T_k^\ast$ for $k>0$. 
Our objective is to investigate whether, under unitary evolution of the full system, subsystem $A$ evolves into the mixed Gibbs state $\rho_\mathrm{th}^A\propto\exp(-\beta(\mathcal{E}_i)H_\mathrm{eff}^A)$. $H_\mathrm{eff}^A$ is the effective Hamiltonian restricted to subsystem $A$, and $\beta(\mathcal{E}_i)$ is the temperature, corresponding to the energy density of the initial state. 

In this section we use open boundary conditions (OBC) for subsystem $A$; hence, we can equally write $H_\mathrm{eff}^A=\mathrm{tr}_{\bar{A}}(H_\mathrm{eff})$, where the trace is evaluated on the complementary subsystem $\bar{A}$. A discussion devoted to the role of boundary conditions for the effective subsystem Hamiltonian can be found in Sec.~\ref{subsec:thermal}.

\begin{figure}[t!]
	\includegraphics[scale=0.65]{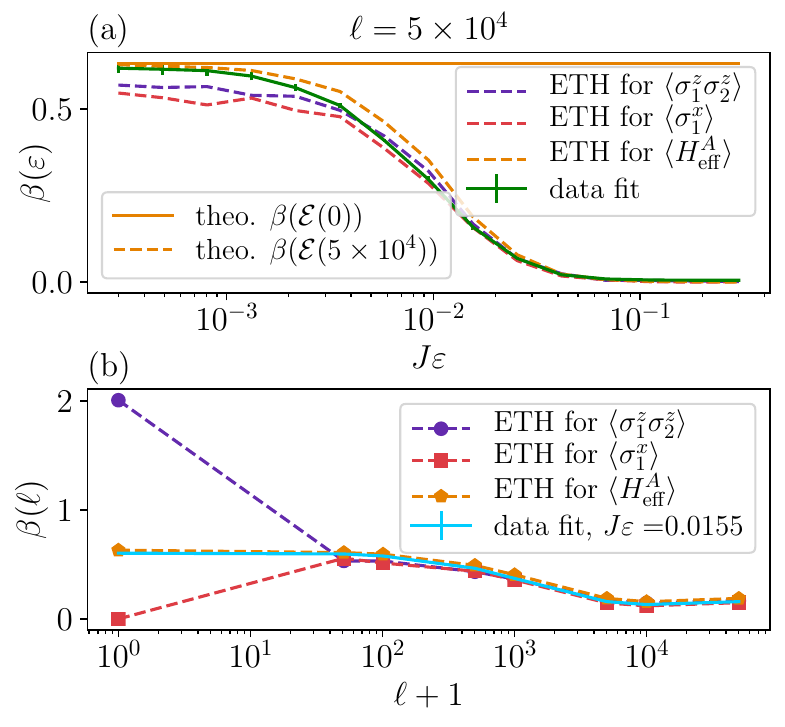}
	\caption{
		\label{fig:temp_obs}
		Verifying ETH for different observables for $H_1$ for the three observables of interest $\langle \sigma^z_1\sigma^z_2\rangle$, $\langle\sigma^x\rangle$, and $\langle H_{\mathrm{eff}}^A\rangle$. 
		\textbf{(a)} $\varepsilon$-dependence of temperature  at $\ell=5\times10^4$, cf.~Fig.~\ref{fig:beta_pure}b. 
		\textbf{(b)} $\ell$-dependence of temperature at $\varepsilon=0.0155$, cf.~Fig.~\ref{fig:beta_pure}c. We apply OBCs to the subsystems effective Hamiltonian $H_{\mathrm{eff}}^A$.
		The parameters are the same as in Fig.~\ref{fig:E_vs_ell_pure}.
	}
\end{figure}

Because we do not have a handy analytical expression for $H_\mathrm{eff}$ in the nonintegrable mixed-field Ising model, Eq.~\eqref{eq:Ising_mixed}, [cf.~Sec.~\ref{subsec:clIM} for a model amenable to the IFE], we work to leading order in $\varepsilon$. coincidentally, this provides a sufficient description for the range of $\varepsilon$ values that exhibit a prethermal plateau. 
Thus, close to $T_k^\ast$, we have 
$$H_\mathrm{eff}\approx H_\mathrm{ave}+\mathcal{O}(\varepsilon)$$
which is nonintegrable by construction, and hence we expect ETH to apply w.r.t.~$H_\mathrm{eff}$. One can, of course, add higher-order corrections whenever they can be computed [cf.~Sec.~\ref{subsec:clIM}].

We demonstrate the applicability of ETH in two steps. First, we compute the energy density of the initial pure state $\mathcal{E}_i = \langle\psi_i|H_\mathrm{eff}|\psi_i\rangle/L$, defined on the full system of $L$ sites. We can associate an inverse temperature $\beta(\mathcal{E}_i)$ to the initial energy density, by solving the implicit equation 
\begin{equation}
\label{eq:beta_of_E}
\mathcal{E}_i = \frac{1}{L_A}\mathrm{tr}_A \left(\rho_\mathrm{th}^A H_\mathrm{eff}^A\right),\quad
\rho^A_\mathrm{th}=\frac{\mathrm{e}^{-\beta H_\mathrm{eff}^A}}{\mathrm{tr}_A \mathrm{e}^{-\beta H_\mathrm{eff}^A}}
\end{equation}
for $\beta$. This provides us with a theoretically predicted reference value for the inverse temperature of the prethermal plateau.

Independently, as a second step, we also extract a value for $\beta$ from our exact numerical simulations. To do this, we first construct an approximation to the density matrix of the diagonal ensemble $\rho_d$~\cite{polkovnikov2011colloquium} empirically from the time series of the evolved state $|\psi(\ell)\rangle$:  
\begin{equation}
\label{eq:rho_d}
\rho_d \approx \rho_d(\ell)  = \frac{1}{M}\sum_{m=\ell}^{M} |\psi(\ell+m)\rangle\langle\psi(\ell+m)|,
\end{equation}
where $m=1,\dots,M$ are consecutive stroboscopic times. Note that, unlike the exact definition $\rho_d = \sum_{n} |\langle\psi_i|n_F\rangle|^2 |n_F\rangle\langle n_F|$, 
(i) the empirical definition in Eq.~\eqref{eq:rho_d} gives the diagonal density matrix in the computational basis and hence it does not involve/require knowledge of the exact Floquet eigenstates $|n_F\rangle$. Moreover, 
(ii) ensemble averages using $\rho_d(\ell)$ correspond in a natural way to experimental measurements in the system~\cite{neill2016ergodic}. 
(iii) the time- or $\ell$-dependence of $\rho_d(\ell)$ allows us to monitor the time evolution of the diagonal ensemble. In practice, we use $M=20$ consecutive stroboscopic states to construct the diagonal ensemble, but one should be careful that the system does not deviate from its steady state physically during this window, e.g., by monitoring the expectation values of local observables. 

To extract a numerical value for $\beta$, we first compute the reduced diagonal ensemble density matrix
\begin{equation}
\rho_d^A(\ell)=\mathrm{tr}_{\bar A}\rho_d(\ell).
\end{equation} 
After that, we plot the spectrum of $\rho_d^A$ against the spectrum of $H^A_\mathrm{eff}$ on a semi-log scale. A perfect straight line would indicate that $\rho_d^A(\ell)$ defines a thermal state w.r.t.~$H^A_\mathrm{eff}$. This would imply that the system is in a perfect thermal equilibrium. However, away from infinite temperature, deviations from a perfectly straight line are expected: they quantify to what extend the system is driven away from its thermal equilibrium. Using a least-squares fit, we can associate a (varying in time) temperature to the time-evolved state, which we call approximately thermal provided the corresponding least-squares fit errors remain small.

A systematic comparison of the full density matrices (not just their spectra) to quantify the deviation of the numerically-extracted diagonal ensemble from the thermal ensemble, is presented in App.~\ref{app:errors}.

\begin{figure*}[t!]
	\includegraphics[width=1\textwidth]{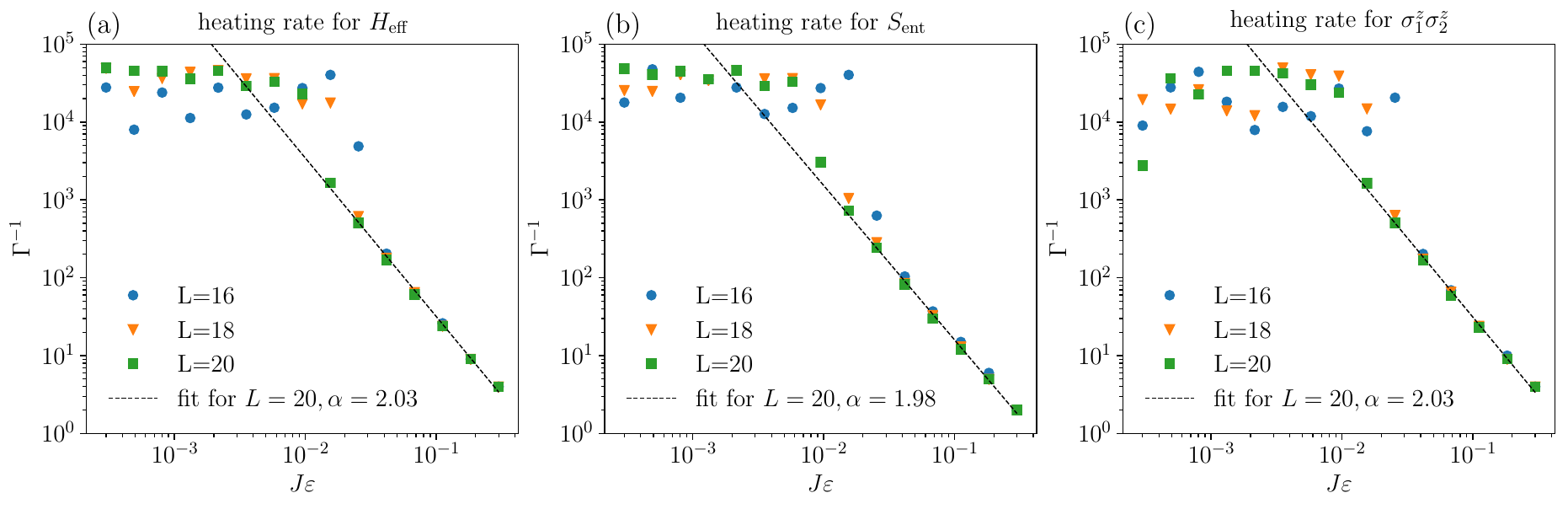}
	\caption{\label{fig:heating_rate}
		Numerically extracted heating rates $\Gamma^{-1}$ for $H_1$ as a function of the periodic kick strength $\varepsilon$ show a quadratic dependence, characteristic for Fermi's Golden Rule physics [see text]. 
		\textbf{(a)} energy density, 
		\textbf{(b)} entanglement entropy density, and 
		\textbf{(c)} a local observable. 
		The heating rates are extracted from the numerical data using Eq.~\eqref{Eq:scaling_values}.
		We fit the seven largest $\varepsilon$ data points using a least square fit [dashed black line], with the resulting exponent $\alpha$ shown in the legend.
		Different colors/markers show different system sizes.  
		The parameters are the same as in Fig.~\ref{fig:E_vs_ell_pure}. 
	}
\end{figure*}

Figure~\ref{fig:beta_pure}a indicates that, starting from a pure state on the full system, the subsystem evolves into a thermal state (to an excellent approximation), whose temperature matches well the value predicted by ETH w.r.t.~$H_\mathrm{eff}$. In particular, for $\varepsilon\lesssim10^{-3}$, the long-lived prethermal plateau appears to be well described by a thermal density matrix with inverse temperature $\beta(\mathcal{E}_i)$.
Likewise, Fig.~\ref{fig:beta_pure}b shows the dynamically extracted values for the inverse temperature $\beta$ as a function of the energy conservation breaking parameter $\varepsilon$; the error bars show the least square fit uncertainty.
This curve depends on the time $\ell$ at which the diagonal ensemble is constructed, since all states are expected to reach infinite temperature at sufficiently long times in the thermodynamic limit. 

Our data is fully consistent with ETH predictions for the prethermal plateau [Fig.~\ref{fig:beta_pure}b, solid orange line]; however, it contains more information. Fig.~\ref{fig:beta_pure}c shows the dynamically extracted values of $\beta$ at different times $\ell$ during the evolution.
The dashed line marks the solution to Eq.~\eqref{eq:beta_of_E}, where we replaced $\mathcal{E}_i$ by its value at a later time $\mathcal{E}(\ell)$. 
Although heating processes cause the system to leave the prethermal plateau, the state of the system at subsequent times is still well-described (to a good approximation) by a thermal state w.r.t.~the approximate $H_\mathrm{eff}$ [dashed lines in Fig.~\ref{fig:beta_pure}(b-c)]. 
Notice that, although thermal states are universal, in the sense that they maximize the thermodynamic entropy, at a finite temperature they are only well-defined if the Hamiltonian is known, with respect to which the state is thermal; this is highly non-trivial in time-dependent systems.

In interpreting these results, one should keep in mind that the effective Hamiltonian $H_{\mathrm{eff}}^A\approx H_{\mathrm{ave}}^A$, that we assume to describe the physics along the path of thermalization to infinite temperature, is not the exact Floquet Hamiltonian. Additionally, the diagonal ensemble averages are performed using a finite number of states $M$, and we work with relatively small system sizes (compared to the thermodynamic limit).
Hence, it is natural to expect that the temperature fits contain some uncertainty, which can be quantified by the least square fit error shown in Fig.~\ref{fig:beta_pure}. 
The resulting deviation from the perfect thermal state results in a gradual change in the associated temperature and eventually dies out as the system approaches the infinite temperature state. 
To quantify the deviation from the perfect thermal state, beside the uncertainty of the least square fit presented here, a number of different measures can be deployed. While the least square fit uncertainty compares the density matrices on the level of the eigenvalues, quantities such as the Kullback-Leibler divergence and the Uhlmann fidelity directly measure the similarity of density matrices in operator space, cf.~App.~\ref{app:errors}.

In Fig.~\ref{fig:temp_obs} we demonstrate that the observed thermalization behavior is generic: we present the same comparison between ETH-predicted and fitted inverse temperature but for a few different observables; this is performed using Eq.~\eqref{eq:beta_of_E} and replacing $H_{\mathrm{eff}}^A$ and $\mathcal{E}(\ell)$ with the corresponding local observable and its expectation value, respectively.
Initially, as the system is not thermal, large deviations appear between ETH predicted and fitted value for non-energy related quantities. Yet, as soon as the system evolves into a prethermal state, ETH predicts the expectation value of local observables, given their instantaneous expectation value.

The above finding may come as a surprise, since the heating processes that drive the system out of the prethermal state are the same which cause the failure of the IFE to converge, and which have been shown to arise from non-analytic terms (in $\Omega^{-1}$) present in $H_F$ but not in $H_\mathrm{eff}$ to any order~\cite{bukov_15_res}. 
We find that, although the IFE fails to predict the exact value of the energy density $\mathcal{E}(\ell)$ in `unconstrained thermalization'  stage (III) of the dynamics, given $\mathcal{E}(\ell)$ and $H_\mathrm{eff}$ one can reconstruct the thermal state that characterizes the system at that point of time. 
This is reminiscent of the observation that the IFE describes well ensemble expectation values in classical many-body Floquet systems, but not the precise dynamics of observables in isolated evolved configurations (due to classical chaos)~\cite{mori2018floquet}. This result is remarkable, because it hints at the existence of a simple hydrodynamic effective description for the dynamics of closed many-body Floquet systems all the way up to the infinite-temperature state at sufficiently long times~\cite{mori2018thermalization};  attempts to do this have already been made in static open systems~\cite{lange2018time,mori2020thermalization}; recently, experimental protocols to measure temperature in systems undergoing a slowly-changing equilibrium were also proposed~\cite{schuckert2020probing}. 
To the best of our knowledge, the law that governs the time-dependence of $\beta(\ell)$ is currently unknown.

Finally, note that the data in Fig.~\ref{fig:beta_pure}c provides numerical evidence in support of the Hypothesis we stated at the end of Sec.~\ref{sec:setup}.

\subsection{\label{subsec:heating_rate}Qualitative Heating Rates}

We now turn our attention to the heating rates in the vicinity of the commensurate points $T^\ast_k$. In generic Floquet systems, heating in the vicinity of the
infinite-frequency point ($k = 0$) is exponentially suppressed in the drive
frequency for both classical and quantum systems~\cite{abadal2020floquet,mori_15,howell2019asymptotic,mori2018floquet}. More
precisely, for one-dimensional systems, energy absorption acquires an
additional logarithmic correction, and is superexponentially suppressed~\cite{Avdoshkin2020}.
In contrast, here we show that for $k>0$, heating w.r.t.~$H_1$ is algebraically suppressed. 

We define the heating rate $\Gamma(\varepsilon)$ empirically, as the inverse time at which the value of an observable (or the entanglement entropy density) drops to half of its prethermal plateau value. Conversely, we call $\Gamma^{-1}$ that heating `time', which solves the equation 
\begin{equation}
\label{Eq:scaling_values}
O(\ell) =O_{\mathrm{prethermal}} \pm \frac{\vert O_\mathrm{prethermal} - O(\beta=0)\vert }{2}.
\end{equation}
This definition allows us to extract the $\varepsilon$-dependence of $\Gamma^{-1}$ from the numerical data, up to a pre-factor which depends on the model parameters.

Figure~\ref{fig:heating_rate} demonstrates a power-law scaling $\Gamma^{-1}\propto\varepsilon^{-\alpha}$ of the heating rates for $k=2$. The data is fully consistent with applying Fermi's Golden Rule (FGR) to the periodically-kicked problem~\eqref{eq:UF_ast}, which predicts $\alpha=2$~\cite{mallaya2019heating}. This represents a major difference compared to the $k=0$ point, where $\Gamma^{-1}\propto\exp(\varepsilon/\xi)$~\cite{howell2019asymptotic}. Thus, heating close to $T^\ast_k$ is only power-law suppressed for $H_1$, which explains the relatively small values of $\varepsilon$ required for a prethermal plateau to form. Note that the power-law scaling of the heating rate is universally seen in the dynamics of all observables and the entanglement entropy density. We also checked that this behavior appears for all $k>0$, not just $k=2$ (see App. \ref{app:k-dep}). 

We mention in passing that observing an exponentially-suppressed heating for $k>0$ is likely possible for $H_1$ if $\gamma/J\gtrsim1$ is large enough. This becomes plausible when the first commensurate point $T^\ast_1$ at $k=1$ corresponds to a drive frequency larger than the single-particle energy scale of the problem. In this case, however, the system falls outside the low-frequency driving regime for $k=1$. Moreover, even in such a case, for a large enough $k$, we expect a power-law scaling of the heating rates.

\begin{figure}[t!]
	\includegraphics[width=1.0\columnwidth]{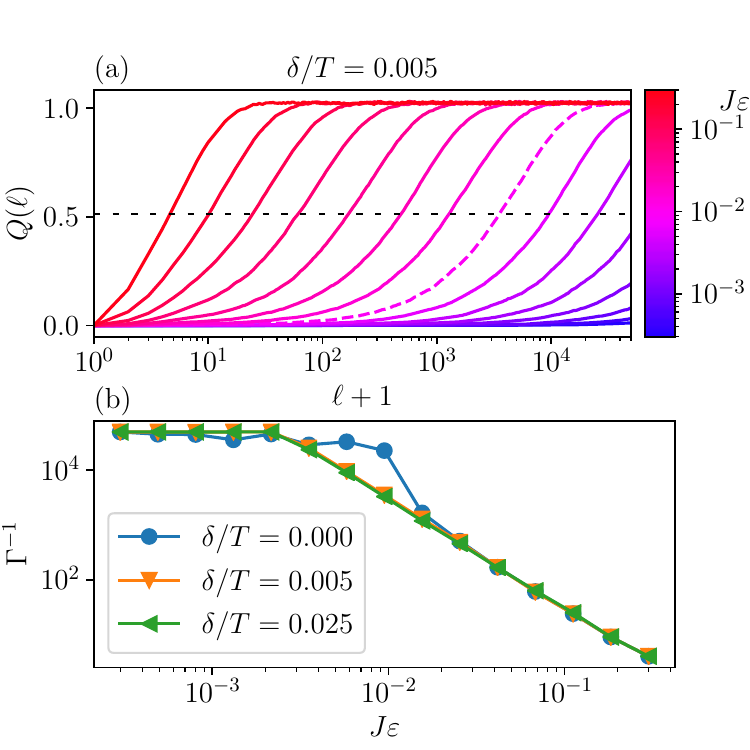}
	\caption{\label{fig:noise}
		Perturbing the periodic dynamics by a small random noise $\delta$ [cf.~Eq.~\eqref{eq:UF_ast}], restores ergodicity in the dynamics and reduces finite-size effects for all values of $\varepsilon$. 
		\textbf{(a)} time-evolution of $\mathcal{Q}(\ell)$ for $\delta/T=0.005$ [compare to Fig.~\ref{fig:E_vs_ell_pure}a]. 
		The purple dashed curve highlights the effect of noise in a direct comparison to the purple dashed curve in Fig.~\ref{fig:E_vs_ell_pure}a.
		\textbf{(b)} heating time $\Gamma^{-1}$ vs.~$\varepsilon$ for the noise-perturbed and noise-free dynamics. 
		The parameters are the same as in Fig.~\ref{fig:E_vs_ell_pure}. 
	}
\end{figure}

\subsection{\label{subsec:noise}Robustness to Drive Noise}

While the observed quadratic scaling provides evidence that FGR underlies the heating behavior for $k>0$, our simulations show that the dynamics of the periodically kicked system may not be fully ergodic out to very long times. This can be seen by noticing that for some values of $\varepsilon$ (e.g., $\varepsilon=0.016, 0.025$) the curves showing the time-dependence of observables get stuck before reaching their infinite temperature values [Fig.~\eqref{fig:E_vs_ell_pure}]. 
Moreover, the presence of a similar feature in the entanglement entropy density curves [Fig.~\eqref{fig:E_vs_ell_pure}b], which by ETH is related to the thermal entropy, suggests that the system does not explore ergodically the entire available Hilbert space.
This secondary plateau occurs at high energy densities long after the system has left the prethermal plateau. 
The phenomenon appears in the behavior of merely all quantities of interest, and is puzzling because $H_\mathrm{eff}$ is a completely ergodic, nonintegrable Hamiltonian.  
In time-independent systems, lack of ergodicity typically suggests the existence of hidden (left-over) conservation laws; these are, however, ruled out both for $H_\mathrm{eff}$ and for the exact Floquet Hamiltonian generated by Eq.~\eqref{eq:Ising_mixed}. Therefore, we look for an explanation related to the nonequilibrium dynamics of the system. 

To investigate this non-ergodic feature in detail, we perturb the periodicity of the drive: we keep the strength $\varepsilon$ of the small kick fixed, while adding a small random number $\delta\in[0,0.05T]$ to the duration $T/2$ the Hamiltonian $H$ is applied for:
\begin{eqnarray}
\label{eq:noisy_drive}
	U_\delta(2(T^\ast_k+\varepsilon) ) =  \mathrm e^{-i (T+\delta) H/4}\mathrm e^{-i \varepsilon V}\mathrm e^{-i (T+\delta) H/4}.
\end{eqnarray}
We consider the regime of small perturbations $\delta\lesssim T/2$, irrespective of the value of $\varepsilon$ [which itself is a perturbation around $T^\ast_k$ that controls breaking of energy conservation]. Since this procedure destroys the perfect periodicity of the Floquet drive, any drive-induced synchronization effects~\cite{howell2019asymptotic} would be destroyed as well. Therefore, by comparing the perturbed and perturbation-free Floquet dynamics, we can infer whether synchronization effects occur in our system. This is intimately related to the Markovian properties of the Floquet dynamics which tells if the latter retains memory of its evolution.

In Fig.~\ref{fig:noise}a we show the time evolution of the energy in the kicked system subject to small perturbation strength $\delta/T=0.005$. Comparing the curves to the perturbation-free case [cf., in particular, the color dashed lines in Fig.~\ref{fig:noise}a and Fig.~\ref{fig:E_vs_ell_pure}a], we clearly see that the random perturbation in the drive period helps restore ergodicity: all curves in the noise-perturbed dynamics approach the infinite-temperature value at sufficiently long times. Moreover, we also find that adding the noise-perturbation does not change the time it takes for the system to leave the prethermal plateau: in Fig.~\ref{fig:noise}b, we show the heating time curves $\Gamma^{-1}(\varepsilon)$. 

Finite $\delta$ breaks the periodicity of the drive, and hence $H_{\mathrm{eff}}$ changes from period to period. Naively, one should render Floquet theory inapplicable. However, for kicked systems we can still apply the more general Baker-Campbell-Hausdorff formula. Hence, in the present case, where the thermalization dynamics is mainly driven by the leading order $H_{\mathrm{eff}}=H_1/2+\mathcal{O}(\varepsilon)$, the finite perturbation strength results in an additive correction of the order $\delta/(2T)$:
\begin{eqnarray}
\label{eq:delta}
H_{\mathrm{eff}}\rightarrow H_{\mathrm{eff}}\left(1+\frac{\delta}{2T}\right).
\end{eqnarray}
As a result, small value of $\delta/T$ have negligible effects on the observed prethermalization. The more prominent effect of adding the perturbation is that all crossover values of $\varepsilon$ shift to align perfectly on the straight line with increasing $\delta/T$ [Fig.~\ref{fig:noise}b]. 

The simulation data demonstrates that the periodically-driven system ($\delta=0$) is not fully ergodic at the finite system sizes $L$ within the reach of reliable simulations. Finite-size scaling indicates that ergodicity is restored as the system size approaches the thermodynamic limit even in the periodically-driven system [App.~\ref{app:L-dep}]. Yet, at finite system size, adding the perturbation provides a useful technique to simulate ergodic behavior. 
Curiously, the observed dynamics shares similar features with many-body dynamical localization~\cite{Keser2016,Rozenbaum2017,Rylands2020,Fava2020}. 
It is currently an open question what mechanism causes this drive-induced synchronization at long times for finite system sizes, and whether this can be interpreted as a collective phenomenon induced by the Floquet drive in a finite-size system.

\begin{figure*}[t!]
	\includegraphics[width=1\textwidth]{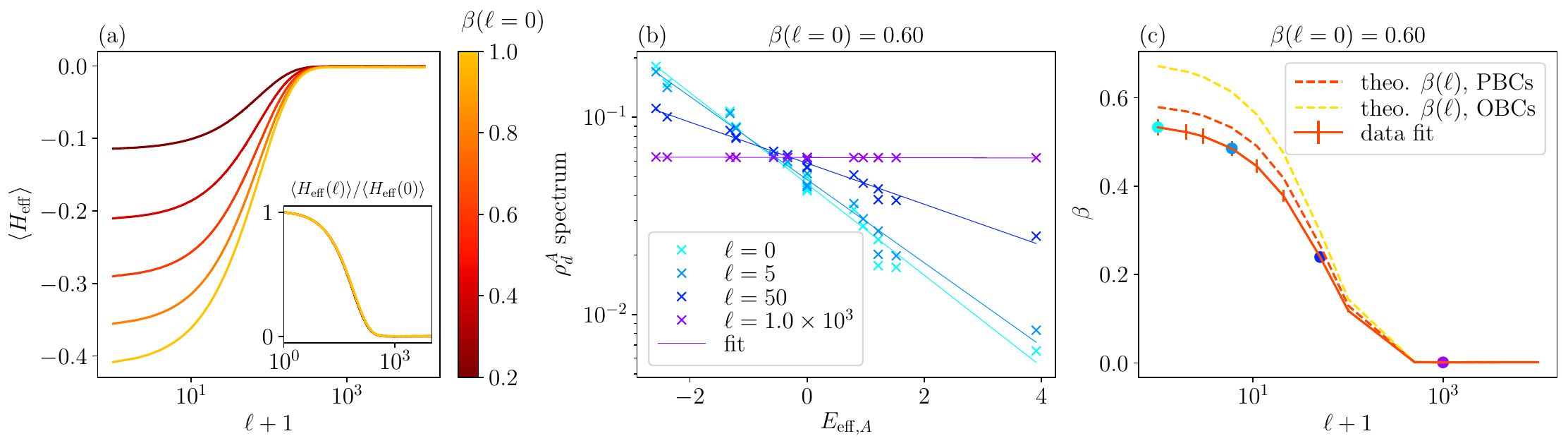}
	\caption{\label{fig:beta_thermal}
	Stroboscopic evolution of a thermal state for $H_1$: heating follows the ETH predictions w.r.t $H_{\mathrm{eff}}=H_1/2$ from the initial $\beta(\ell\!=\!0)=0.6$ up to infinite temperature, where thermal properties are kept throughout the evolution. 
	\textbf{(a)} dynamics of the energy density starting from a thermal state for a few different values of the initial inverse temperature $\beta$; the inset shows rescaled data and demonstrates universality of the Floquet dynamics as a function of the initial temperature (i.e.~energy density). 
	\textbf{(b)} spectrum of the reduced density matrix against the eigenvalues of $H_{\mathrm{eff}}^A$ at $\beta(\ell\!=\!0)=0.6$ for four different values of $\ell$ during the evolution. Crosses indicate the eigenvalues of $\rho_\beta^A(\ell)$ with solid lines and error bars indicating the best least-squares fit to the data.
	\textbf{(c)} time-dependence of the inverse temperature $\beta(\ell)$ at $\beta(\ell\!=\!0)=0.6$. The solid line marks the least-square fit; the dashed lines are the solution to Eq.~\eqref{eq:beta_of_E} using the instantaneous value of the energy density $\mathcal{E}(\ell)$ for periodic boundary conditions (PBCs) and OBCs of the subsystem [see text]. The solid dots correspond to the data sets shown in (b). 
	The subsystem size is $L_A=4$ and $\varepsilon=0.08$, and we use $N=20$ states to approximate the initial thermal ensemble; the rest of the simulation parameters are the same as in Fig.~\ref{fig:E_vs_ell_pure}.
}  
\end{figure*}

\subsection{\label{subsec:thermal}Thermal Initial Ensemble and Dependence on the Energy Density of the Initial State}

The discussion on ergodicity in Sec.~\ref{subsec:noise} raises the question whether the state of the periodically-driven system in the prethermal plateau is fully thermal w.r.t.~$H_\mathrm{eff}$. While the results in Sec.~\ref{subsec:subsys} already provide a strong indication for this claim, they do show small deviations from the expected thermal behavior [quantified in App.~\ref{app:errors}]. To settle this question, and to show that the results from the previous sections are not sensitive to the energy density of the initial state, we simulate the dynamics of a thermal ensemble, and compare the behavior of the time-evolved thermal state to that of the evolved pure state [Sec.~\ref{subsec:subsys}]. 

Simulating exactly the dynamics of a thermal ensemble would require solving the von Neumann equation for the density matrix of the full system, starting from a thermal initial state. Unfortunately, with the computational power at our disposal, this proves to be infeasible for spin chains of size $L=20$. The reasons for this are the exponentially large (in $L$) Hilbert space size, and the long evolution times required in our study.

Therefore, we resort to an approximate approach, based on quantum typicality~\cite{bartsch2009dynamical,reimann2018dynamical,reimann2019typicality,richter2019combining,weinberg2021enhanced}. Typicality, which is unrelated to integrability, states that the trace of an operator $\mathcal{O}$ defined on a Hilbert space $\mathcal{H}$ can be approximated as
\begin{equation}
\label{eq:typicality}
\frac{1}{\mathrm{dim}(\mathcal{H})}\mathrm{tr}\;\mathcal{O} \approx \frac{1}{N}\sum_{n=1}^N \langle r_n|\mathcal{O}|r_n\rangle,
\end{equation}
where $|r_n\rangle$ are  Haar-random states. The approximation becomes exact in the limit $N\to\infty$. 
Hence, thermal expectation values w.r.t.~$H_\mathrm{eff}$ in the full system, at temperature $\beta^{-1}$, can be approximated as~\cite{prelovsek2013ground}
\begin{equation}
\langle\mathcal{O}\rangle_\beta \approx \frac{ \frac{\mathrm{dim}(\mathcal{H})}{N}\sum_{n=1}^N \langle r_n|\mathrm{e}^{-\frac{\beta}{2} H_\mathrm{eff}} \mathcal{O} \mathrm{e}^{-\frac{\beta}{2} H_\mathrm{eff}}|r_n\rangle }{ \frac{\mathrm{dim}(\mathcal{H})}{N} \sum_{n=1}^N \langle r_n|\mathrm{e}^{-\frac{\beta}{2} H_\mathrm{eff}}  \mathrm{e}^{-\frac{\beta}{2} H_\mathrm{eff}}|r_n\rangle }.
\nonumber
\end{equation}
Interpreting the expression on the right-hand side as an ensemble average, the thermal density matrix is approximately equal to
\begin{eqnarray}
\label{eq:rho_th_full}
\rho_\beta &\approx& \frac{1}{Z_\beta}\frac{\mathrm{dim}(\mathcal{H})}{N}\sum_{n=1}^N  \mathrm{e}^{-\frac{\beta}{2} H_\mathrm{eff}}|r_n\rangle\langle r_n|\mathrm{e}^{-\frac{\beta}{2} H_\mathrm{eff}}, \nonumber\\
Z_\beta&\approx& \frac{\mathrm{dim}(\mathcal{H})}{N}\sum_{n=1}^N  \langle r_n|\mathrm{e}^{-\frac{\beta}{2} H_\mathrm{eff}} \mathrm{e}^{-\frac{\beta}{2} H_\mathrm{eff}}|r_n\rangle,
\end{eqnarray}
where the subscript $\beta$ denotes the inverse temperature of the thermal state.
Notice that this definition requires $N$ \emph{pure} states $|\psi_n\rangle=\mathrm{e}^{-\frac{\beta}{2} H_\mathrm{eff}}|r_n\rangle$. Therefore, to compute the time evolution of the thermal ensemble, by linearity of the ensemble average, it suffices to evolve each state $|\psi_n\rangle$ separately and then build:
\begin{equation}
\rho_\beta(\ell) \approx \frac{1}{Z_\beta}\frac{\mathrm{dim}(\mathcal{H})}{N}\sum_{n=1}^N U_F^\ell |\psi_n\rangle\langle\psi_n|\left[U_F^\ell\right]^\dagger.
\end{equation} 

Note the difference of this approximate thermal ensemble $\rho_\beta(\ell)$ to the empirical diagonal ensemble $\rho_d(\ell)$ we introduced in Eq.~\eqref{eq:rho_d}: we construct the diagonal ensemble out of a time series of quantum states [at sufficiently long times when the initial transients have died out], starting from a single initial state.
In contrast, in the approximate thermal ensemble, we have a set of initial states which we evolve up to some time $\ell$ before taking a measurement. While the two ensembles may seem different, for dynamics governed by nonintegrable Hamiltonians, they become equivalent in the thermodynamic limit: in fact, it is within the sense of the diagonal ensemble, that thermal expectation values, as defined in statistical mechanics, are to be carried out in practice [since experimentalists typically do not have many copies of the many-body system to build a proper statistical ensemble]. 

The random states $|r_n\rangle$  are defined in the full Hilbert space; in practice, we decompose the simulation over various symmetry sectors of $H_\mathrm{eff}$ for efficiency. In order to avoid building and diagonalizing the $2^L\times 2^L$ matrix $\rho_\beta$, we first reduce the evolved pure states to subsystem $A$. Noting that the partial trace and the ensemble average are mutually commuting linear operations, we obtain
\begin{equation}
\rho_\beta^A(\ell) = \frac{1}{Z_\beta}\frac{\mathrm{dim}(\mathcal{H})}{N}\sum_{n=1}^N  \mathrm{tr}_{\bar A} \left(U_F^\ell |\psi_n\rangle\langle\psi_n|\left[U_F^\ell\right]^\dagger\right).
\end{equation}
This computation of $\rho_\beta^A(\ell)$ can be trivially parallelized over the ensemble to gain speed.

Figure~\ref{fig:beta_thermal}a shows the time evolution of the energy density of the time-evolved approximate thermal ensemble with $N=20$. Because the system starts already in a thermal state w.r.t.~$H_\mathrm{eff}$, its dynamics does not feature the initial constrained thermalization stage (I), in contrast to starting from a pure state. 
Repeating the simulation for a few different initial inverse temperatures $\beta$, we see that the Floquet dynamics is insensitive to the energy density of the initial state (provided the assumptions of ETH for $H_\mathrm{eff}$ are satisfied, see inset in Fig.~\ref{fig:beta_thermal}a).

Figure~\ref{fig:beta_thermal}b shows snapshots of the spectrum of $\rho_\beta^A(\ell)$ at a few different times $\ell$. Once initialized in a thermal state, the system remains thermal throughout the time evolution to an excellent approximation. This is a trivial observation but it rules out the possibility for the system to enter a nonequilibrium state during the evolution before reaching infinite-temperature at long times; instead, we see that the state can be described well by a thermal ensemble with a slowly varying temperature. This provides additional evidence for the Hypothesis laid out in Sec.~\ref{sec:setup}: indeed, irrespective of the energy density of the initial state, once a  nonintegrable high-frequency Floquet system enters a thermal state, it will remain thermal [to an excellent approximation] under continued exposure to the periodic drive; its temperature increases slowly as the system heats up to infinite temperature~\cite{lange2018time}. Similar to Sec.~\ref{subsec:subsys}, we find once again that the state of the system during the evolution is (approximately) thermal w.r.t.~$H_\mathrm{eff}\neq H_F$ even at times past the prethermal plateau; this confirms that, although $H_\mathrm{eff}$ is insufficient to capture the heating dynamics of the system, given the energy density of the evolved state at some time $\ell$, $H_\mathrm{eff}\neq H_F$ contains the necessary information to effectively describe the thermal state the Floquet system is in, at any point during the evolution. However, what $H_\mathrm{eff}$ misses, are the very processes that cause the system to heat up in the first place. 

Finally, in Figure~\ref{fig:beta_thermal}c we show the evolution of the inverse temperature $\beta$, and compare it to the theoretical prediction according to Eq.~\eqref{eq:beta_of_E}. We attribute the mismatch at short times $\ell$ to finite-size effects, finite-$\varepsilon$ corrections to the effective Hamiltonian $H_\mathrm{eff}$, and to the relatively small number of $N=20$ states used for the ensemble average. Moreover, note also that the boundary conditions that we select for the effective Hamiltonian of the subsystem have an influence on the ETH-predicted value for $\beta$. In fact, despite being physically unrealistic, periodic boundary conditions (PBCs) seem to describe better the fitted values of $\beta$ as compared to (the more natural) OBCs.  
However, such differences are not expected to persist in the thermodynamic limit.
This trend can be already observed with the system size scaling we did (see App. \ref{app:non-int}, Fig.~\ref{fig:scaling_2}).

Based on the data in Fig~\ref{fig:beta_thermal}, we conclude that, besides the initial constrained thermalization transient, there is no difference (within the limits of finite-size simulations) in the later stages of the Floquet evolution of a pure, as compared to a thermal initial state. In both cases, we observe a qualitatively and quantitatively similar behavior.

\begin{figure}
	\centering
	\includegraphics[scale=0.38]{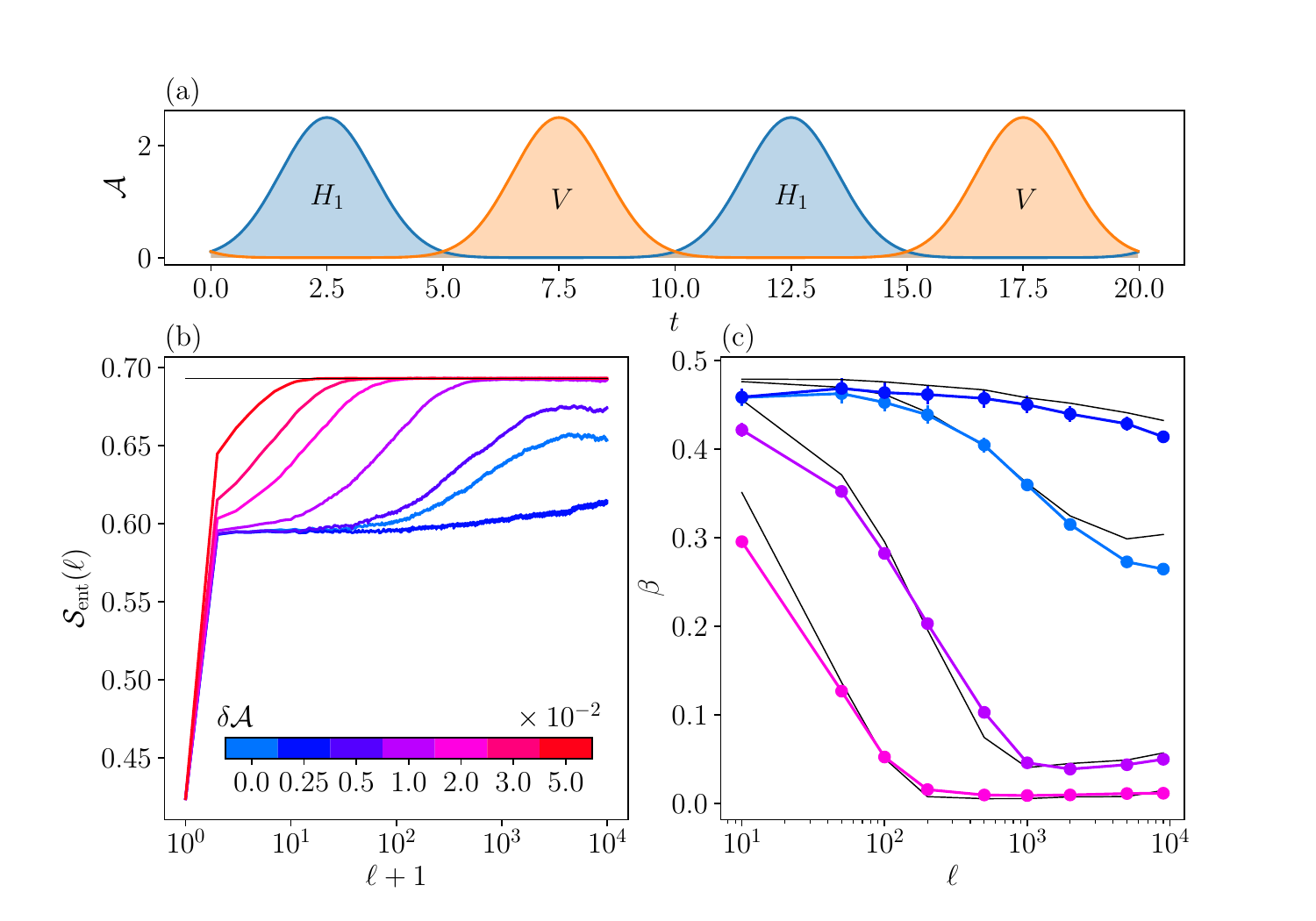}
	\caption{$H_1$ driven according to the protocol of Eq.~\eqref{eq:non_step_drive_full}: 
		{\bf (a)} schematics of the driving protocol with $\sigma=1/\gamma$, $d=10\sigma$ and $d_0=d/4$.
		{\bf (b)} time evolution of the entanglement entropy density for different amplitude shifts $\delta\mathcal{A}$.  The solid black line represent the Page-corrected value of maximum entropy 
		{\bf (c)} ETH predicted (solid black lines) and fitted (color lines) temperature values for a subsystem of size $L_A=4$. The colors in (c) are associated with the amplitude shifts of (b). Error bars mark least square fit errors. The system size is $L=18$. The initial state and the remaining parameters are the same as in Fig.~\ref{fig:E_vs_ell_pure}.}
	\label{fig:non_step_drive_spins}
\end{figure}

\subsection{\label{subsec:cont_drive}Continuous drives}

While choosing piecewise constant driving protocols with appropriate frequency is a convenient way to generate low-frequency prethermalization, it is certainly not the only possibility. In fact, the determining requirement is not the step-drive; rather, it is the commensurability of $H(t)$ for a given time period of the driving protocol. Then, the time-ordered integral of the time-evolution operator can be simplified so that suitable parameter choices map parts of the driving protocol to unity. 

To demonstrate this mechanism in more detail, consider a sequential Gaussian drive
\begin{eqnarray}
	\label{eq:cont_drive}
	z_{\sigma}^{(d_0,d)}(t)= \sum_{j \in \mathbb{N}} \mathcal{A}\exp\left(- 
	\frac{(t-d_0 - j\times d)^2}{(2\sigma^2)}\right),
\end{eqnarray}
where the parameter $d$ determines the frequency. An analogue of the step-driven models of the former sections can then be found by shifting the different parts of the drive ($H_1$ and $V$) w.r.t.~one another by $d/2$ (see Fig.~\ref{fig:non_step_drive_spins}a). Since now $H_1$ and $V$ are no longer applied in disjoint regions of time, commensurability is generically lost and the Floquet time-evolution operator is given by a complicated time-ordered exponential:
\begin{equation}
	\label{eq:non_step_drive_full}
	U_F = \mathcal{T} \!\exp\left(-i\!\int_0^d\!\mathrm{d}t\left[ \!z_\sigma^{(d_0,d)}(t)H_1\!+\!z_\sigma^{(d_0+1/2d,d)}(t)V\right]\right)\!,
\end{equation}
where $\mathcal{T}$ is the time-ordering operator. Importantly however, for $\sigma \ll d$, $H_1$ and $V$ act in approximately disjoint time intervals so that we find, up to small corrections, 
\begin{eqnarray}
	U_F & \approx &  \exp\left(-i\!\int_0^d\!\mathrm{d}t ~\!z_\sigma^{(d_0+1/2d,d)}(t)V\right)\\\nonumber
	&&\times \exp\left(-i\!\int_0^d\!\mathrm{d}t ~~\!z_\sigma^{(d_0,d)}(t)H_n\!\right).
\end{eqnarray}
Note that the above equation does not require a time-ordering operator. Thus, the time-integrals can be carried out explicitly, which yields
\begin{eqnarray}
	\label{eq:cont_drive_UF}
	U_F & \simeq &  \exp\left(-i \mathcal{A}\sqrt{2\pi} \sigma V\right)\exp\left(-i \mathcal{A}\sqrt{2\pi}\sigma H_1\!\right).
\end{eqnarray}
Hence, in a direct analogy to the step drives, setting $\mathcal{A}\sigma =\sqrt{2\pi}k/(2\gamma)$ maps the $V$-part of the drive to unity, so that heating processes are suppressed.

Numerically, the Floquet unitary of Eq.~\eqref{eq:non_step_drive_full} can be integrated exactly over one period, and used to obtain the exact time evolution of the system. Figure~\ref{fig:non_step_drive_spins}b shows the corresponding results for the entanglement entropy density evolved up to $10^4$ driving cycles for different amplitudes $\mathcal{A}=(1+\delta\mathcal{A})\sqrt{2\pi}k/2$. Here and in the following, we choose $k=2$, $\sigma=1/\gamma$ and $d=10 \sigma$. $\delta\mathcal{A}$ takes over the role of $\varepsilon$ in the step-driven models. Interestingly, $\delta\mathcal{A}=0$ does not yield optimal suppression of thermalization. Instead, small deviations of $\delta \mathcal{A}\sim 0.0025$ yield much longer prethermal plateaus as compared to $\delta\mathcal{A}=0$. This effect can be traced back to the non-zero temporal overlap of $H_1$ and $V$, i.e.~to the time-ordered corrections that need to be added to Eq.~\eqref{eq:cont_drive_UF} in order to obtain the exact Floquet unitary of Eq.~\eqref{eq:non_step_drive_full}. Yet, as long as these corrections remain small, prethermalization is expected to appear around $\mathcal{A}\sigma =\sqrt{2\pi}k/(2\gamma)$. 

In fact, the time-evolved states show the same thermalization dynamics as observed for step-drive protocols: after a short constraint transient, subsystems thermalize w.r.t.~an effective subsystem Hamiltonian $H_{\mathrm{eff}}^A$ so that the inverse temperature $\beta$ matches (to a good approximation) the value predicted by the instantaneous energy expectation values (Fig.~\ref{fig:non_step_drive_spins}c). Up to a multiplicative constant, we find that the effective Hamiltonian $H_{\mathrm{eff}}^A$ is, to leading order, the same as for step-drive protocols: $H_{\mathrm{eff}}^A=\pi k/(\gamma d) H_1$.

\section{\label{sec:integrable}Integrable Drives}

\begin{figure*}[t!]
	\includegraphics[width=1\textwidth]{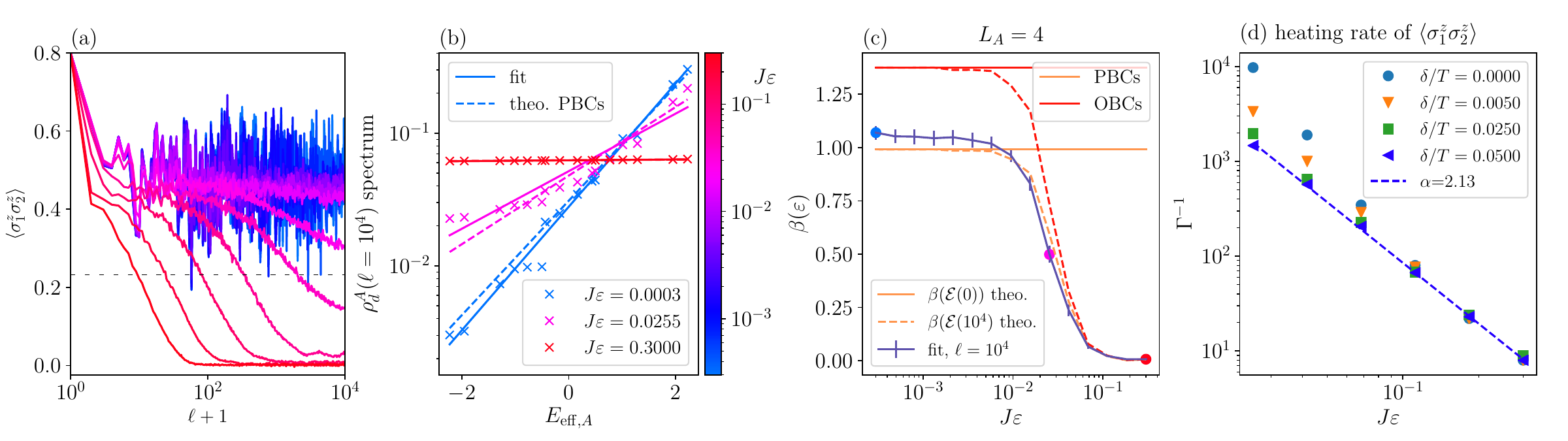}
	\caption{\label{fig:data_TFIM}
		Heating in the vicinity of the commensurate points $T^\ast_k$ for the transverse-field Ising model defined by $H_2$: 
		\textbf{(a)} heating behaviour for different values of $J\varepsilon$ as a function of stroboscopic times $\ell$ (we choose the same $\varepsilon$ values as in Fig.~\ref{fig:E_vs_ell_pure}), 
		\textbf{(b)} spectrum of the reduced density matrix at $\ell=10^4$ for three different values of $J\varepsilon$ [cf.~Fig.~\ref{fig:beta_pure}], 
		\textbf{(c)} $\beta$ values obtained from fitting the spectrum of the reduced density matrix [blue line with errorbars], and computation with the instantaneous energy at $\ell=10^4$ [dashed curves]. 
		\textbf{(d)} $\varepsilon$-dependence of the heating rates for the effective Hamiltonian $H_\mathrm{eff}$ for different values of the noise strength parameter $\delta$. 
		The simulation parameters are the same as in Fig.~\ref{fig:E_vs_ell_pure}.
	}
\end{figure*}

\subsection{\label{subsec:TFIM}Transverse-field Ising Model}

Let us now turn to integrable drives. Following a global spin rotation, an integrable limit of the Hamiltonian $H_1$ is given by the transverse-field Ising model
\begin{eqnarray}
\label{Eq:Ising_transvers}
H_2= \sum_{j} J \sigma^x_{j+1}\sigma^x_j+ h_z \sigma^z_j,
\end{eqnarray}
where we set $h_z/J=0.9045$. We drive the system according to the protocol of Eqs.~\eqref{eq:U_F} and~\eqref{eq:kick}. The initial state is the domain wall pure state in the $z$-basis, projected to the zero momentum sector of positive parity. The corresponding effective Hamiltonian to leading order in $\varepsilon$ is $H_\mathrm{eff} = H_2 /2+ \mathcal{O}(\varepsilon)$. Despite the integrability of $H_2$, and similar to $H_1$, it is infeasible to obtain a closed-form analytical expression for the higher-order correction terms to the Floquet Hamiltonian. 

Investigating the heating behaviour of Eq.~\eqref{Eq:Ising_transvers} around $T^\ast_k$ is particularly interesting from two perspectives: 
(i) unlike the nonintegrable Ising model, where $\varepsilon$ breaks only the remaining energy conservation law, here the same $\varepsilon$ also breaks integrability~\cite{Bertini2015,reimann2019typicality}. Recently, it was proposed that quantum chaotic behavior, set out by infinitesimal integrability breaking, can be sensitively detected using adiabatic gauge potentials~\cite{pandey2020adiabatic}.
Exactly at $T^\ast_k$, integrability is restored and, in the vicinity of these points, we can study how the integrability breaking parameter $\varepsilon$ influences the thermalizing dynamics. 
(ii) despite being an integrable model, $H_2$ possesses a non-commensurate spectrum, which can be found by virtue of the Jordan-Wigner mapping to free fermions. This is a prerequisite for the proliferation of resonances, once the drive is turned on; away from $T^\ast_k$, resonances are expected to facilitate thermalization.

In general, integrable models do not obey ETH. Instead, they often thermalize to a Generalized Gibbs ensemble with a Lagrange multiplier associated to each conserved quantity of the system~\cite{Kollar2011,Calabrese2011,dalessio2016quantum}. Nonetheless, quenches from specific initial states may occasionally lead to thermalization to a Gibbs ensembles in integrable models~\cite{deutsch2018eigenstate}. 
Figure~\ref{fig:data_TFIM}a shows that, in the regime of small $\varepsilon$, the dynamics of the periodically kicked system $H_2$ forms a prethermal plateau; however the expectation values of observables in the plateau are marked by large fluctuations, reminiscent of revivals~\cite{Russomanno2012}, whose origin can be traced back to the integrable character of $H_2$.

\begin{figure*}[t!]
	\includegraphics[width=1\textwidth]{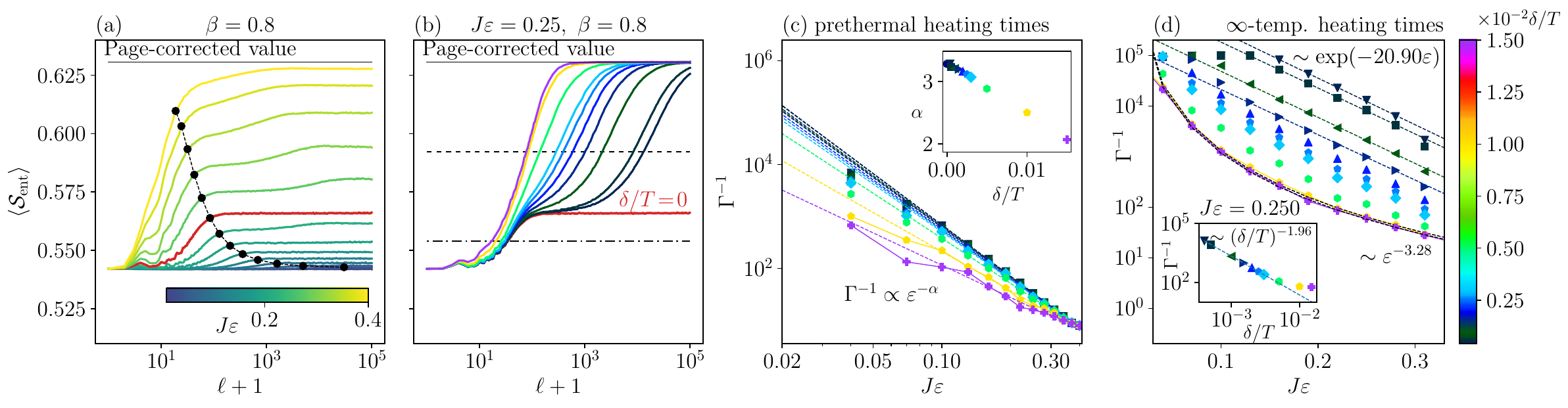}
	\caption{\label{fig:scaling_case_2a_3} Heating behaviour of a thermal initial state (w.r.t.~$H_\mathrm{eff}\! \approx \!H_3/2$) for the driven Ising model without quantum fluctuations, shown by the entanglement entropy density of the half chain: 
	\textbf{(a)} Time evolution for a selection of $\varepsilon $ values [cf.~colorbar]. The black dots connected by the dashed line indicate the timescale necessary to reach the prethermal plateau.
	\textbf{(b)} effect of finite noise strength $\delta$ on the time evolution at $J\varepsilon=0.25$. The dashed and dashed-dotted line are used to extract relevant heating times as described in the text. 
	\textbf{(c)} Prethermal heating times extracted from the horizontal dashed-dotted line in (b) as a function of $\varepsilon$ for different values of $\delta/T$. The dashed lines in (c) show the corresponding least-square fits: the slopes determine the scaling of heating times with $\delta$, shown in the inset in (c). 
	\textbf{(d)} Infinite-temperature heating times extracted from the horizontal dashed line in (b). The dashed lines represent the corresponding least-square fits: the slopes determine the scaling of heating times with $\delta$, shown in the inset in (d).
	The simulation parameters are the same as for Fig.~\ref{fig:E_vs_ell_pure}, except for $L=16$. 
}
\end{figure*}

Figure~\ref{fig:data_TFIM}b shows that the system prethermalizes approximately for small values of $\varepsilon$, as becomes evident from the eigenvalues of the reduced density matrix. Moreover, for the given initial energy density, we find that ETH is satisfied, provided PBCs are applied to the subsystem effective Hamiltonian $H_{\mathrm{eff}}^A$ [Fig.~\ref{fig:data_TFIM}c] (note that significant deviations appear when OBCs are applied, yet they cannot survive in the thermodynamic limit). Similar as compared to the nonintegrable drive generated by $H_1$, also here our findings reach beyond those of ETH as the inverse temperature follows the theoretical prediction obtained from Eq.~\eqref{eq:beta_of_E}. However, in contrast to the nonintegrable drive $H_1$, in the transition regime, $\varepsilon\sim 10^{-3}\div 10^{-2}$, before reaching infinite temperature, the eigenvalues of the reduced density matrix show significant deviations from the expected exponential dependence, which survive with increasing the subsystem size [cf.~App.~\ref{app:int} and Fig.~\ref{fig:data_TFIM_2_0} upper row]. A plausible explanation for this behavior is that the state of the system is not fully thermal in this $\varepsilon$-regime [cf.~App.~\ref{app:errors}]. This implies that, for the Floquet dynamics generated by $H_2$, a substantial number of states in the Hilbert space are restrained from participating in the thermalization process even for $L=20$ spins [cf. App.~\ref{app:int}, Fig.~\ref{fig:scaling_2c_1}]. 

Interestingly, the thermal character of the state in the transition regime can be restored by adding a small noise $\delta $ to the driving period which breaks periodicity [cf.~Sec.~\ref{sec:nonintegrable}] [cf.~App.~\ref{app:int}, Fig.~\ref{fig:data_TFIM_2_0} lower panel]. Importantly, $\delta>0$ results in a thermal state well before the system reaches infinite temperature: the fit values for $\beta$ shift systematically towards the ones obtained from Eq.~\eqref{eq:beta_of_E} using the instantaneous energy densities [cf. App.~\ref{app:int}, Fig.~\ref{fig:data_TFIM_2}(b-d)]. This corroborates our Hypothesis also for integrable Hamiltonians [cf.~Fig.~\ref{fig:data_TFIM_2}(a-d)].
Remarkably, finite noise restores ergodicity only in the unconstrained thermalization stage (III) of the dynamics between the prethermal regime and the featureless  infinite-temperature state at long times; it hardly affects the prethermal properties of the dynamics, e.g.~the expectation values of observables, and the time required to leave the prethermal plateau [Fig.~\ref{fig:data_TFIM}d]. In turn, this implies that the relevant effective Hamiltonian is not drastically affected by the addition of small noise.

\subsection{\label{subsec:clIM}Ising Model without Quantum Fluctuations}

Last, let us discuss an Ising drive without quantum fluctuations, modeled by the Hamiltonian
\begin{eqnarray}
\label{Eq:Ising3}
H_3= \sum_{j} J \sigma^z_{j+1}\sigma^z_j+ h_z \sigma^z_j,
\end{eqnarray}
where $J=1.0$ and $h_z=0.809$. Quantum fluctuations in the driven system are introduced by the kicks $V$, cf.~Eq.~\eqref{eq:kick}, so that the leading-order approximation to $H_F(\varepsilon)$ is non-integrable~\cite{Talia2019}. In this section, we consider a thermal initial state at $\beta(\ell=0)=0.8$ [cf.~Sec.~\ref{subsec:thermal}]. 

The results below can be summarized in the following two points: 
(i) we provide numerical evidence that the drive generated by the Hamiltonian of Eq.~\eqref{Eq:Ising3} does not obey a Fermi Golden Rule scaling for the heating times. Instead the heating times cross over from a power-law scaling with an anomalous exponent $\alpha$ at large perturbation strength [cf.~Sec.~\ref{subsec:noise}] to an exponential scaling for infinitesimal noise strengths.
(ii) an IFE based on the Replica trick allows us to compute higher-order corrections to the effective Hamiltonian. We demonstrate that these higher order corrections are important to capture the physics in the vicinity of $T^\ast_k$. 

Fig.~\ref{fig:scaling_case_2a_3}a displays the time evolution of the entanglement entropy density for the dynamics generated by $H_3$ using the kicks from Eq.~\eqref{eq:UF_ast}. Already from this figure it becomes evident that $H_3$ behaves quite different as compared to $H_1$ and $H_2$: Instead of showing one stable prethermal level for different $\varepsilon$ [as is the case for $H_1$, cf.~Fig.~\ref{fig:E_vs_ell_pure}], here, different $\varepsilon$ values result in different saturation levels at prethermal times (we checked that these plateaus are not a finite-size effects, see App.~\ref{app:case_2a}, Fig.~\ref{fig:finite_size_case_2a}). Moreover, the required $\varepsilon$ values to observe prethermal dynamics for $H_3$, $\varepsilon \sim 10^{-1}$, are about two orders of magnitude larger compared to the previous two drives $H_1, H_2$, which correlates with the lack of quantum fluctuations in $H_3$. 

The varying saturation levels of the prethermal plateau complicate extracting the heating times. Yet, it is easy to recognize that the dynamics features two times scales: the first one captures the time needed to reach the prethermal plateau; it carries a clear dependence on $\varepsilon$ as evident from Fig.~\ref{fig:scaling_case_2a_3}a (black dots connected by solid line). The second describes the time required to heat up to infinite temperature [Fig.~\ref{fig:scaling_case_2a_3}b]. For the periodic perturbation-free dynamics ($\delta=0$), the second time scales is intractable within the evolution time range of our simulations, which points to a much longer heating time. Interestingly, as opposed to the previously discussed drives, finite periodicity-breaking noise leads to a significant reduction of the heating times so that a clear pattern becomes tractable.

To separate well the two timescales from each other, we apply Eq.~\eqref{Eq:scaling_values} iteratively: first, we replace $O_{\mathrm{prethermal}}$ by the initial expectation value and $O_{\beta=0}$ by the prethermal value. The solution to the corresponding equation provides an estimate $\ell_{p}$ for the time required to reach the prethermal plateau, i.e., the timescale of constrained thermalization, cf.~the time required to cross the dashed-dotted horizontal line in Fig.~\ref{fig:scaling_case_2a_3}b. 
Independently, we attempt to solve Eq.~\eqref{Eq:scaling_values} once again, yet this time we replace $O_{\mathrm{prethermal}}$ with $O(\ell_p)$, which yields the time scale for unconstrained thermalization, cf.~the time required to cross the dashed horizontal line in Fig.~\ref{fig:scaling_case_2a_3}b. The results of this analysis are depicted in Fig.~\ref{fig:scaling_case_2a_3}(c-d). The constrained thermalization time is clearly described by a power-law which survives a finite weak noise strength $\delta>0$, see Fig.~\ref{fig:scaling_case_2a_3}c, inset. In contrast, the unconstrained heating time follows an exponential law over two decades for small $\delta/T$, i.e., $\Gamma^{-1}\sim \exp(-\xi \epsilon)$. We note that the exponential scaling clearly cannot persist as $\varepsilon \rightarrow 0$, since this would imply finite heating times at $\varepsilon=0$, where heating is inhibited by the restored energy conservation. Increasing the noise strength $\delta$ leads to increasingly shorter heating times until eventually the prethermal plateau disappears and, therefore, the scaling crosses over to the power-law scaling of the unconstrained thermalization timescale.

These findings are intriguing, because they imply the existence of refined estimates for the scaling of the heating rates with $\varepsilon$ [so far, Floquet systems have mostly been treated on equal footing to derive a generic upper bound on the hating rate~\cite{abanin_15, mori_15}]. One can even speculate about the existence of a wider class of Floquet models with suppressed heating behavior.

\subsubsection*{\label{subsubsec:replica}Replica Resummation and Thermalization}

The significant change of the prethermal plateau level within the range of $\varepsilon$ values we investigate [Fig.~\ref{fig:scaling_case_2a_3}a], implies that higher order corrections (in $\varepsilon$) to the effective Hamiltonian are of increased importance for understanding the dynamics of the system (as opposed to the models discussed in Secs.~\ref{sec:nonintegrable} and \ref{subsec:TFIM}).
The drive $H_3$ was chosen to allow for an analytical treatment of the leading-order correction to the average Hamiltonian using the replica expansion~\cite{vajna2018replica}, cf.~App.~\ref{app:analytical}. Note that the replica trick is needed at the commensurate points $k>0$ for the kicked system, where higher-order nested commutator terms in the Baker-Campbell-Hausdorff series also contain terms to first order in $\varepsilon$, and hence one is required to re-sum an infinite subseries to correctly identify the first-order correction. 

\begin{figure}
	\includegraphics[scale=0.5]{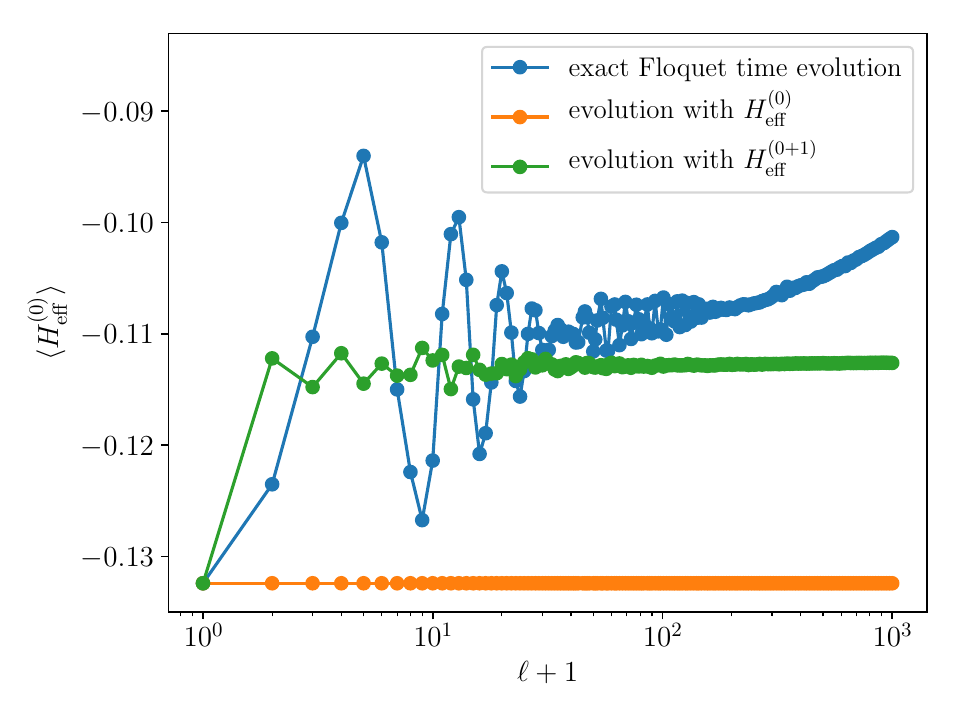}
	\caption{\label{fig:scaling_case_2a} Time evolution using $H_3$ of a thermal initial state (w.r.t.~$H_\mathrm{eff}^{(0)}\!=\!H_3/2$) at $\beta(\ell\!=\!0)=0.8$ according to Eq.~\eqref{eq:rho_th_full} using (i) the exact Floquet driving protocol (blue), (ii) the effective Hamiltonian in zeroth order of $\varepsilon$ (orange), and (iii) the effective Hamiltonian in first order $\varepsilon$ (green). The parameters are the same as in Fig.~\ref{fig:E_vs_ell_pure}, except for $J=0.6$ and $J\varepsilon=0.0943$.
}
\end{figure}

To facilitate the analytical computation, we switch back to a two-step protocol:
\begin{eqnarray}
U_F(2(T^{\ast}_k+\varepsilon) )= \mathrm{e}^{-iT H_3/2}\mathrm{e}^{-i\varepsilon V}.
\end{eqnarray}
Then, for $k>0$, the replica expansion gives [cf.~App.~\ref{app:analytical}]:

\begin{figure*}[t!]
	\includegraphics[width=1\textwidth]{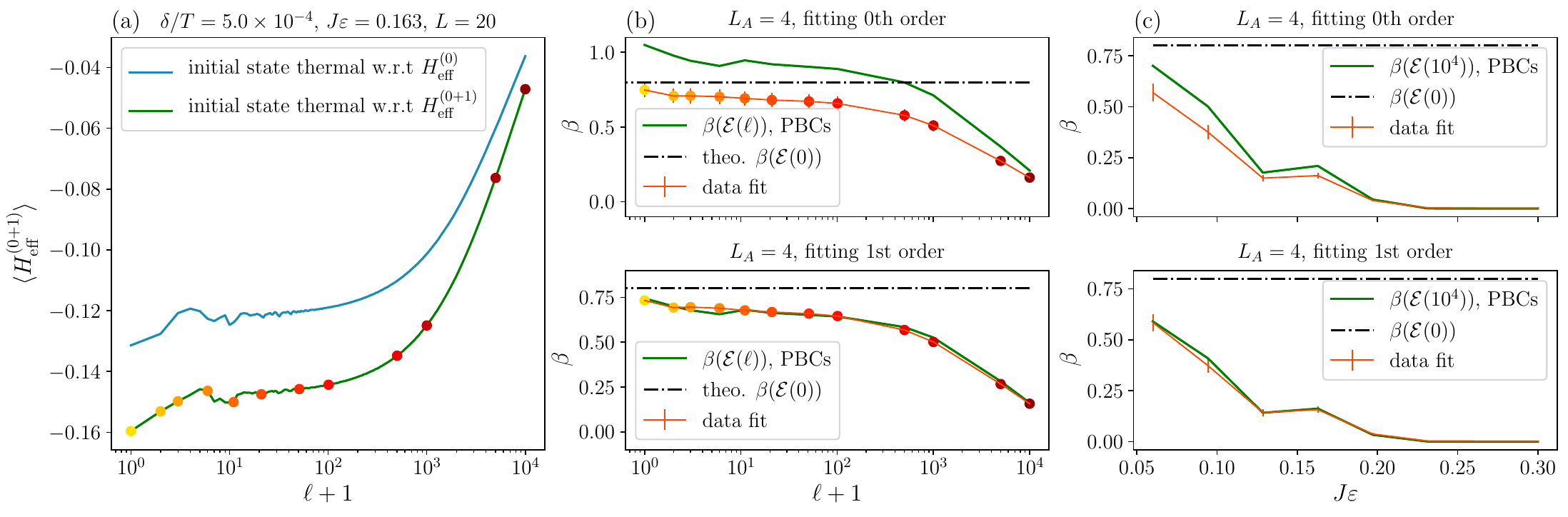}
	\caption{\label{fig:thermal_case_2a} Heating behavior for $H_3$ starting from a thermal initial state: 
		\textbf{(a)} stroboscopic time evolution for fixed $J\varepsilon=0.163$, where the corresponding initial state is thermal w.r.t. $H_{\mathrm{eff}}^{(0)}$ (blue), and $H_{\mathrm{eff}}^{(0+1)}$ (green) at $\beta(\ell\!=\!0)=0.8$. 
		\textbf{(b)} inverse temperature $\beta$ as a function of stroboscopic driving times $\ell$: The solid red line with error bars marks the values extracted from the spectrum of the reduced density matrix for a subsystem size $L_A=4$. Thereby, we use instantaneous energies $\mathcal{E}(\ell)$ from (a), where the initial state is thermal w.r.t $H_{\mathrm{eff}}^{(0+1)}$. The colored dots serve as a guide to the eye. The solid green line is the solution $\beta(\mathcal{E}(\ell))$ to Eq.~\eqref{eq:beta_of_E}. The horizontal dashed-dotted line marks the ETH prediction for the prethermal plateau.
		In the upper panel, we apply Eq.~\eqref{eq:beta_of_E} using $H_\mathrm{eff}^{(0)}$, while in the lower panel we use $H_\mathrm{eff}^{(0+1)}$. 
		\textbf{(c)} Same as (b), but for a fixed $\ell=10^4$ as a function of different values of $\varepsilon$. Furthermore, to avoid divergences in the first order replica expansion of $H_\mathrm{eff}$, we set $J=0.6$. The remaining parameters are as in Fig.~\ref{fig:E_vs_ell_pure}.
	}
\end{figure*}

\begin{eqnarray}
H_\mathrm{eff} &=& H_\mathrm{eff} ^{(0)} + H_\mathrm{eff} ^{(1)} + \mathcal{O}(\varepsilon^2)\equiv H_{\mathrm{eff}}^{(0+1)}+\mathcal{O}(\varepsilon^2), \nonumber\\
H_\mathrm{eff} ^{(0)} &=& \frac{1}{2}H_3, \nonumber \\
H_\mathrm{eff} ^{(1)} &=& \!\frac{\varepsilon \gamma}{2}\bigg[J\sum_j (\sigma^z_j\sigma_{j+1}^y+\sigma^y_j\sigma_{j+1}^z)\nonumber\\
&+&\! \frac{1}{4}\!\left[f_{2J\!+\! h_z}(T)\!-\! f_{2J-h_z}(T)\right]\sum_j \sigma^x_j\sigma_{j+1}^z+\sigma^z_j\sigma_{j+1}^x\nonumber\\
&+& \! \frac{1}{4}\!\left[f_{2J\! +\!  h_z}\! ( T )\!+\! f_{2\! J\! - \!h_z}\!( T )\! -\! 2\! f_{h_z}\!( T )\!\right]\sum_j \sigma^z_{j\! -\! 1}\sigma_j^x\sigma_{j\! +\!1}^z\nonumber\\
&+&\! \frac{1}{4}\!\left[f_{2J\! + \! h_z}(T)\!+\! f_{2J\!-\! h_z}(T)\!+\! 2f_{h_z}(T)\right]\sum_j \sigma^x_{j}\nonumber\\
&+&\! h_z\sum_j \sigma^y_{j}\bigg],
\label{eq:replica_H3}
\end{eqnarray}
where $f_{\chi}(T)=\chi \cot(\chi T/2)$. Note that $f_{\chi}$ carries the only $k$-dependence up to first order via its argument $T=2(T^{\ast}_k+\varepsilon) $. Since it constitutes a periodic function, $f_\chi(T+2n\pi/\chi)=f_{\chi}(T)$, in the regime $\pi/\gamma \approx n\pi/\chi$, this implies a very weak dependence of the effective Hamiltonian on the value of $k$ to first order in $\varepsilon$, consistent with our numerical observations. 

Note that the cotangent function present in the first-order terms, can lead to divergences, which likely persist also in higher-order terms~\cite{vajna2018replica}. Indeed, for the range of $\varepsilon$ values we discuss, it turns out that $J=1.0$ might occasionally lead to a divergence as we vary $\varepsilon$. Thus, subsequently we set $J=0.6$.

More importantly, notice that $H_\mathrm{eff} ^{(0)}\sim  \mathcal{O}(1)$ is integrable, while $H_\mathrm{eff} ^{(1)} \sim  \mathcal{O}(\varepsilon)$ is nonintegrable, and hence the exact Floquet Hamiltonian is (most likely) also nonintegrable, which explains the origin of slow thermalization. Indeed, we observed that a number of pure initial states do not thermalize well within the accessible system sizes and evolution cycles, which is why we choose to prepare the system in a thermal initial state.

Figure \ref{fig:scaling_case_2a} shows the importance of the first-order corrections to properly capture the dynamics of the prethermal plateau for the Hamiltonian $H_3$.  The system is prepared in a thermal initial state w.r.t.~$H_\mathrm{eff}^{(0)}$, and two quantum quenches are performed: a quench to $H_{\mathrm{eff}}^{(0+1)}$ results in dynamics which saturates closer to the prethermal plateau obtained from the exact time evolution, as compared to a quench to $H_{\mathrm{eff}}^{(0)}$ (which obeys no dynamics by construction).

The above check exhibits only the most obvious implication of the first-order correction. More subtly, higher-order corrections to the effective Hamiltonian also improve the description of the thermal state at all stages of the time evolution. To demonstrate this, we investigate the dynamics of two different thermal initial states: (i) an initial state, thermal w.r.t.~$H_\mathrm{eff}^{(0)}$, and (ii) another initial state, thermal w.r.t.~$H_\mathrm{eff}^{(0+1)}$ [cf.~Fig.~\ref{fig:thermal_case_2a}a]. In the vicinity of $T_k^\ast$, we analyze the thermal properties of the associated reduced density matrix along the way up to infinite temperature. 
Because we want to compare the two leading-order corrections, we now have two ways of computing an associated inverse temperature: we can either use $H_\mathrm{eff}^{(0)}$ or $H_\mathrm{eff}^{(0+1)}$. 
Using $H_\mathrm{eff}^{(0+1)}$, we consistently obtain a very good agreement of the fitted  and the ETH-predicted inverse temperature for various values of $\varepsilon$ and different times $\ell$ [Fig.~\ref{fig:thermal_case_2a}(b-c) lower panel]. By contrast, using $H_\mathrm{eff}^{(0)}$ produces sizable deviations in the fitted values vs.~the ETH predictions, until the system is close enough to the infinite-temperature state [Fig.~\ref{fig:thermal_case_2a}(b-c) upper panel]. We note in passing that the deviations we observe between the fitted and the ETH-predicted inverse temperature can be observable-dependent [cf.~App. \ref{app:case_2a}].

Considered altogether, the above analysis implies that $H_\mathrm{eff}^{(0)}$ is not the correct effective Hamiltonian the system thermalizes to, in both the prethermal plateau and the subsequent unconstrained thermalization regime. Instead, higher-order corrections are indeed required to properly capture the thermalizing Floquet dynamics generated by $H_3$, and the Replica trick provides a useful expansion to compute them. Moreover, despite the different scaling of the heating times, even for nonintegrable effective Hamiltonians which obey ETH, we see that the IFE, supplemented with the instantaneous energy density, can provide a good description of the thermalizing dynamics of the evolved state throughout the entire evolution cycle all the way up to infinite temperature.

Interestingly, the Replica expansion also proves useful to better understand the sensitivity of the heating times to the perturbation strength in the case of noisy drives. As for $H_1$ and $H_2$, the leading-order effective Hamiltonian $H_\mathrm{eff}^{(0)}$ only acquires an overall multiplicative factor
\begin{eqnarray}
H_{\mathrm{eff}}^{(0)}\rightarrow H_{\mathrm{eff}}^{(0)}\left(1+\frac{\delta}{2T}\right).
\end{eqnarray}
Given the applicability of ETH, this translates into small energy density, and with it -- to temperature fluctuations of the order of $\delta/(2T)$. However, the first order (in $\varepsilon$) correction $H_{\mathrm{eff}}^{(1)}$ is subject to more drastic changes. To linear order in $\delta$, we find
\begin{eqnarray}
f_{\chi}(T)\rightarrow f_\chi(T)\left(1-\frac{ \chi\delta}{\sin(\chi T)}\right).
\end{eqnarray}
Besides the fact that ${\chi\delta}/{\sin(\chi T)}$ can become large, finite $\delta/T$ also changes the relative weights of the different terms appearing in $H_{\mathrm{eff}}^{(1)}$, cf.~Eq.~\eqref{eq:replica_H3}. This results in applying a substantially different effective Hamiltonian for each period of the drive; thus, the resulting effect on the dynamics cannot be interpreted as small temperature fluctuations since also the eigenspectrum of $H_{\mathrm{eff}}$ is subject to non-negligible changes from one period to the next. 
Hence, due to lack of periodicity in the drive, we can no longer define prethermalization w.r.t. $H_{\mathrm{eff}}^{(0+1)}$; as a result, we observe an increased sensitivity to the noise strength $\delta$.

\section{\label{sec:discussion}Conclusion}

In summary, we investigated a class of step-driven Floquet systems, with the help of which it is possible to extend the high-frequency prethermal physics to low drive frequencies, on the order of the couplings in the Hamiltonian. The dynamics in this class features a frequency axis which, by construction, contains isolated stable points, where energy is conserved exactly. We demonstrated that these points come with prethermal regimes, whose width as a function of the deviation $\varepsilon$ from the commensurate point, varies between a power-law and an exponential, depending on the drive. Surprisingly, Fermi's Golden Rule is not universally applicable.
Intriguing open questions are whether one can bridge these windows to enhance stability, and whether there exist models with a single continuous stable window all the way down to the low frequency regime. 

Throughout the paper, we studied the thermalization behavior of three integrable and nonintegrable drives $H_j$: the mixed-field Ising drive and the transverse-field Ising drive generate Floquet dynamics with quadratic in $\varepsilon$ heating rates that follow Fermi's Golden Rule; in contrast, the Ising drive without quantum fluctuations, exhibits a more refined heating behaviour that violates the Fermi Golden Rule scaling. Thus, intuition based on the equilibrium integrable-nonintegrable classification does not carry over to non-equilibrium systems in a straightforward manner; instead, we find an interesting correlation between whether Floquet heating is non-quadratically or quadratically suppressed in $\varepsilon$ on the one hand, and whether it is feasible to re-sum a subseries of the inverse-frequency expansion, on the other. 

During the study, we introduced a new technique to facilitate Floquet systems to explore the entire underlying Hilbert space: 
we apply small random perturbation/noise in the duration of the Hamiltonian $H_j$. We showed that this procedure minimizes finite-size effects and allows us to look for a proper parametric dependence of the heating rates in the curves for the time evolution of physical quantities. Additionally, we observed that the $k>0$ commensurate points also enhance the ergodic properties of Floquet systems, as compared to the infinite-frequency point $k=0$, since the smaller drive frequencies at $k>0$ provide the required spectrum folding for Floquet many-body resonances to occur. These technical advances allowed us to obtain clean scaling of the numerical data required for a proper study of thermalization in finite-size many-body systems, cf.~App.~\ref{app:L-dep}. 
As long as the prethermal physics is described by the leading-order in $\varepsilon$ effective Hamiltonian, finite noise strengths do not reduce the duration of the associated prethermal plateau, as they effectively translate (in accord with ETH) into small temperature fluctuations.
In contrast, when higher order corrections in $\varepsilon$ become important for the thermalization dynamics, the prethermal physics becomes sensitive to the noise strength.

We also provided numerical evidence in favor of the following conjecture: 
Consider a pure state subject to a periodic drive generated by a nonintegrable Floquet Hamiltonian, whose dynamics features a prethermal plateau. We observe that, upon leaving this plateau, a subsystem of the original system remains (to an excellent approximation) thermal w.r.t~an effective Hamiltonian as defined by the IFE [to the order it can be computed/defined]. Although the expansion represents a divergent asymptotic series and does not capture the heating process itself, given the instantaneous value of the energy density, $H_\mathrm{eff}$ is sufficient to construct a thermal ensemble which captures the dynamics of the system as it continues to heat up. The temperature of the subsystem increases gradually with time, and can be obtained from the energy density w.r.t.~$H_\mathrm{eff}$.
In contrast to this behavior, whenever the system heats up straight to infinite temperature (i.e., no prethermal plateau can form), we distinguish two scenarios: 
(i) in the high driving frequency limit ($k=0$) the system is not in a thermal state until it reaches infinite temperature. 
(ii) for finite intermediate frequencies ($k>0$), the system can still (approximately) thermalize w.r.t.~the ergodic $H_j$ but only if the corresponding intrinsic thermalization timescale for $H_j$ is smaller than $T^\ast_k$. 
It remains open whether the crossover as a function of $\varepsilon$ between a (pre-)thermal state w.r.t.~$H_\mathrm{eff}$, and a non-thermal state can become a sharp transition, and what conditions would be required for this to happen. Such a behavior could be detectable in the behavior of the thermodynamic entropy of the system which is maximal for a thermal state and smaller for any other state.

Our study also bears relevance to experiments. Recently, Floquet prethermal physics has been observed in driven cold atomic systems~\cite{singh2019quantifying,abadal2020floquet}. A straightforward application of our analysis is to use periodic drives to continuously fine-tune the temperature of cold-atomic systems in time. This could be used, e.g.~to trigger temperature-driven phase transitions, such as the Mott insulator transition or the Kosterlitz-Thouelss transition. Related ideas can potentially prove useful to design a new temperature knob in quantum simulators which are well isolated from external reservoirs.
Finally, we mention that it may soon be within the scope of present day quantum gas microscopes to reconstruct the density matrix of a subsystem via density matrix tomography, and verify or negate the conclusions and predictions of our work. The main quantity -- the reduced diagonal density matrix -- defines a natural measurement ensemble in experiments~\cite{neill2016ergodic}.  


\emph{Acknowledgments.---}We wish to thank J.~Bardarson, A.~Das, W.~W.~Ho,  F.~Huveneers, V.~Khemani, A.~Polkovnikov, F.~Pollmann, T.~Prosen, W.~De Roeck, B.~Trauzettel and P.~Weinberg for valuable discussions. 
C. F. acknowledges financial support from the DFG (SPP1666, SFB1170 ToCoTronics), the Wüzburg-Dresden Cluster of Excellence ct.qmat, EXC2147, project-id 39085490 and the Elitenetzwerk Bayern Graduate School on Topological insulators and the ERC Starting Grant No. 679722.
M.B.~was supported by the U.S. Department of Energy, Office of Science, Office of Advanced Scientific Computing Research, under the Accelerated Research in Quantum Computing (ARQC) program, the U.S. Department of Energy under cooperative research agreement DE-SC0009919, the Emergent Phenomena in Quantum Systems initiative of the Gordon and Betty Moore Foundation, and the Bulgarian National Science Fund within National Science Program VIHREN, contract number KP-06-DV-5.
This research was supported in part by the International Centre for Theoretical Sciences (ICTS) during a visit for participating in the program -  Thermalization, Many body localization and Hydrodynamics (Code: ICTS/hydrodynamics2019/11).
We used \href{https://github.com/weinbe58/QuSpin#quspin}{Quspin} for simulating the dynamics of the quantum systems~\cite{weinberg2017quspin,weinberg2019quspin}.
The authors are pleased to acknowledge that the computational work reported on in this paper was performed on the Shared Computing Cluster which is administered by Boston University’s Research Computing Services and on the W\"urzburg HPC cluster.

\bibliographystyle{apsrev4-1}
\bibliography{./bibliography}

\begin{thebibliography}{120}%
\makeatletter
\providecommand \@ifxundefined [1]{%
 \@ifx{#1\undefined}
}%
\providecommand \@ifnum [1]{%
 \ifnum #1\expandafter \@firstoftwo
 \else \expandafter \@secondoftwo
 \fi
}%
\providecommand \@ifx [1]{%
 \ifx #1\expandafter \@firstoftwo
 \else \expandafter \@secondoftwo
 \fi
}%
\providecommand \natexlab [1]{#1}%
\providecommand \enquote  [1]{``#1''}%
\providecommand \bibnamefont  [1]{#1}%
\providecommand \bibfnamefont [1]{#1}%
\providecommand \citenamefont [1]{#1}%
\providecommand \href@noop [0]{\@secondoftwo}%
\providecommand \href [0]{\begingroup \@sanitize@url \@href}%
\providecommand \@href[1]{\@@startlink{#1}\@@href}%
\providecommand \@@href[1]{\endgroup#1\@@endlink}%
\providecommand \@sanitize@url [0]{\catcode `\\12\catcode `\$12\catcode
  `\&12\catcode `\#12\catcode `\^12\catcode `\_12\catcode `\%12\relax}%
\providecommand \@@startlink[1]{}%
\providecommand \@@endlink[0]{}%
\providecommand \url  [0]{\begingroup\@sanitize@url \@url }%
\providecommand \@url [1]{\endgroup\@href {#1}{\urlprefix }}%
\providecommand \urlprefix  [0]{URL }%
\providecommand \Eprint [0]{\href }%
\providecommand \doibase [0]{http://dx.doi.org/}%
\providecommand \selectlanguage [0]{\@gobble}%
\providecommand \bibinfo  [0]{\@secondoftwo}%
\providecommand \bibfield  [0]{\@secondoftwo}%
\providecommand \translation [1]{[#1]}%
\providecommand \BibitemOpen [0]{}%
\providecommand \bibitemStop [0]{}%
\providecommand \bibitemNoStop [0]{.\EOS\space}%
\providecommand \EOS [0]{\spacefactor3000\relax}%
\providecommand \BibitemShut  [1]{\csname bibitem#1\endcsname}%
\let\auto@bib@innerbib\@empty
\bibitem [{\citenamefont {Goldman}\ and\ \citenamefont
  {Dalibard}(2014)}]{goldman2014periodically}%
  \BibitemOpen
  \bibfield  {author} {\bibinfo {author} {\bibfnamefont {N.}~\bibnamefont
  {Goldman}}\ and\ \bibinfo {author} {\bibfnamefont {J.}~\bibnamefont
  {Dalibard}},\ }\href {\doibase 10.1103/PhysRevX.4.031027} {\bibfield
  {journal} {\bibinfo  {journal} {Phys. Rev. X}\ }\textbf {\bibinfo {volume}
  {4}},\ \bibinfo {pages} {031027} (\bibinfo {year} {2014})}\BibitemShut
  {NoStop}%
\bibitem [{\citenamefont {Goldman}\ \emph {et~al.}(2015)\citenamefont
  {Goldman}, \citenamefont {Dalibard}, \citenamefont {Aidelsburger},\ and\
  \citenamefont {Cooper}}]{goldman2015case}%
  \BibitemOpen
  \bibfield  {author} {\bibinfo {author} {\bibfnamefont {N.}~\bibnamefont
  {Goldman}}, \bibinfo {author} {\bibfnamefont {J.}~\bibnamefont {Dalibard}},
  \bibinfo {author} {\bibfnamefont {M.}~\bibnamefont {Aidelsburger}}, \ and\
  \bibinfo {author} {\bibfnamefont {N.~R.}\ \bibnamefont {Cooper}},\ }\href
  {\doibase 10.1103/PhysRevA.91.033632} {\bibfield  {journal} {\bibinfo
  {journal} {Phys. Rev. A}\ }\textbf {\bibinfo {volume} {91}},\ \bibinfo
  {pages} {033632} (\bibinfo {year} {2015})}\BibitemShut {NoStop}%
\bibitem [{\citenamefont {Eckardt}(2017)}]{eckardt2017atomic}%
  \BibitemOpen
  \bibfield  {author} {\bibinfo {author} {\bibfnamefont {A.}~\bibnamefont
  {Eckardt}},\ }\href {\doibase 10.1103/RevModPhys.89.011004} {\bibfield
  {journal} {\bibinfo  {journal} {Rev. Mod. Phys.}\ }\textbf {\bibinfo {volume}
  {89}},\ \bibinfo {pages} {011004} (\bibinfo {year} {2017})}\BibitemShut
  {NoStop}%
\bibitem [{\citenamefont {Bukov}\ \emph
  {et~al.}(2015{\natexlab{a}})\citenamefont {Bukov}, \citenamefont
  {D'Alessio},\ and\ \citenamefont {Polkovnikov}}]{bukov2015universal}%
  \BibitemOpen
  \bibfield  {author} {\bibinfo {author} {\bibfnamefont {M.}~\bibnamefont
  {Bukov}}, \bibinfo {author} {\bibfnamefont {L.}~\bibnamefont {D'Alessio}}, \
  and\ \bibinfo {author} {\bibfnamefont {A.}~\bibnamefont {Polkovnikov}},\
  }\href {http://www.tandfonline.com/doi/full/10.1080/00018732.2015.1055918}
  {\bibfield  {journal} {\bibinfo  {journal} {Advances in Physics}\ }\textbf
  {\bibinfo {volume} {64}},\ \bibinfo {pages} {139} (\bibinfo {year}
  {2015}{\natexlab{a}})}\BibitemShut {NoStop}%
\bibitem [{\citenamefont {Rodriguez-Vega}\ \emph {et~al.}(2020)\citenamefont
  {Rodriguez-Vega}, \citenamefont {Vogl},\ and\ \citenamefont
  {Fiete}}]{rodriguez2020moir}%
  \BibitemOpen
  \bibfield  {author} {\bibinfo {author} {\bibfnamefont {M.}~\bibnamefont
  {Rodriguez-Vega}}, \bibinfo {author} {\bibfnamefont {M.}~\bibnamefont
  {Vogl}}, \ and\ \bibinfo {author} {\bibfnamefont {G.~A.}\ \bibnamefont
  {Fiete}},\ }\href {https://arxiv.org/abs/2011.11079} {\bibfield  {journal}
  {\bibinfo  {journal} {arXiv preprint arXiv:2011.11079}\ } (\bibinfo {year}
  {2020})}\BibitemShut {NoStop}%
\bibitem [{\citenamefont {G{\"o}rg}\ \emph {et~al.}(2018)\citenamefont
  {G{\"o}rg}, \citenamefont {Messer}, \citenamefont {Sandholzer}, \citenamefont
  {Jotzu}, \citenamefont {Desbuquois},\ and\ \citenamefont
  {Esslinger}}]{gorg2018enhancement}%
  \BibitemOpen
  \bibfield  {author} {\bibinfo {author} {\bibfnamefont {F.}~\bibnamefont
  {G{\"o}rg}}, \bibinfo {author} {\bibfnamefont {M.}~\bibnamefont {Messer}},
  \bibinfo {author} {\bibfnamefont {K.}~\bibnamefont {Sandholzer}}, \bibinfo
  {author} {\bibfnamefont {G.}~\bibnamefont {Jotzu}}, \bibinfo {author}
  {\bibfnamefont {R.}~\bibnamefont {Desbuquois}}, \ and\ \bibinfo {author}
  {\bibfnamefont {T.}~\bibnamefont {Esslinger}},\ }\href {\doibase
  https://doi.org/10.1038/nature25135} {\bibfield  {journal} {\bibinfo
  {journal} {Nature}\ }\textbf {\bibinfo {volume} {553}},\ \bibinfo {pages}
  {481} (\bibinfo {year} {2018})}\BibitemShut {NoStop}%
\bibitem [{\citenamefont {Struck}\ \emph {et~al.}(2013)\citenamefont {Struck},
  \citenamefont {Weinberg}, \citenamefont {{\"O}lschl{\"a}ger}, \citenamefont
  {Windpassinger}, \citenamefont {Simonet}, \citenamefont {Sengstock},
  \citenamefont {H{\"o}ppner}, \citenamefont {Hauke}, \citenamefont {Eckardt},
  \citenamefont {Lewenstein},\ and\ \citenamefont {Mathey}}]{struck_13}%
  \BibitemOpen
  \bibfield  {author} {\bibinfo {author} {\bibfnamefont {J.}~\bibnamefont
  {Struck}}, \bibinfo {author} {\bibfnamefont {M.}~\bibnamefont {Weinberg}},
  \bibinfo {author} {\bibfnamefont {C.}~\bibnamefont {{\"O}lschl{\"a}ger}},
  \bibinfo {author} {\bibfnamefont {P.}~\bibnamefont {Windpassinger}}, \bibinfo
  {author} {\bibfnamefont {J.}~\bibnamefont {Simonet}}, \bibinfo {author}
  {\bibfnamefont {K.}~\bibnamefont {Sengstock}}, \bibinfo {author}
  {\bibfnamefont {R.}~\bibnamefont {H{\"o}ppner}}, \bibinfo {author}
  {\bibfnamefont {P.}~\bibnamefont {Hauke}}, \bibinfo {author} {\bibfnamefont
  {A.}~\bibnamefont {Eckardt}}, \bibinfo {author} {\bibfnamefont
  {M.}~\bibnamefont {Lewenstein}}, \ and\ \bibinfo {author} {\bibfnamefont
  {L.}~\bibnamefont {Mathey}},\ }\href
  {http://www.nature.com/nphys/journal/v9/n11/full/nphys2750.html} {\bibfield
  {journal} {\bibinfo  {journal} {Nature Physics}\ }\textbf {\bibinfo {volume}
  {9}},\ \bibinfo {pages} {738} (\bibinfo {year} {2013})}\BibitemShut {NoStop}%
\bibitem [{\citenamefont {Aidelsburger}\ \emph {et~al.}(2013)\citenamefont
  {Aidelsburger}, \citenamefont {Atala}, \citenamefont {Lohse}, \citenamefont
  {Barreiro}, \citenamefont {Paredes},\ and\ \citenamefont
  {Bloch}}]{aidelsburger_13}%
  \BibitemOpen
  \bibfield  {author} {\bibinfo {author} {\bibfnamefont {M.}~\bibnamefont
  {Aidelsburger}}, \bibinfo {author} {\bibfnamefont {M.}~\bibnamefont {Atala}},
  \bibinfo {author} {\bibfnamefont {M.}~\bibnamefont {Lohse}}, \bibinfo
  {author} {\bibfnamefont {J.~T.}\ \bibnamefont {Barreiro}}, \bibinfo {author}
  {\bibfnamefont {B.}~\bibnamefont {Paredes}}, \ and\ \bibinfo {author}
  {\bibfnamefont {I.}~\bibnamefont {Bloch}},\ }\href
  {http://link.aps.org/doi/10.1103/PhysRevLett.111.185301} {\bibfield
  {journal} {\bibinfo  {journal} {Phys. Rev. Lett.}\ }\textbf {\bibinfo
  {volume} {111}},\ \bibinfo {pages} {185301} (\bibinfo {year}
  {2013})}\BibitemShut {NoStop}%
\bibitem [{\citenamefont {Miyake}\ \emph {et~al.}(2013)\citenamefont {Miyake},
  \citenamefont {Siviloglou}, \citenamefont {Kennedy}, \citenamefont {Burton},\
  and\ \citenamefont {Ketterle}}]{miyake_13}%
  \BibitemOpen
  \bibfield  {author} {\bibinfo {author} {\bibfnamefont {H.}~\bibnamefont
  {Miyake}}, \bibinfo {author} {\bibfnamefont {G.~A.}\ \bibnamefont
  {Siviloglou}}, \bibinfo {author} {\bibfnamefont {C.~J.}\ \bibnamefont
  {Kennedy}}, \bibinfo {author} {\bibfnamefont {W.~C.}\ \bibnamefont {Burton}},
  \ and\ \bibinfo {author} {\bibfnamefont {W.}~\bibnamefont {Ketterle}},\
  }\href {http://link.aps.org/doi/10.1103/PhysRevLett.111.185302} {\bibfield
  {journal} {\bibinfo  {journal} {Phys. Rev. Lett.}\ }\textbf {\bibinfo
  {volume} {111}},\ \bibinfo {pages} {185302} (\bibinfo {year}
  {2013})}\BibitemShut {NoStop}%
\bibitem [{\citenamefont {Jotzu}\ \emph {et~al.}(2015)\citenamefont {Jotzu},
  \citenamefont {Messer}, \citenamefont {G{\"o}rg}, \citenamefont {Greif},
  \citenamefont {Desbuquois},\ and\ \citenamefont {Esslinger}}]{jotzu_15}%
  \BibitemOpen
  \bibfield  {author} {\bibinfo {author} {\bibfnamefont {G.}~\bibnamefont
  {Jotzu}}, \bibinfo {author} {\bibfnamefont {M.}~\bibnamefont {Messer}},
  \bibinfo {author} {\bibfnamefont {F.}~\bibnamefont {G{\"o}rg}}, \bibinfo
  {author} {\bibfnamefont {D.}~\bibnamefont {Greif}}, \bibinfo {author}
  {\bibfnamefont {R.}~\bibnamefont {Desbuquois}}, \ and\ \bibinfo {author}
  {\bibfnamefont {T.}~\bibnamefont {Esslinger}},\ }\href
  {http://link.aps.org/doi/10.1103/PhysRevLett.115.073002} {\bibfield
  {journal} {\bibinfo  {journal} {Phys. Rev. Lett.}\ }\textbf {\bibinfo
  {volume} {115}},\ \bibinfo {pages} {073002} (\bibinfo {year}
  {2015})}\BibitemShut {NoStop}%
\bibitem [{\citenamefont {Nascimbene}\ \emph {et~al.}(2015)\citenamefont
  {Nascimbene}, \citenamefont {Goldman}, \citenamefont {Cooper},\ and\
  \citenamefont {Dalibard}}]{nascimbene2015dynamic}%
  \BibitemOpen
  \bibfield  {author} {\bibinfo {author} {\bibfnamefont {S.}~\bibnamefont
  {Nascimbene}}, \bibinfo {author} {\bibfnamefont {N.}~\bibnamefont {Goldman}},
  \bibinfo {author} {\bibfnamefont {N.~R.}\ \bibnamefont {Cooper}}, \ and\
  \bibinfo {author} {\bibfnamefont {J.}~\bibnamefont {Dalibard}},\ }\href
  {\doibase 10.1103/PhysRevLett.115.140401} {\bibfield  {journal} {\bibinfo
  {journal} {Phys. Rev. Lett.}\ }\textbf {\bibinfo {volume} {115}},\ \bibinfo
  {pages} {140401} (\bibinfo {year} {2015})}\BibitemShut {NoStop}%
\bibitem [{\citenamefont {Price}\ \emph {et~al.}(2017)\citenamefont {Price},
  \citenamefont {Ozawa},\ and\ \citenamefont {Goldman}}]{price2017synthetic}%
  \BibitemOpen
  \bibfield  {author} {\bibinfo {author} {\bibfnamefont {H.~M.}\ \bibnamefont
  {Price}}, \bibinfo {author} {\bibfnamefont {T.}~\bibnamefont {Ozawa}}, \ and\
  \bibinfo {author} {\bibfnamefont {N.}~\bibnamefont {Goldman}},\ }\href
  {\doibase 10.1103/PhysRevA.95.023607} {\bibfield  {journal} {\bibinfo
  {journal} {Phys. Rev. A}\ }\textbf {\bibinfo {volume} {95}},\ \bibinfo
  {pages} {023607} (\bibinfo {year} {2017})}\BibitemShut {NoStop}%
\bibitem [{\citenamefont {Tarnowski}\ \emph {et~al.}(2019)\citenamefont
  {Tarnowski}, \citenamefont {{\"U}nal}, \citenamefont {Fl{\"a}schner},
  \citenamefont {Rem}, \citenamefont {Eckardt}, \citenamefont {Sengstock},\
  and\ \citenamefont {Weitenberg}}]{tarnowski2019measuring}%
  \BibitemOpen
  \bibfield  {author} {\bibinfo {author} {\bibfnamefont {M.}~\bibnamefont
  {Tarnowski}}, \bibinfo {author} {\bibfnamefont {F.~N.}\ \bibnamefont
  {{\"U}nal}}, \bibinfo {author} {\bibfnamefont {N.}~\bibnamefont
  {Fl{\"a}schner}}, \bibinfo {author} {\bibfnamefont {B.~S.}\ \bibnamefont
  {Rem}}, \bibinfo {author} {\bibfnamefont {A.}~\bibnamefont {Eckardt}},
  \bibinfo {author} {\bibfnamefont {K.}~\bibnamefont {Sengstock}}, \ and\
  \bibinfo {author} {\bibfnamefont {C.}~\bibnamefont {Weitenberg}},\ }\href
  {https://www.nature.com/articles/s41467-019-09668-y} {\bibfield  {journal}
  {\bibinfo  {journal} {Nature communications}\ }\textbf {\bibinfo {volume}
  {10}},\ \bibinfo {pages} {1} (\bibinfo {year} {2019})}\BibitemShut {NoStop}%
\bibitem [{\citenamefont {G{\"o}rg}\ \emph {et~al.}(2019)\citenamefont
  {G{\"o}rg}, \citenamefont {Sandholzer}, \citenamefont {Minguzzi},
  \citenamefont {Desbuquois}, \citenamefont {Messer},\ and\ \citenamefont
  {Esslinger}}]{gorg2019realization}%
  \BibitemOpen
  \bibfield  {author} {\bibinfo {author} {\bibfnamefont {F.}~\bibnamefont
  {G{\"o}rg}}, \bibinfo {author} {\bibfnamefont {K.}~\bibnamefont
  {Sandholzer}}, \bibinfo {author} {\bibfnamefont {J.}~\bibnamefont
  {Minguzzi}}, \bibinfo {author} {\bibfnamefont {R.}~\bibnamefont
  {Desbuquois}}, \bibinfo {author} {\bibfnamefont {M.}~\bibnamefont {Messer}},
  \ and\ \bibinfo {author} {\bibfnamefont {T.}~\bibnamefont {Esslinger}},\
  }\href {https://www.nature.com/articles/s41567-019-0615-4} {\bibfield
  {journal} {\bibinfo  {journal} {Nature Physics}\ }\textbf {\bibinfo {volume}
  {15}},\ \bibinfo {pages} {1161} (\bibinfo {year} {2019})}\BibitemShut
  {NoStop}%
\bibitem [{\citenamefont {Wintersperger}\ \emph
  {et~al.}(2020{\natexlab{a}})\citenamefont {Wintersperger}, \citenamefont
  {Braun}, \citenamefont {{\"U}nal}, \citenamefont {Eckardt}, \citenamefont
  {Di~Liberto}, \citenamefont {Goldman}, \citenamefont {Bloch},\ and\
  \citenamefont {Aidelsburger}}]{wintersperger2020realization}%
  \BibitemOpen
  \bibfield  {author} {\bibinfo {author} {\bibfnamefont {K.}~\bibnamefont
  {Wintersperger}}, \bibinfo {author} {\bibfnamefont {C.}~\bibnamefont
  {Braun}}, \bibinfo {author} {\bibfnamefont {F.~N.}\ \bibnamefont {{\"U}nal}},
  \bibinfo {author} {\bibfnamefont {A.}~\bibnamefont {Eckardt}}, \bibinfo
  {author} {\bibfnamefont {M.}~\bibnamefont {Di~Liberto}}, \bibinfo {author}
  {\bibfnamefont {N.}~\bibnamefont {Goldman}}, \bibinfo {author} {\bibfnamefont
  {I.}~\bibnamefont {Bloch}}, \ and\ \bibinfo {author} {\bibfnamefont
  {M.}~\bibnamefont {Aidelsburger}},\ }\href
  {https://doi.org/10.1038/s41567-020-0949-y} {\bibfield  {journal} {\bibinfo
  {journal} {Nature Physics}\ }\textbf {\bibinfo {volume} {16}},\ \bibinfo
  {pages} {1058} (\bibinfo {year} {2020}{\natexlab{a}})}\BibitemShut {NoStop}%
\bibitem [{\citenamefont {Quelle}\ \emph {et~al.}(2017)\citenamefont {Quelle},
  \citenamefont {Weitenberg}, \citenamefont {Sengstock},\ and\ \citenamefont
  {Smith}}]{quelle2017driving}%
  \BibitemOpen
  \bibfield  {author} {\bibinfo {author} {\bibfnamefont {A.}~\bibnamefont
  {Quelle}}, \bibinfo {author} {\bibfnamefont {C.}~\bibnamefont {Weitenberg}},
  \bibinfo {author} {\bibfnamefont {K.}~\bibnamefont {Sengstock}}, \ and\
  \bibinfo {author} {\bibfnamefont {C.~M.}\ \bibnamefont {Smith}},\ }\href
  {https://iopscience.iop.org/article/10.1088/1367-2630/aa8646} {\bibfield
  {journal} {\bibinfo  {journal} {New Journal of Physics}\ }\textbf {\bibinfo
  {volume} {19}},\ \bibinfo {pages} {113010} (\bibinfo {year}
  {2017})}\BibitemShut {NoStop}%
\bibitem [{\citenamefont {Schweizer}\ \emph {et~al.}(2019)\citenamefont
  {Schweizer}, \citenamefont {Grusdt}, \citenamefont {Berngruber},
  \citenamefont {Barbiero}, \citenamefont {Demler}, \citenamefont {Goldman},
  \citenamefont {Bloch},\ and\ \citenamefont
  {Aidelsburger}}]{schweizer2019floquet}%
  \BibitemOpen
  \bibfield  {author} {\bibinfo {author} {\bibfnamefont {C.}~\bibnamefont
  {Schweizer}}, \bibinfo {author} {\bibfnamefont {F.}~\bibnamefont {Grusdt}},
  \bibinfo {author} {\bibfnamefont {M.}~\bibnamefont {Berngruber}}, \bibinfo
  {author} {\bibfnamefont {L.}~\bibnamefont {Barbiero}}, \bibinfo {author}
  {\bibfnamefont {E.}~\bibnamefont {Demler}}, \bibinfo {author} {\bibfnamefont
  {N.}~\bibnamefont {Goldman}}, \bibinfo {author} {\bibfnamefont
  {I.}~\bibnamefont {Bloch}}, \ and\ \bibinfo {author} {\bibfnamefont
  {M.}~\bibnamefont {Aidelsburger}},\ }\href {\doibase
  https://doi.org/10.1038/s41567-019-0649-7} {\bibfield  {journal} {\bibinfo
  {journal} {Nature Physics}\ }\textbf {\bibinfo {volume} {15}},\ \bibinfo
  {pages} {1168} (\bibinfo {year} {2019})}\BibitemShut {NoStop}%
\bibitem [{\citenamefont {Barbiero}\ \emph {et~al.}(2019)\citenamefont
  {Barbiero}, \citenamefont {Schweizer}, \citenamefont {Aidelsburger},
  \citenamefont {Demler}, \citenamefont {Goldman},\ and\ \citenamefont
  {Grusdt}}]{barbiero2019coupling}%
  \BibitemOpen
  \bibfield  {author} {\bibinfo {author} {\bibfnamefont {L.}~\bibnamefont
  {Barbiero}}, \bibinfo {author} {\bibfnamefont {C.}~\bibnamefont {Schweizer}},
  \bibinfo {author} {\bibfnamefont {M.}~\bibnamefont {Aidelsburger}}, \bibinfo
  {author} {\bibfnamefont {E.}~\bibnamefont {Demler}}, \bibinfo {author}
  {\bibfnamefont {N.}~\bibnamefont {Goldman}}, \ and\ \bibinfo {author}
  {\bibfnamefont {F.}~\bibnamefont {Grusdt}},\ }\href {\doibase
  https://10.1126/sciadv.aav7444} {\bibfield  {journal} {\bibinfo  {journal}
  {Science Advances}\ }\textbf {\bibinfo {volume} {5}},\ \bibinfo {pages}
  {eaav7444} (\bibinfo {year} {2019})}\BibitemShut {NoStop}%
\bibitem [{\citenamefont {Sandholzer}\ \emph {et~al.}(2019)\citenamefont
  {Sandholzer}, \citenamefont {Murakami}, \citenamefont {G\"org}, \citenamefont
  {Minguzzi}, \citenamefont {Messer}, \citenamefont {Desbuquois}, \citenamefont
  {Eckstein}, \citenamefont {Werner},\ and\ \citenamefont
  {Esslinger}}]{sandholzer2019quantum}%
  \BibitemOpen
  \bibfield  {author} {\bibinfo {author} {\bibfnamefont {K.}~\bibnamefont
  {Sandholzer}}, \bibinfo {author} {\bibfnamefont {Y.}~\bibnamefont
  {Murakami}}, \bibinfo {author} {\bibfnamefont {F.}~\bibnamefont {G\"org}},
  \bibinfo {author} {\bibfnamefont {J.}~\bibnamefont {Minguzzi}}, \bibinfo
  {author} {\bibfnamefont {M.}~\bibnamefont {Messer}}, \bibinfo {author}
  {\bibfnamefont {R.}~\bibnamefont {Desbuquois}}, \bibinfo {author}
  {\bibfnamefont {M.}~\bibnamefont {Eckstein}}, \bibinfo {author}
  {\bibfnamefont {P.}~\bibnamefont {Werner}}, \ and\ \bibinfo {author}
  {\bibfnamefont {T.}~\bibnamefont {Esslinger}},\ }\href {\doibase
  10.1103/PhysRevLett.123.193602} {\bibfield  {journal} {\bibinfo  {journal}
  {Phys. Rev. Lett.}\ }\textbf {\bibinfo {volume} {123}},\ \bibinfo {pages}
  {193602} (\bibinfo {year} {2019})}\BibitemShut {NoStop}%
\bibitem [{\citenamefont {Rechtsman}\ \emph {et~al.}(2013)\citenamefont
  {Rechtsman}, \citenamefont {Zeuner}, \citenamefont {Plotnik}, \citenamefont
  {Lumer}, \citenamefont {Podolsky}, \citenamefont {Dreisow}, \citenamefont
  {Nolte}, \citenamefont {Segev},\ and\ \citenamefont
  {Szameit}}]{rechtsman_13}%
  \BibitemOpen
  \bibfield  {author} {\bibinfo {author} {\bibfnamefont {M.~C.}\ \bibnamefont
  {Rechtsman}}, \bibinfo {author} {\bibfnamefont {J.~M.}\ \bibnamefont
  {Zeuner}}, \bibinfo {author} {\bibfnamefont {Y.}~\bibnamefont {Plotnik}},
  \bibinfo {author} {\bibfnamefont {Y.}~\bibnamefont {Lumer}}, \bibinfo
  {author} {\bibfnamefont {D.}~\bibnamefont {Podolsky}}, \bibinfo {author}
  {\bibfnamefont {F.}~\bibnamefont {Dreisow}}, \bibinfo {author} {\bibfnamefont
  {S.}~\bibnamefont {Nolte}}, \bibinfo {author} {\bibfnamefont
  {M.}~\bibnamefont {Segev}}, \ and\ \bibinfo {author} {\bibfnamefont
  {A.}~\bibnamefont {Szameit}},\ }\href
  {http://www.nature.com/nature/journal/v496/n7444/abs/nature12066.html}
  {\bibfield  {journal} {\bibinfo  {journal} {Nature}\ }\textbf {\bibinfo
  {volume} {496}},\ \bibinfo {pages} {196} (\bibinfo {year}
  {2013})}\BibitemShut {NoStop}%
\bibitem [{\citenamefont {Hafezi}(2014)}]{hafezi_14}%
  \BibitemOpen
  \bibfield  {author} {\bibinfo {author} {\bibfnamefont {M.}~\bibnamefont
  {Hafezi}},\ }\href {http://link.aps.org/doi/10.1103/PhysRevLett.112.210405}
  {\bibfield  {journal} {\bibinfo  {journal} {Phys. Rev. Lett.}\ }\textbf
  {\bibinfo {volume} {112}},\ \bibinfo {pages} {210405} (\bibinfo {year}
  {2014})}\BibitemShut {NoStop}%
\bibitem [{\citenamefont {Mittal}\ \emph {et~al.}(2014)\citenamefont {Mittal},
  \citenamefont {Fan}, \citenamefont {Faez}, \citenamefont {Migdall},
  \citenamefont {Taylor},\ and\ \citenamefont {Hafezi}}]{mittal_14}%
  \BibitemOpen
  \bibfield  {author} {\bibinfo {author} {\bibfnamefont {S.}~\bibnamefont
  {Mittal}}, \bibinfo {author} {\bibfnamefont {J.}~\bibnamefont {Fan}},
  \bibinfo {author} {\bibfnamefont {S.}~\bibnamefont {Faez}}, \bibinfo {author}
  {\bibfnamefont {A.}~\bibnamefont {Migdall}}, \bibinfo {author} {\bibfnamefont
  {J.~M.}\ \bibnamefont {Taylor}}, \ and\ \bibinfo {author} {\bibfnamefont
  {M.}~\bibnamefont {Hafezi}},\ }\href
  {http://link.aps.org/doi/10.1103/PhysRevLett.113.087403} {\bibfield
  {journal} {\bibinfo  {journal} {Phys. Rev. Lett.}\ }\textbf {\bibinfo
  {volume} {113}},\ \bibinfo {pages} {087403} (\bibinfo {year}
  {2014})}\BibitemShut {NoStop}%
\bibitem [{\citenamefont {Topp}\ \emph {et~al.}(2019)\citenamefont {Topp},
  \citenamefont {Jotzu}, \citenamefont {McIver}, \citenamefont {Xian},
  \citenamefont {Rubio},\ and\ \citenamefont {Sentef}}]{topp2019topological}%
  \BibitemOpen
  \bibfield  {author} {\bibinfo {author} {\bibfnamefont {G.~E.}\ \bibnamefont
  {Topp}}, \bibinfo {author} {\bibfnamefont {G.}~\bibnamefont {Jotzu}},
  \bibinfo {author} {\bibfnamefont {J.~W.}\ \bibnamefont {McIver}}, \bibinfo
  {author} {\bibfnamefont {L.}~\bibnamefont {Xian}}, \bibinfo {author}
  {\bibfnamefont {A.}~\bibnamefont {Rubio}}, \ and\ \bibinfo {author}
  {\bibfnamefont {M.~A.}\ \bibnamefont {Sentef}},\ }\href {\doibase
  10.1103/PhysRevResearch.1.023031} {\bibfield  {journal} {\bibinfo  {journal}
  {Phys. Rev. Research}\ }\textbf {\bibinfo {volume} {1}},\ \bibinfo {pages}
  {023031} (\bibinfo {year} {2019})}\BibitemShut {NoStop}%
\bibitem [{\citenamefont {McIver}\ \emph {et~al.}(2020)\citenamefont {McIver},
  \citenamefont {Schulte}, \citenamefont {Stein}, \citenamefont {Matsuyama},
  \citenamefont {Jotzu}, \citenamefont {Meier},\ and\ \citenamefont
  {Cavalleri}}]{mciver2020light}%
  \BibitemOpen
  \bibfield  {author} {\bibinfo {author} {\bibfnamefont {J.~W.}\ \bibnamefont
  {McIver}}, \bibinfo {author} {\bibfnamefont {B.}~\bibnamefont {Schulte}},
  \bibinfo {author} {\bibfnamefont {F.-U.}\ \bibnamefont {Stein}}, \bibinfo
  {author} {\bibfnamefont {T.}~\bibnamefont {Matsuyama}}, \bibinfo {author}
  {\bibfnamefont {G.}~\bibnamefont {Jotzu}}, \bibinfo {author} {\bibfnamefont
  {G.}~\bibnamefont {Meier}}, \ and\ \bibinfo {author} {\bibfnamefont
  {A.}~\bibnamefont {Cavalleri}},\ }\href
  {https://www.nature.com/articles/s41567-019-0698-y} {\bibfield  {journal}
  {\bibinfo  {journal} {Nature physics}\ }\textbf {\bibinfo {volume} {16}},\
  \bibinfo {pages} {38} (\bibinfo {year} {2020})}\BibitemShut {NoStop}%
\bibitem [{\citenamefont {Nuske}\ \emph {et~al.}(2020)\citenamefont {Nuske},
  \citenamefont {Broers}, \citenamefont {Schulte}, \citenamefont {Jotzu},
  \citenamefont {Sato}, \citenamefont {Cavalleri}, \citenamefont {Rubio},
  \citenamefont {McIver},\ and\ \citenamefont {Mathey}}]{nuske2020floquet}%
  \BibitemOpen
  \bibfield  {author} {\bibinfo {author} {\bibfnamefont {M.}~\bibnamefont
  {Nuske}}, \bibinfo {author} {\bibfnamefont {L.}~\bibnamefont {Broers}},
  \bibinfo {author} {\bibfnamefont {B.}~\bibnamefont {Schulte}}, \bibinfo
  {author} {\bibfnamefont {G.}~\bibnamefont {Jotzu}}, \bibinfo {author}
  {\bibfnamefont {S.}~\bibnamefont {Sato}}, \bibinfo {author} {\bibfnamefont
  {A.}~\bibnamefont {Cavalleri}}, \bibinfo {author} {\bibfnamefont
  {A.}~\bibnamefont {Rubio}}, \bibinfo {author} {\bibfnamefont
  {J.}~\bibnamefont {McIver}}, \ and\ \bibinfo {author} {\bibfnamefont
  {L.}~\bibnamefont {Mathey}},\ }\href {https://arxiv.org/abs/2005.10824}
  {\bibfield  {journal} {\bibinfo  {journal} {arXiv preprint arXiv:2005.10824}\
  } (\bibinfo {year} {2020})}\BibitemShut {NoStop}%
\bibitem [{\citenamefont {Bukov}\ \emph
  {et~al.}(2015{\natexlab{b}})\citenamefont {Bukov}, \citenamefont
  {Gopalakrishnan}, \citenamefont {Knap},\ and\ \citenamefont
  {Demler}}]{bukov2015prethermal}%
  \BibitemOpen
  \bibfield  {author} {\bibinfo {author} {\bibfnamefont {M.}~\bibnamefont
  {Bukov}}, \bibinfo {author} {\bibfnamefont {S.}~\bibnamefont
  {Gopalakrishnan}}, \bibinfo {author} {\bibfnamefont {M.}~\bibnamefont
  {Knap}}, \ and\ \bibinfo {author} {\bibfnamefont {E.}~\bibnamefont
  {Demler}},\ }\href {\doibase 10.1103/PhysRevLett.115.205301} {\bibfield
  {journal} {\bibinfo  {journal} {Phys. Rev. Lett.}\ }\textbf {\bibinfo
  {volume} {115}},\ \bibinfo {pages} {205301} (\bibinfo {year}
  {2015}{\natexlab{b}})}\BibitemShut {NoStop}%
\bibitem [{\citenamefont {Canovi}\ \emph {et~al.}(2016)\citenamefont {Canovi},
  \citenamefont {Kollar},\ and\ \citenamefont
  {Eckstein}}]{canovi2016stroboscopic}%
  \BibitemOpen
  \bibfield  {author} {\bibinfo {author} {\bibfnamefont {E.}~\bibnamefont
  {Canovi}}, \bibinfo {author} {\bibfnamefont {M.}~\bibnamefont {Kollar}}, \
  and\ \bibinfo {author} {\bibfnamefont {M.}~\bibnamefont {Eckstein}},\ }\href
  {\doibase 10.1103/PhysRevE.93.012130} {\bibfield  {journal} {\bibinfo
  {journal} {Phys. Rev. E}\ }\textbf {\bibinfo {volume} {93}},\ \bibinfo
  {pages} {012130} (\bibinfo {year} {2016})}\BibitemShut {NoStop}%
\bibitem [{\citenamefont {Weinberg}\ \emph {et~al.}(2015)\citenamefont
  {Weinberg}, \citenamefont {\"Olschl\"ager}, \citenamefont {Str\"ater},
  \citenamefont {Prelle}, \citenamefont {Eckardt}, \citenamefont {Sengstock},\
  and\ \citenamefont {Simonet}}]{mweinberg_15}%
  \BibitemOpen
  \bibfield  {author} {\bibinfo {author} {\bibfnamefont {M.}~\bibnamefont
  {Weinberg}}, \bibinfo {author} {\bibfnamefont {C.}~\bibnamefont
  {\"Olschl\"ager}}, \bibinfo {author} {\bibfnamefont {C.}~\bibnamefont
  {Str\"ater}}, \bibinfo {author} {\bibfnamefont {S.}~\bibnamefont {Prelle}},
  \bibinfo {author} {\bibfnamefont {A.}~\bibnamefont {Eckardt}}, \bibinfo
  {author} {\bibfnamefont {K.}~\bibnamefont {Sengstock}}, \ and\ \bibinfo
  {author} {\bibfnamefont {J.}~\bibnamefont {Simonet}},\ }\href {\doibase
  10.1103/PhysRevA.92.043621} {\bibfield  {journal} {\bibinfo  {journal} {Phys.
  Rev. A}\ }\textbf {\bibinfo {volume} {92}},\ \bibinfo {pages} {043621}
  (\bibinfo {year} {2015})}\BibitemShut {NoStop}%
\bibitem [{\citenamefont {Lellouch}\ \emph {et~al.}(2017)\citenamefont
  {Lellouch}, \citenamefont {Bukov}, \citenamefont {Demler},\ and\
  \citenamefont {Goldman}}]{lellouch_17}%
  \BibitemOpen
  \bibfield  {author} {\bibinfo {author} {\bibfnamefont {S.}~\bibnamefont
  {Lellouch}}, \bibinfo {author} {\bibfnamefont {M.}~\bibnamefont {Bukov}},
  \bibinfo {author} {\bibfnamefont {E.}~\bibnamefont {Demler}}, \ and\ \bibinfo
  {author} {\bibfnamefont {N.}~\bibnamefont {Goldman}},\ }\href {\doibase
  10.1103/PhysRevX.7.021015} {\bibfield  {journal} {\bibinfo  {journal} {Phys.
  Rev. X}\ }\textbf {\bibinfo {volume} {7}},\ \bibinfo {pages} {021015}
  (\bibinfo {year} {2017})}\BibitemShut {NoStop}%
\bibitem [{\citenamefont {Reitter}\ \emph {et~al.}(2017)\citenamefont
  {Reitter}, \citenamefont {N\"ager}, \citenamefont {Wintersperger},
  \citenamefont {Str\"ater}, \citenamefont {Bloch}, \citenamefont {Eckardt},\
  and\ \citenamefont {Schneider}}]{reitter_17}%
  \BibitemOpen
  \bibfield  {author} {\bibinfo {author} {\bibfnamefont {M.}~\bibnamefont
  {Reitter}}, \bibinfo {author} {\bibfnamefont {J.}~\bibnamefont {N\"ager}},
  \bibinfo {author} {\bibfnamefont {K.}~\bibnamefont {Wintersperger}}, \bibinfo
  {author} {\bibfnamefont {C.}~\bibnamefont {Str\"ater}}, \bibinfo {author}
  {\bibfnamefont {I.}~\bibnamefont {Bloch}}, \bibinfo {author} {\bibfnamefont
  {A.}~\bibnamefont {Eckardt}}, \ and\ \bibinfo {author} {\bibfnamefont
  {U.}~\bibnamefont {Schneider}},\ }\href {\doibase
  10.1103/PhysRevLett.119.200402} {\bibfield  {journal} {\bibinfo  {journal}
  {Phys. Rev. Lett.}\ }\textbf {\bibinfo {volume} {119}},\ \bibinfo {pages}
  {200402} (\bibinfo {year} {2017})}\BibitemShut {NoStop}%
\bibitem [{\citenamefont {Lellouch}\ and\ \citenamefont
  {Goldman}(2018)}]{lellouch_18}%
  \BibitemOpen
  \bibfield  {author} {\bibinfo {author} {\bibfnamefont {S.}~\bibnamefont
  {Lellouch}}\ and\ \bibinfo {author} {\bibfnamefont {N.}~\bibnamefont
  {Goldman}},\ }\href {\doibase 10.1088/2058-9565/aab2b9} {\enquote {\bibinfo
  {title} {Parametric instabilities in resonantly-driven
  bose{\textendash}einstein condensates},}\ } (\bibinfo {year}
  {2018})\BibitemShut {NoStop}%
\bibitem [{\citenamefont {Wintersperger}\ \emph
  {et~al.}(2020{\natexlab{b}})\citenamefont {Wintersperger}, \citenamefont
  {Bukov}, \citenamefont {N\"ager}, \citenamefont {Lellouch}, \citenamefont
  {Demler}, \citenamefont {Schneider}, \citenamefont {Bloch}, \citenamefont
  {Goldman},\ and\ \citenamefont {Aidelsburger}}]{wintersperger2020paramteric}%
  \BibitemOpen
  \bibfield  {author} {\bibinfo {author} {\bibfnamefont {K.}~\bibnamefont
  {Wintersperger}}, \bibinfo {author} {\bibfnamefont {M.}~\bibnamefont
  {Bukov}}, \bibinfo {author} {\bibfnamefont {J.}~\bibnamefont {N\"ager}},
  \bibinfo {author} {\bibfnamefont {S.}~\bibnamefont {Lellouch}}, \bibinfo
  {author} {\bibfnamefont {E.}~\bibnamefont {Demler}}, \bibinfo {author}
  {\bibfnamefont {U.}~\bibnamefont {Schneider}}, \bibinfo {author}
  {\bibfnamefont {I.}~\bibnamefont {Bloch}}, \bibinfo {author} {\bibfnamefont
  {N.}~\bibnamefont {Goldman}}, \ and\ \bibinfo {author} {\bibfnamefont
  {M.}~\bibnamefont {Aidelsburger}},\ }\href {\doibase
  10.1103/PhysRevX.10.011030} {\bibfield  {journal} {\bibinfo  {journal} {Phys.
  Rev. X}\ }\textbf {\bibinfo {volume} {10}},\ \bibinfo {pages} {011030}
  (\bibinfo {year} {2020}{\natexlab{b}})}\BibitemShut {NoStop}%
\bibitem [{\citenamefont {Boulier}\ \emph {et~al.}(2019)\citenamefont
  {Boulier}, \citenamefont {Maslek}, \citenamefont {Bukov}, \citenamefont
  {Bracamontes}, \citenamefont {Magnan}, \citenamefont {Lellouch},
  \citenamefont {Demler}, \citenamefont {Goldman},\ and\ \citenamefont
  {Porto}}]{boulier2019paramteric}%
  \BibitemOpen
  \bibfield  {author} {\bibinfo {author} {\bibfnamefont {T.}~\bibnamefont
  {Boulier}}, \bibinfo {author} {\bibfnamefont {J.}~\bibnamefont {Maslek}},
  \bibinfo {author} {\bibfnamefont {M.}~\bibnamefont {Bukov}}, \bibinfo
  {author} {\bibfnamefont {C.}~\bibnamefont {Bracamontes}}, \bibinfo {author}
  {\bibfnamefont {E.}~\bibnamefont {Magnan}}, \bibinfo {author} {\bibfnamefont
  {S.}~\bibnamefont {Lellouch}}, \bibinfo {author} {\bibfnamefont
  {E.}~\bibnamefont {Demler}}, \bibinfo {author} {\bibfnamefont
  {N.}~\bibnamefont {Goldman}}, \ and\ \bibinfo {author} {\bibfnamefont
  {J.~V.}\ \bibnamefont {Porto}},\ }\href {\doibase 10.1103/PhysRevX.9.011047}
  {\bibfield  {journal} {\bibinfo  {journal} {Phys. Rev. X}\ }\textbf {\bibinfo
  {volume} {9}},\ \bibinfo {pages} {011047} (\bibinfo {year}
  {2019})}\BibitemShut {NoStop}%
\bibitem [{\citenamefont {Moessner}\ and\ \citenamefont
  {Sondhi}(2017)}]{moessner2017equilibration}%
  \BibitemOpen
  \bibfield  {author} {\bibinfo {author} {\bibfnamefont {R.}~\bibnamefont
  {Moessner}}\ and\ \bibinfo {author} {\bibfnamefont {S.}~\bibnamefont
  {Sondhi}},\ }\href {https://www.nature.com/articles/nphys4106} {\bibfield
  {journal} {\bibinfo  {journal} {Nature Physics}\ }\textbf {\bibinfo {volume}
  {13}},\ \bibinfo {pages} {424} (\bibinfo {year} {2017})}\BibitemShut
  {NoStop}%
\bibitem [{\citenamefont {Shirley}(1965)}]{Shirley1965}%
  \BibitemOpen
  \bibfield  {author} {\bibinfo {author} {\bibfnamefont {J.~H.}\ \bibnamefont
  {Shirley}},\ }\href {\doibase 10.1103/PhysRev.138.B979} {\bibfield  {journal}
  {\bibinfo  {journal} {Phys. Rev.}\ }\textbf {\bibinfo {volume} {138}},\
  \bibinfo {pages} {B979} (\bibinfo {year} {1965})}\BibitemShut {NoStop}%
\bibitem [{\citenamefont {Sambe}(1973)}]{Sambe1973}%
  \BibitemOpen
  \bibfield  {author} {\bibinfo {author} {\bibfnamefont {H.}~\bibnamefont
  {Sambe}},\ }\href {\doibase 10.1103/PhysRevA.7.2203} {\bibfield  {journal}
  {\bibinfo  {journal} {Phys. Rev. A}\ }\textbf {\bibinfo {volume} {7}},\
  \bibinfo {pages} {2203} (\bibinfo {year} {1973})}\BibitemShut {NoStop}%
\bibitem [{\citenamefont {D'Alessio}\ and\ \citenamefont
  {Rigol}(2014)}]{dalessio_14}%
  \BibitemOpen
  \bibfield  {author} {\bibinfo {author} {\bibfnamefont {L.}~\bibnamefont
  {D'Alessio}}\ and\ \bibinfo {author} {\bibfnamefont {M.}~\bibnamefont
  {Rigol}},\ }\href {http://link.aps.org/doi/10.1103/PhysRevX.4.041048}
  {\bibfield  {journal} {\bibinfo  {journal} {Phys. Rev. X}\ }\textbf {\bibinfo
  {volume} {4}},\ \bibinfo {pages} {041048} (\bibinfo {year}
  {2014})}\BibitemShut {NoStop}%
\bibitem [{\citenamefont {Lazarides}\ \emph {et~al.}(2014)\citenamefont
  {Lazarides}, \citenamefont {Das},\ and\ \citenamefont
  {Moessner}}]{lazarides_14}%
  \BibitemOpen
  \bibfield  {author} {\bibinfo {author} {\bibfnamefont {A.}~\bibnamefont
  {Lazarides}}, \bibinfo {author} {\bibfnamefont {A.}~\bibnamefont {Das}}, \
  and\ \bibinfo {author} {\bibfnamefont {R.}~\bibnamefont {Moessner}},\ }\href
  {\doibase 10.1103/PhysRevE.90.012110} {\bibfield  {journal} {\bibinfo
  {journal} {Phys. Rev. E}\ }\textbf {\bibinfo {volume} {90}},\ \bibinfo
  {pages} {012110} (\bibinfo {year} {2014})}\BibitemShut {NoStop}%
\bibitem [{\citenamefont {Bar~Lev}\ \emph {et~al.}(2017)\citenamefont
  {Bar~Lev}, \citenamefont {Luitz},\ and\ \citenamefont
  {Lazarides}}]{bar2017absence}%
  \BibitemOpen
  \bibfield  {author} {\bibinfo {author} {\bibfnamefont {Y.}~\bibnamefont
  {Bar~Lev}}, \bibinfo {author} {\bibfnamefont {D.~J.}\ \bibnamefont {Luitz}},
  \ and\ \bibinfo {author} {\bibfnamefont {A.}~\bibnamefont {Lazarides}},\
  }\href {https://scipost.org/10.21468/SciPostPhys.3.4.029} {\bibfield
  {journal} {\bibinfo  {journal} {SciPost Physics}\ }\textbf {\bibinfo {volume}
  {3}},\ \bibinfo {pages} {029} (\bibinfo {year} {2017})}\BibitemShut {NoStop}%
\bibitem [{\citenamefont {Weidinger}\ and\ \citenamefont
  {Knap}(2017)}]{weidinger2017floquet}%
  \BibitemOpen
  \bibfield  {author} {\bibinfo {author} {\bibfnamefont {S.~A.}\ \bibnamefont
  {Weidinger}}\ and\ \bibinfo {author} {\bibfnamefont {M.}~\bibnamefont
  {Knap}},\ }\href {\doibase https://doi.org/10.1038/srep45382} {\bibfield
  {journal} {\bibinfo  {journal} {Scientific Reports}\ }\textbf {\bibinfo
  {volume} {7}},\ \bibinfo {pages} {45382} (\bibinfo {year}
  {2017})}\BibitemShut {NoStop}%
\bibitem [{\citenamefont {Prosen}(1998)}]{prosen_98a}%
  \BibitemOpen
  \bibfield  {author} {\bibinfo {author} {\bibfnamefont {T.}~\bibnamefont
  {Prosen}},\ }\href {http://link.aps.org/doi/10.1103/PhysRevLett.80.1808}
  {\bibfield  {journal} {\bibinfo  {journal} {Phys. Rev. Lett.}\ }\textbf
  {\bibinfo {volume} {80}},\ \bibinfo {pages} {1808} (\bibinfo {year}
  {1998})}\BibitemShut {NoStop}%
\bibitem [{\citenamefont {Prosen}(1999)}]{prosen_99}%
  \BibitemOpen
  \bibfield  {author} {\bibinfo {author} {\bibfnamefont {T.}~\bibnamefont
  {Prosen}},\ }\href {http://link.aps.org/doi/10.1103/PhysRevE.60.3949}
  {\bibfield  {journal} {\bibinfo  {journal} {Phys. Rev. E}\ }\textbf {\bibinfo
  {volume} {60}},\ \bibinfo {pages} {3949} (\bibinfo {year}
  {1999})}\BibitemShut {NoStop}%
\bibitem [{\citenamefont {D’Alessio}\ and\ \citenamefont
  {Polkovnikov}(2013)}]{dalessio2013many}%
  \BibitemOpen
  \bibfield  {author} {\bibinfo {author} {\bibfnamefont {L.}~\bibnamefont
  {D’Alessio}}\ and\ \bibinfo {author} {\bibfnamefont {A.}~\bibnamefont
  {Polkovnikov}},\ }\href {\doibase https://doi.org/10.1016/j.aop.2013.02.011}
  {\bibfield  {journal} {\bibinfo  {journal} {Annals of Physics}\ }\textbf
  {\bibinfo {volume} {333}},\ \bibinfo {pages} {19} (\bibinfo {year}
  {2013})}\BibitemShut {NoStop}%
\bibitem [{\citenamefont {Haldar}\ \emph {et~al.}(2018)\citenamefont {Haldar},
  \citenamefont {Moessner},\ and\ \citenamefont {Das}}]{haldar2018onset}%
  \BibitemOpen
  \bibfield  {author} {\bibinfo {author} {\bibfnamefont {A.}~\bibnamefont
  {Haldar}}, \bibinfo {author} {\bibfnamefont {R.}~\bibnamefont {Moessner}}, \
  and\ \bibinfo {author} {\bibfnamefont {A.}~\bibnamefont {Das}},\ }\href
  {\doibase 10.1103/PhysRevB.97.245122} {\bibfield  {journal} {\bibinfo
  {journal} {Phys. Rev. B}\ }\textbf {\bibinfo {volume} {97}},\ \bibinfo
  {pages} {245122} (\bibinfo {year} {2018})}\BibitemShut {NoStop}%
\bibitem [{\citenamefont {Ji}\ and\ \citenamefont
  {Fine}(2018)}]{kai2018suppression}%
  \BibitemOpen
  \bibfield  {author} {\bibinfo {author} {\bibfnamefont {K.}~\bibnamefont
  {Ji}}\ and\ \bibinfo {author} {\bibfnamefont {B.~V.}\ \bibnamefont {Fine}},\
  }\href {\doibase 10.1103/PhysRevLett.121.050602} {\bibfield  {journal}
  {\bibinfo  {journal} {Phys. Rev. Lett.}\ }\textbf {\bibinfo {volume} {121}},\
  \bibinfo {pages} {050602} (\bibinfo {year} {2018})}\BibitemShut {NoStop}%
\bibitem [{\citenamefont {Berges}\ \emph {et~al.}(2004)\citenamefont {Berges},
  \citenamefont {Bors\'anyi},\ and\ \citenamefont {Wetterich}}]{Berges2004}%
  \BibitemOpen
  \bibfield  {author} {\bibinfo {author} {\bibfnamefont {J.}~\bibnamefont
  {Berges}}, \bibinfo {author} {\bibfnamefont {S.}~\bibnamefont {Bors\'anyi}},
  \ and\ \bibinfo {author} {\bibfnamefont {C.}~\bibnamefont {Wetterich}},\
  }\href {\doibase 10.1103/PhysRevLett.93.142002} {\bibfield  {journal}
  {\bibinfo  {journal} {Phys. Rev. Lett.}\ }\textbf {\bibinfo {volume} {93}},\
  \bibinfo {pages} {142002} (\bibinfo {year} {2004})}\BibitemShut {NoStop}%
\bibitem [{\citenamefont {Abanin}\ \emph {et~al.}(2015)\citenamefont {Abanin},
  \citenamefont {De~Roeck},\ and\ \citenamefont {Huveneers}}]{abanin_15}%
  \BibitemOpen
  \bibfield  {author} {\bibinfo {author} {\bibfnamefont {D.~A.}\ \bibnamefont
  {Abanin}}, \bibinfo {author} {\bibfnamefont {W.}~\bibnamefont {De~Roeck}}, \
  and\ \bibinfo {author} {\bibfnamefont {F.}~\bibnamefont {Huveneers}},\ }\href
  {http://link.aps.org/doi/10.1103/PhysRevLett.115.256803} {\bibfield
  {journal} {\bibinfo  {journal} {Phys. Rev. Lett.}\ }\textbf {\bibinfo
  {volume} {115}},\ \bibinfo {pages} {256803} (\bibinfo {year}
  {2015})}\BibitemShut {NoStop}%
\bibitem [{\citenamefont {Mori}\ \emph {et~al.}(2016)\citenamefont {Mori},
  \citenamefont {Kuwahara},\ and\ \citenamefont {Saito}}]{mori_15}%
  \BibitemOpen
  \bibfield  {author} {\bibinfo {author} {\bibfnamefont {T.}~\bibnamefont
  {Mori}}, \bibinfo {author} {\bibfnamefont {T.}~\bibnamefont {Kuwahara}}, \
  and\ \bibinfo {author} {\bibfnamefont {K.}~\bibnamefont {Saito}},\ }\href
  {http://journals.aps.org/prl/abstract/10.1103/PhysRevLett.116.120401}
  {\bibfield  {journal} {\bibinfo  {journal} {Phys. Rev. Lett.}\ }\textbf
  {\bibinfo {volume} {116}},\ \bibinfo {pages} {120401} (\bibinfo {year}
  {2016})}\BibitemShut {NoStop}%
\bibitem [{\citenamefont {De~Roeck}\ and\ \citenamefont
  {Verreet}(2019)}]{de2019very}%
  \BibitemOpen
  \bibfield  {author} {\bibinfo {author} {\bibfnamefont {W.}~\bibnamefont
  {De~Roeck}}\ and\ \bibinfo {author} {\bibfnamefont {V.}~\bibnamefont
  {Verreet}},\ }\href {https://arxiv.org/abs/1911.01998} {\bibfield  {journal}
  {\bibinfo  {journal} {arXiv preprint arXiv:1911.01998}\ } (\bibinfo {year}
  {2019})}\BibitemShut {NoStop}%
\bibitem [{\citenamefont {Else}\ \emph {et~al.}(2016)\citenamefont {Else},
  \citenamefont {Bauer},\ and\ \citenamefont {Nayak}}]{else_16}%
  \BibitemOpen
  \bibfield  {author} {\bibinfo {author} {\bibfnamefont {D.~V.}\ \bibnamefont
  {Else}}, \bibinfo {author} {\bibfnamefont {B.}~\bibnamefont {Bauer}}, \ and\
  \bibinfo {author} {\bibfnamefont {C.}~\bibnamefont {Nayak}},\ }\href
  {\doibase 10.1103/PhysRevLett.117.090402} {\bibfield  {journal} {\bibinfo
  {journal} {Phys. Rev. Lett.}\ }\textbf {\bibinfo {volume} {117}},\ \bibinfo
  {pages} {090402} (\bibinfo {year} {2016})}\BibitemShut {NoStop}%
\bibitem [{\citenamefont {Khemani}\ \emph {et~al.}(2016)\citenamefont
  {Khemani}, \citenamefont {Lazarides}, \citenamefont {Moessner},\ and\
  \citenamefont {Sondhi}}]{khemani_16}%
  \BibitemOpen
  \bibfield  {author} {\bibinfo {author} {\bibfnamefont {V.}~\bibnamefont
  {Khemani}}, \bibinfo {author} {\bibfnamefont {A.}~\bibnamefont {Lazarides}},
  \bibinfo {author} {\bibfnamefont {R.}~\bibnamefont {Moessner}}, \ and\
  \bibinfo {author} {\bibfnamefont {S.~L.}\ \bibnamefont {Sondhi}},\ }\href
  {\doibase 10.1103/PhysRevLett.116.250401} {\bibfield  {journal} {\bibinfo
  {journal} {Phys. Rev. Lett.}\ }\textbf {\bibinfo {volume} {116}},\ \bibinfo
  {pages} {250401} (\bibinfo {year} {2016})}\BibitemShut {NoStop}%
\bibitem [{\citenamefont {Yao}\ \emph {et~al.}(2017)\citenamefont {Yao},
  \citenamefont {Potter}, \citenamefont {Potirniche},\ and\ \citenamefont
  {Vishwanath}}]{yao_17}%
  \BibitemOpen
  \bibfield  {author} {\bibinfo {author} {\bibfnamefont {N.~Y.}\ \bibnamefont
  {Yao}}, \bibinfo {author} {\bibfnamefont {A.~C.}\ \bibnamefont {Potter}},
  \bibinfo {author} {\bibfnamefont {I.-D.}\ \bibnamefont {Potirniche}}, \ and\
  \bibinfo {author} {\bibfnamefont {A.}~\bibnamefont {Vishwanath}},\ }\href
  {\doibase 10.1103/PhysRevLett.118.030401} {\bibfield  {journal} {\bibinfo
  {journal} {Phys. Rev. Lett.}\ }\textbf {\bibinfo {volume} {118}},\ \bibinfo
  {pages} {030401} (\bibinfo {year} {2017})}\BibitemShut {NoStop}%
\bibitem [{\citenamefont {Else}\ \emph {et~al.}(2020)\citenamefont {Else},
  \citenamefont {Ho},\ and\ \citenamefont {Dumitrescu}}]{else2020long}%
  \BibitemOpen
  \bibfield  {author} {\bibinfo {author} {\bibfnamefont {D.~V.}\ \bibnamefont
  {Else}}, \bibinfo {author} {\bibfnamefont {W.~W.}\ \bibnamefont {Ho}}, \ and\
  \bibinfo {author} {\bibfnamefont {P.~T.}\ \bibnamefont {Dumitrescu}},\ }\href
  {\doibase 10.1103/PhysRevX.10.021032} {\bibfield  {journal} {\bibinfo
  {journal} {Phys. Rev. X}\ }\textbf {\bibinfo {volume} {10}},\ \bibinfo
  {pages} {021032} (\bibinfo {year} {2020})}\BibitemShut {NoStop}%
\bibitem [{\citenamefont {Machado}\ \emph {et~al.}(2020)\citenamefont
  {Machado}, \citenamefont {Else}, \citenamefont {Kahanamoku-Meyer},
  \citenamefont {Nayak},\ and\ \citenamefont {Yao}}]{machado2020Long}%
  \BibitemOpen
  \bibfield  {author} {\bibinfo {author} {\bibfnamefont {F.}~\bibnamefont
  {Machado}}, \bibinfo {author} {\bibfnamefont {D.~V.}\ \bibnamefont {Else}},
  \bibinfo {author} {\bibfnamefont {G.~D.}\ \bibnamefont {Kahanamoku-Meyer}},
  \bibinfo {author} {\bibfnamefont {C.}~\bibnamefont {Nayak}}, \ and\ \bibinfo
  {author} {\bibfnamefont {N.~Y.}\ \bibnamefont {Yao}},\ }\href {\doibase
  10.1103/PhysRevX.10.011043} {\bibfield  {journal} {\bibinfo  {journal} {Phys.
  Rev. X}\ }\textbf {\bibinfo {volume} {10}},\ \bibinfo {pages} {011043}
  (\bibinfo {year} {2020})}\BibitemShut {NoStop}%
\bibitem [{\citenamefont {Notarnicola}\ \emph {et~al.}(2018)\citenamefont
  {Notarnicola}, \citenamefont {Iemini}, \citenamefont {Rossini}, \citenamefont
  {Fazio}, \citenamefont {Silva},\ and\ \citenamefont
  {Russomanno}}]{notarnicola2018from}%
  \BibitemOpen
  \bibfield  {author} {\bibinfo {author} {\bibfnamefont {S.}~\bibnamefont
  {Notarnicola}}, \bibinfo {author} {\bibfnamefont {F.}~\bibnamefont {Iemini}},
  \bibinfo {author} {\bibfnamefont {D.}~\bibnamefont {Rossini}}, \bibinfo
  {author} {\bibfnamefont {R.}~\bibnamefont {Fazio}}, \bibinfo {author}
  {\bibfnamefont {A.}~\bibnamefont {Silva}}, \ and\ \bibinfo {author}
  {\bibfnamefont {A.}~\bibnamefont {Russomanno}},\ }\href {\doibase
  10.1103/PhysRevE.97.022202} {\bibfield  {journal} {\bibinfo  {journal} {Phys.
  Rev. E}\ }\textbf {\bibinfo {volume} {97}},\ \bibinfo {pages} {022202}
  (\bibinfo {year} {2018})}\BibitemShut {NoStop}%
\bibitem [{\citenamefont {Rajak}\ \emph {et~al.}(2018)\citenamefont {Rajak},
  \citenamefont {Citro},\ and\ \citenamefont
  {Dalla~Torre}}]{rajak2018stability}%
  \BibitemOpen
  \bibfield  {author} {\bibinfo {author} {\bibfnamefont {A.}~\bibnamefont
  {Rajak}}, \bibinfo {author} {\bibfnamefont {R.}~\bibnamefont {Citro}}, \ and\
  \bibinfo {author} {\bibfnamefont {E.~G.}\ \bibnamefont {Dalla~Torre}},\
  }\href {https://iopscience.iop.org/article/10.1088/1751-8121/aae294}
  {\bibfield  {journal} {\bibinfo  {journal} {Journal of Physics A:
  Mathematical and Theoretical}\ }\textbf {\bibinfo {volume} {51}},\ \bibinfo
  {pages} {465001} (\bibinfo {year} {2018})}\BibitemShut {NoStop}%
\bibitem [{\citenamefont {Howell}\ \emph {et~al.}(2019)\citenamefont {Howell},
  \citenamefont {Weinberg}, \citenamefont {Sels}, \citenamefont {Polkovnikov},\
  and\ \citenamefont {Bukov}}]{howell2019asymptotic}%
  \BibitemOpen
  \bibfield  {author} {\bibinfo {author} {\bibfnamefont {O.}~\bibnamefont
  {Howell}}, \bibinfo {author} {\bibfnamefont {P.}~\bibnamefont {Weinberg}},
  \bibinfo {author} {\bibfnamefont {D.}~\bibnamefont {Sels}}, \bibinfo {author}
  {\bibfnamefont {A.}~\bibnamefont {Polkovnikov}}, \ and\ \bibinfo {author}
  {\bibfnamefont {M.}~\bibnamefont {Bukov}},\ }\href {\doibase
  10.1103/PhysRevLett.122.010602} {\bibfield  {journal} {\bibinfo  {journal}
  {Phys. Rev. Lett.}\ }\textbf {\bibinfo {volume} {122}},\ \bibinfo {pages}
  {010602} (\bibinfo {year} {2019})}\BibitemShut {NoStop}%
\bibitem [{\citenamefont {Mori}(2018)}]{mori2018floquet}%
  \BibitemOpen
  \bibfield  {author} {\bibinfo {author} {\bibfnamefont {T.}~\bibnamefont
  {Mori}},\ }\href {\doibase 10.1103/PhysRevB.98.104303} {\bibfield  {journal}
  {\bibinfo  {journal} {Phys. Rev. B}\ }\textbf {\bibinfo {volume} {98}},\
  \bibinfo {pages} {104303} (\bibinfo {year} {2018})}\BibitemShut {NoStop}%
\bibitem [{\citenamefont {Rajak}\ \emph {et~al.}(2019)\citenamefont {Rajak},
  \citenamefont {Dana},\ and\ \citenamefont
  {Dalla~Torre}}]{rajak2019characterizations}%
  \BibitemOpen
  \bibfield  {author} {\bibinfo {author} {\bibfnamefont {A.}~\bibnamefont
  {Rajak}}, \bibinfo {author} {\bibfnamefont {I.}~\bibnamefont {Dana}}, \ and\
  \bibinfo {author} {\bibfnamefont {E.~G.}\ \bibnamefont {Dalla~Torre}},\
  }\href {\doibase 10.1103/PhysRevB.100.100302} {\bibfield  {journal} {\bibinfo
   {journal} {Phys. Rev. B}\ }\textbf {\bibinfo {volume} {100}},\ \bibinfo
  {pages} {100302} (\bibinfo {year} {2019})}\BibitemShut {NoStop}%
\bibitem [{\citenamefont {Huveneers}\ and\ \citenamefont
  {Lukkarinen}(2020)}]{huveneers2020prethermalization}%
  \BibitemOpen
  \bibfield  {author} {\bibinfo {author} {\bibfnamefont {F.}~\bibnamefont
  {Huveneers}}\ and\ \bibinfo {author} {\bibfnamefont {J.}~\bibnamefont
  {Lukkarinen}},\ }\href {\doibase 10.1103/PhysRevResearch.2.022034} {\bibfield
   {journal} {\bibinfo  {journal} {Phys. Rev. Research}\ }\textbf {\bibinfo
  {volume} {2}},\ \bibinfo {pages} {022034} (\bibinfo {year}
  {2020})}\BibitemShut {NoStop}%
\bibitem [{\citenamefont {Torre}(2020)}]{torre2020statistical}%
  \BibitemOpen
  \bibfield  {author} {\bibinfo {author} {\bibfnamefont {E.~G.~D.}\
  \bibnamefont {Torre}},\ }\href {https://arxiv.org/abs/2005.07207} {\bibfield
  {journal} {\bibinfo  {journal} {arXiv preprint arXiv:2005.07207}\ } (\bibinfo
  {year} {2020})}\BibitemShut {NoStop}%
\bibitem [{\citenamefont {Zhao}\ \emph {et~al.}(2020)\citenamefont {Zhao},
  \citenamefont {Mintert}, \citenamefont {Moessner},\ and\ \citenamefont
  {Knolle}}]{zhao2020random}%
  \BibitemOpen
  \bibfield  {author} {\bibinfo {author} {\bibfnamefont {H.}~\bibnamefont
  {Zhao}}, \bibinfo {author} {\bibfnamefont {F.}~\bibnamefont {Mintert}},
  \bibinfo {author} {\bibfnamefont {R.}~\bibnamefont {Moessner}}, \ and\
  \bibinfo {author} {\bibfnamefont {J.}~\bibnamefont {Knolle}},\ }\href
  {https://arxiv.org/abs/2007.07301} {\bibfield  {journal} {\bibinfo  {journal}
  {arXiv preprint arXiv:2007.07301}\ } (\bibinfo {year} {2020})}\BibitemShut
  {NoStop}%
\bibitem [{\citenamefont {Kuhlenkamp}\ and\ \citenamefont
  {Knap}(2020)}]{kuhlenkamp2020periodically}%
  \BibitemOpen
  \bibfield  {author} {\bibinfo {author} {\bibfnamefont {C.}~\bibnamefont
  {Kuhlenkamp}}\ and\ \bibinfo {author} {\bibfnamefont {M.}~\bibnamefont
  {Knap}},\ }\href {\doibase 10.1103/PhysRevLett.124.106401} {\bibfield
  {journal} {\bibinfo  {journal} {Phys. Rev. Lett.}\ }\textbf {\bibinfo
  {volume} {124}},\ \bibinfo {pages} {106401} (\bibinfo {year}
  {2020})}\BibitemShut {NoStop}%
\bibitem [{\citenamefont {Mukherjee}\ \emph
  {et~al.}(2020{\natexlab{a}})\citenamefont {Mukherjee}, \citenamefont {Nandy},
  \citenamefont {Sen}, \citenamefont {Sen},\ and\ \citenamefont
  {Sengupta}}]{Mukherjee2020a}%
  \BibitemOpen
  \bibfield  {author} {\bibinfo {author} {\bibfnamefont {B.}~\bibnamefont
  {Mukherjee}}, \bibinfo {author} {\bibfnamefont {S.}~\bibnamefont {Nandy}},
  \bibinfo {author} {\bibfnamefont {A.}~\bibnamefont {Sen}}, \bibinfo {author}
  {\bibfnamefont {D.}~\bibnamefont {Sen}}, \ and\ \bibinfo {author}
  {\bibfnamefont {K.}~\bibnamefont {Sengupta}},\ }\href {\doibase
  10.1103/PhysRevB.101.245107} {\bibfield  {journal} {\bibinfo  {journal}
  {Phys. Rev. B}\ }\textbf {\bibinfo {volume} {101}},\ \bibinfo {pages}
  {245107} (\bibinfo {year} {2020}{\natexlab{a}})}\BibitemShut {NoStop}%
\bibitem [{\citenamefont {Mukherjee}\ \emph
  {et~al.}(2020{\natexlab{b}})\citenamefont {Mukherjee}, \citenamefont {Sen},
  \citenamefont {Sen},\ and\ \citenamefont {Sengupta}}]{Mukherjee2020b}%
  \BibitemOpen
  \bibfield  {author} {\bibinfo {author} {\bibfnamefont {B.}~\bibnamefont
  {Mukherjee}}, \bibinfo {author} {\bibfnamefont {A.}~\bibnamefont {Sen}},
  \bibinfo {author} {\bibfnamefont {D.}~\bibnamefont {Sen}}, \ and\ \bibinfo
  {author} {\bibfnamefont {K.}~\bibnamefont {Sengupta}},\ }\href {\doibase
  10.1103/PhysRevB.102.075123} {\bibfield  {journal} {\bibinfo  {journal}
  {Phys. Rev. B}\ }\textbf {\bibinfo {volume} {102}},\ \bibinfo {pages}
  {075123} (\bibinfo {year} {2020}{\natexlab{b}})}\BibitemShut {NoStop}%
\bibitem [{\citenamefont {Dumitrescu}\ \emph {et~al.}(2018)\citenamefont
  {Dumitrescu}, \citenamefont {Vasseur},\ and\ \citenamefont
  {Potter}}]{dumitrescu2018logarithmically}%
  \BibitemOpen
  \bibfield  {author} {\bibinfo {author} {\bibfnamefont {P.~T.}\ \bibnamefont
  {Dumitrescu}}, \bibinfo {author} {\bibfnamefont {R.}~\bibnamefont {Vasseur}},
  \ and\ \bibinfo {author} {\bibfnamefont {A.~C.}\ \bibnamefont {Potter}},\
  }\href {\doibase 10.1103/PhysRevLett.120.070602} {\bibfield  {journal}
  {\bibinfo  {journal} {Phys. Rev. Lett.}\ }\textbf {\bibinfo {volume} {120}},\
  \bibinfo {pages} {070602} (\bibinfo {year} {2018})}\BibitemShut {NoStop}%
\bibitem [{\citenamefont {Martin}\ \emph {et~al.}(2017)\citenamefont {Martin},
  \citenamefont {Refael},\ and\ \citenamefont
  {Halperin}}]{martin2019topological}%
  \BibitemOpen
  \bibfield  {author} {\bibinfo {author} {\bibfnamefont {I.}~\bibnamefont
  {Martin}}, \bibinfo {author} {\bibfnamefont {G.}~\bibnamefont {Refael}}, \
  and\ \bibinfo {author} {\bibfnamefont {B.}~\bibnamefont {Halperin}},\ }\href
  {\doibase 10.1103/PhysRevX.7.041008} {\bibfield  {journal} {\bibinfo
  {journal} {Phys. Rev. X}\ }\textbf {\bibinfo {volume} {7}},\ \bibinfo {pages}
  {041008} (\bibinfo {year} {2017})}\BibitemShut {NoStop}%
\bibitem [{\citenamefont {Crowley}\ \emph {et~al.}(2020)\citenamefont
  {Crowley}, \citenamefont {Martin},\ and\ \citenamefont
  {Chandran}}]{crowley2019halfinteger}%
  \BibitemOpen
  \bibfield  {author} {\bibinfo {author} {\bibfnamefont {P.~J.~D.}\
  \bibnamefont {Crowley}}, \bibinfo {author} {\bibfnamefont {I.}~\bibnamefont
  {Martin}}, \ and\ \bibinfo {author} {\bibfnamefont {A.}~\bibnamefont
  {Chandran}},\ }\href {\doibase 10.1103/PhysRevLett.125.100601} {\bibfield
  {journal} {\bibinfo  {journal} {Phys. Rev. Lett.}\ }\textbf {\bibinfo
  {volume} {125}},\ \bibinfo {pages} {100601} (\bibinfo {year}
  {2020})}\BibitemShut {NoStop}%
\bibitem [{\citenamefont {Moeckel}\ and\ \citenamefont
  {Kehrein}(2008)}]{Moeckel2008}%
  \BibitemOpen
  \bibfield  {author} {\bibinfo {author} {\bibfnamefont {M.}~\bibnamefont
  {Moeckel}}\ and\ \bibinfo {author} {\bibfnamefont {S.}~\bibnamefont
  {Kehrein}},\ }\href {\doibase 10.1103/PhysRevLett.100.175702} {\bibfield
  {journal} {\bibinfo  {journal} {Phys. Rev. Lett.}\ }\textbf {\bibinfo
  {volume} {100}},\ \bibinfo {pages} {175702} (\bibinfo {year}
  {2008})}\BibitemShut {NoStop}%
\bibitem [{\citenamefont {Eckstein}\ \emph {et~al.}(2009)\citenamefont
  {Eckstein}, \citenamefont {Kollar},\ and\ \citenamefont
  {Werner}}]{Eckstein2009}%
  \BibitemOpen
  \bibfield  {author} {\bibinfo {author} {\bibfnamefont {M.}~\bibnamefont
  {Eckstein}}, \bibinfo {author} {\bibfnamefont {M.}~\bibnamefont {Kollar}}, \
  and\ \bibinfo {author} {\bibfnamefont {P.}~\bibnamefont {Werner}},\ }\href
  {\doibase 10.1103/PhysRevLett.103.056403} {\bibfield  {journal} {\bibinfo
  {journal} {Phys. Rev. Lett.}\ }\textbf {\bibinfo {volume} {103}},\ \bibinfo
  {pages} {056403} (\bibinfo {year} {2009})}\BibitemShut {NoStop}%
\bibitem [{\citenamefont {Moeckel}\ and\ \citenamefont
  {Kehrein}(2010)}]{Moeckel_2010}%
  \BibitemOpen
  \bibfield  {author} {\bibinfo {author} {\bibfnamefont {M.}~\bibnamefont
  {Moeckel}}\ and\ \bibinfo {author} {\bibfnamefont {S.}~\bibnamefont
  {Kehrein}},\ }\href {\doibase 10.1088/1367-2630/12/5/055016} {\bibfield
  {journal} {\bibinfo  {journal} {New Journal of Physics}\ }\textbf {\bibinfo
  {volume} {12}},\ \bibinfo {pages} {055016} (\bibinfo {year}
  {2010})}\BibitemShut {NoStop}%
\bibitem [{\citenamefont {Bukov}\ \emph {et~al.}(2016)\citenamefont {Bukov},
  \citenamefont {Heyl}, \citenamefont {Huse},\ and\ \citenamefont
  {Polkovnikov}}]{bukov_15_res}%
  \BibitemOpen
  \bibfield  {author} {\bibinfo {author} {\bibfnamefont {M.}~\bibnamefont
  {Bukov}}, \bibinfo {author} {\bibfnamefont {M.}~\bibnamefont {Heyl}},
  \bibinfo {author} {\bibfnamefont {D.~A.}\ \bibnamefont {Huse}}, \ and\
  \bibinfo {author} {\bibfnamefont {A.}~\bibnamefont {Polkovnikov}},\ }\href
  {http://link.aps.org/doi/10.1103/PhysRevB.93.155132} {\bibfield  {journal}
  {\bibinfo  {journal} {Phys. Rev. B}\ }\textbf {\bibinfo {volume} {93}},\
  \bibinfo {pages} {155132} (\bibinfo {year} {2016})}\BibitemShut {NoStop}%
\bibitem [{\citenamefont {Vogl}\ \emph {et~al.}(2020)\citenamefont {Vogl},
  \citenamefont {Rodriguez-Vega},\ and\ \citenamefont
  {Fiete}}]{vogl2020effective}%
  \BibitemOpen
  \bibfield  {author} {\bibinfo {author} {\bibfnamefont {M.}~\bibnamefont
  {Vogl}}, \bibinfo {author} {\bibfnamefont {M.}~\bibnamefont
  {Rodriguez-Vega}}, \ and\ \bibinfo {author} {\bibfnamefont {G.~A.}\
  \bibnamefont {Fiete}},\ }\href {\doibase 10.1103/PhysRevB.101.024303}
  {\bibfield  {journal} {\bibinfo  {journal} {Phys. Rev. B}\ }\textbf {\bibinfo
  {volume} {101}},\ \bibinfo {pages} {024303} (\bibinfo {year}
  {2020})}\BibitemShut {NoStop}%
\bibitem [{\citenamefont {Verdeny}\ \emph {et~al.}(2013)\citenamefont
  {Verdeny}, \citenamefont {Mielke},\ and\ \citenamefont
  {Mintert}}]{verdeny2013accurate}%
  \BibitemOpen
  \bibfield  {author} {\bibinfo {author} {\bibfnamefont {A.}~\bibnamefont
  {Verdeny}}, \bibinfo {author} {\bibfnamefont {A.}~\bibnamefont {Mielke}}, \
  and\ \bibinfo {author} {\bibfnamefont {F.}~\bibnamefont {Mintert}},\ }\href
  {\doibase 10.1103/PhysRevLett.111.175301} {\bibfield  {journal} {\bibinfo
  {journal} {Phys. Rev. Lett.}\ }\textbf {\bibinfo {volume} {111}},\ \bibinfo
  {pages} {175301} (\bibinfo {year} {2013})}\BibitemShut {NoStop}%
\bibitem [{\citenamefont {Vogl}\ \emph {et~al.}(2019)\citenamefont {Vogl},
  \citenamefont {Laurell}, \citenamefont {Barr},\ and\ \citenamefont
  {Fiete}}]{vogl2019flow}%
  \BibitemOpen
  \bibfield  {author} {\bibinfo {author} {\bibfnamefont {M.}~\bibnamefont
  {Vogl}}, \bibinfo {author} {\bibfnamefont {P.}~\bibnamefont {Laurell}},
  \bibinfo {author} {\bibfnamefont {A.~D.}\ \bibnamefont {Barr}}, \ and\
  \bibinfo {author} {\bibfnamefont {G.~A.}\ \bibnamefont {Fiete}},\ }\href
  {\doibase 10.1103/PhysRevX.9.021037} {\bibfield  {journal} {\bibinfo
  {journal} {Phys. Rev. X}\ }\textbf {\bibinfo {volume} {9}},\ \bibinfo {pages}
  {021037} (\bibinfo {year} {2019})}\BibitemShut {NoStop}%
\bibitem [{\citenamefont {Rodriguez-Vega}\ \emph {et~al.}(2018)\citenamefont
  {Rodriguez-Vega}, \citenamefont {Lentz},\ and\ \citenamefont
  {Seradjeh}}]{rodriguez2018floquet}%
  \BibitemOpen
  \bibfield  {author} {\bibinfo {author} {\bibfnamefont {M.}~\bibnamefont
  {Rodriguez-Vega}}, \bibinfo {author} {\bibfnamefont {M.}~\bibnamefont
  {Lentz}}, \ and\ \bibinfo {author} {\bibfnamefont {B.}~\bibnamefont
  {Seradjeh}},\ }\href
  {https://iopscience.iop.org/article/10.1088/1367-2630/aade37} {\bibfield
  {journal} {\bibinfo  {journal} {New Journal of Physics}\ }\textbf {\bibinfo
  {volume} {20}},\ \bibinfo {pages} {093022} (\bibinfo {year}
  {2018})}\BibitemShut {NoStop}%
\bibitem [{\citenamefont {Sen}\ \emph {et~al.}(2021)\citenamefont {Sen},
  \citenamefont {Sen},\ and\ \citenamefont {Sengupta}}]{sen2021analytic}%
  \BibitemOpen
  \bibfield  {author} {\bibinfo {author} {\bibfnamefont {A.}~\bibnamefont
  {Sen}}, \bibinfo {author} {\bibfnamefont {D.}~\bibnamefont {Sen}}, \ and\
  \bibinfo {author} {\bibfnamefont {K.}~\bibnamefont {Sengupta}},\ }\href
  {https://arxiv.org/abs/2102.00793} {\bibfield  {journal} {\bibinfo  {journal}
  {arXiv preprint arXiv:2102.00793}\ } (\bibinfo {year} {2021})}\BibitemShut
  {NoStop}%
\bibitem [{\citenamefont {Vajna}\ \emph {et~al.}(2018)\citenamefont {Vajna},
  \citenamefont {Klobas}, \citenamefont {Prosen},\ and\ \citenamefont
  {Polkovnikov}}]{vajna2018replica}%
  \BibitemOpen
  \bibfield  {author} {\bibinfo {author} {\bibfnamefont {S.}~\bibnamefont
  {Vajna}}, \bibinfo {author} {\bibfnamefont {K.}~\bibnamefont {Klobas}},
  \bibinfo {author} {\bibfnamefont {T.~c.~v.}\ \bibnamefont {Prosen}}, \ and\
  \bibinfo {author} {\bibfnamefont {A.}~\bibnamefont {Polkovnikov}},\ }\href
  {\doibase 10.1103/PhysRevLett.120.200607} {\bibfield  {journal} {\bibinfo
  {journal} {Phys. Rev. Lett.}\ }\textbf {\bibinfo {volume} {120}},\ \bibinfo
  {pages} {200607} (\bibinfo {year} {2018})}\BibitemShut {NoStop}%
\bibitem [{\citenamefont {Lindner}\ \emph {et~al.}(2017)\citenamefont
  {Lindner}, \citenamefont {Berg},\ and\ \citenamefont {Rudner}}]{Lindner2017}%
  \BibitemOpen
  \bibfield  {author} {\bibinfo {author} {\bibfnamefont {N.~H.}\ \bibnamefont
  {Lindner}}, \bibinfo {author} {\bibfnamefont {E.}~\bibnamefont {Berg}}, \
  and\ \bibinfo {author} {\bibfnamefont {M.~S.}\ \bibnamefont {Rudner}},\
  }\href {\doibase 10.1103/PhysRevX.7.011018} {\bibfield  {journal} {\bibinfo
  {journal} {Phys. Rev. X}\ }\textbf {\bibinfo {volume} {7}},\ \bibinfo {pages}
  {011018} (\bibinfo {year} {2017})}\BibitemShut {NoStop}%
\bibitem [{\citenamefont {Gulden}\ \emph {et~al.}(2020)\citenamefont {Gulden},
  \citenamefont {Berg}, \citenamefont {Rudner},\ and\ \citenamefont
  {Lindner}}]{Gulden2020}%
  \BibitemOpen
  \bibfield  {author} {\bibinfo {author} {\bibfnamefont {T.}~\bibnamefont
  {Gulden}}, \bibinfo {author} {\bibfnamefont {E.}~\bibnamefont {Berg}},
  \bibinfo {author} {\bibfnamefont {M.~S.}\ \bibnamefont {Rudner}}, \ and\
  \bibinfo {author} {\bibfnamefont {N.~H.}\ \bibnamefont {Lindner}},\ }\href
  {\doibase 10.21468/SciPostPhys.9.1.015} {\bibfield  {journal} {\bibinfo
  {journal} {SciPost Phys.}\ }\textbf {\bibinfo {volume} {9}},\ \bibinfo
  {pages} {15} (\bibinfo {year} {2020})}\BibitemShut {NoStop}%
\bibitem [{\citenamefont {Gawatz}\ \emph {et~al.}(2021)\citenamefont {Gawatz},
  \citenamefont {Balram}, \citenamefont {Berg}, \citenamefont {Lindner},\ and\
  \citenamefont {Rudner}}]{gawatz2021prethermalization}%
  \BibitemOpen
  \bibfield  {author} {\bibinfo {author} {\bibfnamefont {R.}~\bibnamefont
  {Gawatz}}, \bibinfo {author} {\bibfnamefont {A.~C.}\ \bibnamefont {Balram}},
  \bibinfo {author} {\bibfnamefont {E.}~\bibnamefont {Berg}}, \bibinfo {author}
  {\bibfnamefont {N.~H.}\ \bibnamefont {Lindner}}, \ and\ \bibinfo {author}
  {\bibfnamefont {M.~S.}\ \bibnamefont {Rudner}},\ }\href
  {https://arxiv.org/abs/2103.15831} {\bibfield  {journal} {\bibinfo  {journal}
  {arXiv preprint arXiv:2103.15831}\ } (\bibinfo {year} {2021})}\BibitemShut
  {NoStop}%
\bibitem [{\citenamefont {Fleckenstein}\ and\ \citenamefont
  {Bukov}(2021)}]{fleckenstein_short}%
  \BibitemOpen
  \bibfield  {author} {\bibinfo {author} {\bibfnamefont {C.}~\bibnamefont
  {Fleckenstein}}\ and\ \bibinfo {author} {\bibfnamefont {M.}~\bibnamefont
  {Bukov}},\ }\href {\doibase 10.1103/PhysRevB.103.L140302} {\bibfield
  {journal} {\bibinfo  {journal} {Phys. Rev. B}\ }\textbf {\bibinfo {volume}
  {103}},\ \bibinfo {pages} {L140302} (\bibinfo {year} {2021})}\BibitemShut
  {NoStop}%
\bibitem [{\citenamefont {Rozenbaum}\ and\ \citenamefont
  {Galitski}(2017)}]{Rozenbaum2017}%
  \BibitemOpen
  \bibfield  {author} {\bibinfo {author} {\bibfnamefont {E.~B.}\ \bibnamefont
  {Rozenbaum}}\ and\ \bibinfo {author} {\bibfnamefont {V.}~\bibnamefont
  {Galitski}},\ }\href {\doibase 10.1103/PhysRevB.95.064303} {\bibfield
  {journal} {\bibinfo  {journal} {Phys. Rev. B}\ }\textbf {\bibinfo {volume}
  {95}},\ \bibinfo {pages} {064303} (\bibinfo {year} {2017})}\BibitemShut
  {NoStop}%
\bibitem [{\citenamefont {Rylands}\ \emph {et~al.}(2020)\citenamefont
  {Rylands}, \citenamefont {Rozenbaum}, \citenamefont {Galitski},\ and\
  \citenamefont {Konik}}]{Rylands2020}%
  \BibitemOpen
  \bibfield  {author} {\bibinfo {author} {\bibfnamefont {C.}~\bibnamefont
  {Rylands}}, \bibinfo {author} {\bibfnamefont {E.~B.}\ \bibnamefont
  {Rozenbaum}}, \bibinfo {author} {\bibfnamefont {V.}~\bibnamefont {Galitski}},
  \ and\ \bibinfo {author} {\bibfnamefont {R.}~\bibnamefont {Konik}},\ }\href
  {\doibase 10.1103/PhysRevLett.124.155302} {\bibfield  {journal} {\bibinfo
  {journal} {Phys. Rev. Lett.}\ }\textbf {\bibinfo {volume} {124}},\ \bibinfo
  {pages} {155302} (\bibinfo {year} {2020})}\BibitemShut {NoStop}%
\bibitem [{\citenamefont {Fava}\ \emph {et~al.}(2020)\citenamefont {Fava},
  \citenamefont {Fazio},\ and\ \citenamefont {Russomanno}}]{Fava2020}%
  \BibitemOpen
  \bibfield  {author} {\bibinfo {author} {\bibfnamefont {M.}~\bibnamefont
  {Fava}}, \bibinfo {author} {\bibfnamefont {R.}~\bibnamefont {Fazio}}, \ and\
  \bibinfo {author} {\bibfnamefont {A.}~\bibnamefont {Russomanno}},\ }\href
  {\doibase 10.1103/PhysRevB.101.064302} {\bibfield  {journal} {\bibinfo
  {journal} {Phys. Rev. B}\ }\textbf {\bibinfo {volume} {101}},\ \bibinfo
  {pages} {064302} (\bibinfo {year} {2020})}\BibitemShut {NoStop}%
\bibitem [{Note1()}]{Note1}%
  \BibitemOpen
  \bibinfo {note} {While we do not provide a proof for this statement, we
  strongly believe it to be plausible for generic enough systems.}\BibitemShut
  {Stop}%
\bibitem [{\citenamefont {Pizzi}\ \emph {et~al.}(2020)\citenamefont {Pizzi},
  \citenamefont {Malz}, \citenamefont {De~Tomasi}, \citenamefont {Knolle},\
  and\ \citenamefont {Nunnenkamp}}]{pizzi2020time}%
  \BibitemOpen
  \bibfield  {author} {\bibinfo {author} {\bibfnamefont {A.}~\bibnamefont
  {Pizzi}}, \bibinfo {author} {\bibfnamefont {D.}~\bibnamefont {Malz}},
  \bibinfo {author} {\bibfnamefont {G.}~\bibnamefont {De~Tomasi}}, \bibinfo
  {author} {\bibfnamefont {J.}~\bibnamefont {Knolle}}, \ and\ \bibinfo {author}
  {\bibfnamefont {A.}~\bibnamefont {Nunnenkamp}},\ }\href {\doibase
  10.1103/PhysRevB.102.214207} {\bibfield  {journal} {\bibinfo  {journal}
  {Phys. Rev. B}\ }\textbf {\bibinfo {volume} {102}},\ \bibinfo {pages}
  {214207} (\bibinfo {year} {2020})}\BibitemShut {NoStop}%
\bibitem [{\citenamefont {Page}(1993)}]{Page1993}%
  \BibitemOpen
  \bibfield  {author} {\bibinfo {author} {\bibfnamefont {D.~N.}\ \bibnamefont
  {Page}},\ }\href {\doibase 10.1103/PhysRevLett.71.1291} {\bibfield  {journal}
  {\bibinfo  {journal} {Phys. Rev. Lett.}\ }\textbf {\bibinfo {volume} {71}},\
  \bibinfo {pages} {1291} (\bibinfo {year} {1993})}\BibitemShut {NoStop}%
\bibitem [{\citenamefont {Nekhoroshev}(1971)}]{nekhoroshev1971behavior}%
  \BibitemOpen
  \bibfield  {author} {\bibinfo {author} {\bibfnamefont {N.~N.}\ \bibnamefont
  {Nekhoroshev}},\ }\href
  {http://www.mathnet.ru/php/archive.phtml?wshow=paper&jrnid=faa&paperid=2626&option_lang=eng}
  {\bibfield  {journal} {\bibinfo  {journal} {Functional Analysis and Its
  Applications}\ }\textbf {\bibinfo {volume} {5}},\ \bibinfo {pages} {338}
  (\bibinfo {year} {1971})}\BibitemShut {NoStop}%
\bibitem [{\citenamefont {Kaneko}\ and\ \citenamefont
  {Konishi}(1989)}]{Kunihiko1989}%
  \BibitemOpen
  \bibfield  {author} {\bibinfo {author} {\bibfnamefont {K.}~\bibnamefont
  {Kaneko}}\ and\ \bibinfo {author} {\bibfnamefont {T.}~\bibnamefont
  {Konishi}},\ }\href {\doibase 10.1103/PhysRevA.40.6130} {\bibfield  {journal}
  {\bibinfo  {journal} {Phys. Rev. A}\ }\textbf {\bibinfo {volume} {40}},\
  \bibinfo {pages} {6130} (\bibinfo {year} {1989})}\BibitemShut {NoStop}%
\bibitem [{\citenamefont {Konishi}\ and\ \citenamefont
  {Kaneko}(1990)}]{Konishi_1990}%
  \BibitemOpen
  \bibfield  {author} {\bibinfo {author} {\bibfnamefont {T.}~\bibnamefont
  {Konishi}}\ and\ \bibinfo {author} {\bibfnamefont {K.}~\bibnamefont
  {Kaneko}},\ }\href {\doibase 10.1088/0305-4470/23/15/004} {\bibfield
  {journal} {\bibinfo  {journal} {Journal of Physics A: Mathematical and
  General}\ }\textbf {\bibinfo {volume} {23}},\ \bibinfo {pages} {L715}
  (\bibinfo {year} {1990})}\BibitemShut {NoStop}%
\bibitem [{\citenamefont {Mallayya}\ and\ \citenamefont
  {Rigol}(2019)}]{mallaya2019heating}%
  \BibitemOpen
  \bibfield  {author} {\bibinfo {author} {\bibfnamefont {K.}~\bibnamefont
  {Mallayya}}\ and\ \bibinfo {author} {\bibfnamefont {M.}~\bibnamefont
  {Rigol}},\ }\href {\doibase 10.1103/PhysRevLett.123.240603} {\bibfield
  {journal} {\bibinfo  {journal} {Phys. Rev. Lett.}\ }\textbf {\bibinfo
  {volume} {123}},\ \bibinfo {pages} {240603} (\bibinfo {year}
  {2019})}\BibitemShut {NoStop}%
\bibitem [{\citenamefont {D'Alessio}\ \emph {et~al.}(2016)\citenamefont
  {D'Alessio}, \citenamefont {Kafri}, \citenamefont {Polkovnikov},\ and\
  \citenamefont {Rigol}}]{dalessio2016quantum}%
  \BibitemOpen
  \bibfield  {author} {\bibinfo {author} {\bibfnamefont {L.}~\bibnamefont
  {D'Alessio}}, \bibinfo {author} {\bibfnamefont {Y.}~\bibnamefont {Kafri}},
  \bibinfo {author} {\bibfnamefont {A.}~\bibnamefont {Polkovnikov}}, \ and\
  \bibinfo {author} {\bibfnamefont {M.}~\bibnamefont {Rigol}},\ }\href
  {\doibase https://doi.org/10.1080/00018732.2016.1198134} {\bibfield
  {journal} {\bibinfo  {journal} {Advances in Physics}\ }\textbf {\bibinfo
  {volume} {65}},\ \bibinfo {pages} {239} (\bibinfo {year} {2016})}\BibitemShut
  {NoStop}%
\bibitem [{\citenamefont {Deutsch}(2018)}]{deutsch2018eigenstate}%
  \BibitemOpen
  \bibfield  {author} {\bibinfo {author} {\bibfnamefont {J.~M.}\ \bibnamefont
  {Deutsch}},\ }\href {\doibase 10.1088/1361-6633/aac9f1} {\bibfield  {journal}
  {\bibinfo  {journal} {Reports on Progress in Physics}\ }\textbf {\bibinfo
  {volume} {81}},\ \bibinfo {pages} {082001} (\bibinfo {year}
  {2018})}\BibitemShut {NoStop}%
\bibitem [{\citenamefont {Garrison}\ and\ \citenamefont
  {Grover}(2018)}]{garrison2018does}%
  \BibitemOpen
  \bibfield  {author} {\bibinfo {author} {\bibfnamefont {J.~R.}\ \bibnamefont
  {Garrison}}\ and\ \bibinfo {author} {\bibfnamefont {T.}~\bibnamefont
  {Grover}},\ }\href {\doibase 10.1103/PhysRevX.8.021026} {\bibfield  {journal}
  {\bibinfo  {journal} {Phys. Rev. X}\ }\textbf {\bibinfo {volume} {8}},\
  \bibinfo {pages} {021026} (\bibinfo {year} {2018})}\BibitemShut {NoStop}%
\bibitem [{\citenamefont {Dymarsky}\ \emph {et~al.}(2018)\citenamefont
  {Dymarsky}, \citenamefont {Lashkari},\ and\ \citenamefont
  {Liu}}]{dymarsky2018subsystem}%
  \BibitemOpen
  \bibfield  {author} {\bibinfo {author} {\bibfnamefont {A.}~\bibnamefont
  {Dymarsky}}, \bibinfo {author} {\bibfnamefont {N.}~\bibnamefont {Lashkari}},
  \ and\ \bibinfo {author} {\bibfnamefont {H.}~\bibnamefont {Liu}},\ }\href
  {\doibase 10.1103/PhysRevE.97.012140} {\bibfield  {journal} {\bibinfo
  {journal} {Phys. Rev. E}\ }\textbf {\bibinfo {volume} {97}},\ \bibinfo
  {pages} {012140} (\bibinfo {year} {2018})}\BibitemShut {NoStop}%
\bibitem [{\citenamefont {Polkovnikov}\ \emph {et~al.}(2011)\citenamefont
  {Polkovnikov}, \citenamefont {Sengupta}, \citenamefont {Silva},\ and\
  \citenamefont {Vengalattore}}]{polkovnikov2011colloquium}%
  \BibitemOpen
  \bibfield  {author} {\bibinfo {author} {\bibfnamefont {A.}~\bibnamefont
  {Polkovnikov}}, \bibinfo {author} {\bibfnamefont {K.}~\bibnamefont
  {Sengupta}}, \bibinfo {author} {\bibfnamefont {A.}~\bibnamefont {Silva}}, \
  and\ \bibinfo {author} {\bibfnamefont {M.}~\bibnamefont {Vengalattore}},\
  }\href {\doibase 10.1103/RevModPhys.83.863} {\bibfield  {journal} {\bibinfo
  {journal} {Rev. Mod. Phys.}\ }\textbf {\bibinfo {volume} {83}},\ \bibinfo
  {pages} {863} (\bibinfo {year} {2011})}\BibitemShut {NoStop}%
\bibitem [{\citenamefont {Neill}\ \emph {et~al.}(2016)\citenamefont {Neill},
  \citenamefont {Roushan}, \citenamefont {Fang}, \citenamefont {Chen},
  \citenamefont {Kolodrubetz}, \citenamefont {Chen}, \citenamefont {Megrant},
  \citenamefont {Barends}, \citenamefont {Campbell}, \citenamefont {Chiaro}
  \emph {et~al.}}]{neill2016ergodic}%
  \BibitemOpen
  \bibfield  {author} {\bibinfo {author} {\bibfnamefont {C.}~\bibnamefont
  {Neill}}, \bibinfo {author} {\bibfnamefont {P.}~\bibnamefont {Roushan}},
  \bibinfo {author} {\bibfnamefont {M.}~\bibnamefont {Fang}}, \bibinfo {author}
  {\bibfnamefont {Y.}~\bibnamefont {Chen}}, \bibinfo {author} {\bibfnamefont
  {M.}~\bibnamefont {Kolodrubetz}}, \bibinfo {author} {\bibfnamefont
  {Z.}~\bibnamefont {Chen}}, \bibinfo {author} {\bibfnamefont {A.}~\bibnamefont
  {Megrant}}, \bibinfo {author} {\bibfnamefont {R.}~\bibnamefont {Barends}},
  \bibinfo {author} {\bibfnamefont {B.}~\bibnamefont {Campbell}}, \bibinfo
  {author} {\bibfnamefont {B.}~\bibnamefont {Chiaro}},  \emph {et~al.},\ }\href
  {\doibase https://doi.org/10.1038/nphys3830} {\bibfield  {journal} {\bibinfo
  {journal} {Nature Physics}\ }\textbf {\bibinfo {volume} {12}},\ \bibinfo
  {pages} {1037} (\bibinfo {year} {2016})}\BibitemShut {NoStop}%
\bibitem [{\citenamefont {Mori}\ \emph {et~al.}(2018)\citenamefont {Mori},
  \citenamefont {Ikeda}, \citenamefont {Kaminishi},\ and\ \citenamefont
  {Ueda}}]{mori2018thermalization}%
  \BibitemOpen
  \bibfield  {author} {\bibinfo {author} {\bibfnamefont {T.}~\bibnamefont
  {Mori}}, \bibinfo {author} {\bibfnamefont {T.~N.}\ \bibnamefont {Ikeda}},
  \bibinfo {author} {\bibfnamefont {E.}~\bibnamefont {Kaminishi}}, \ and\
  \bibinfo {author} {\bibfnamefont {M.}~\bibnamefont {Ueda}},\ }\href {\doibase
  10.1088/1361-6455/aabcdf} {\bibfield  {journal} {\bibinfo  {journal} {Journal
  of Physics B: Atomic, Molecular and Optical Physics}\ }\textbf {\bibinfo
  {volume} {51}},\ \bibinfo {pages} {112001} (\bibinfo {year}
  {2018})}\BibitemShut {NoStop}%
\bibitem [{\citenamefont {Lange}\ \emph {et~al.}(2018)\citenamefont {Lange},
  \citenamefont {Lenar\ifmmode \check{c}\else
  \v{c}\fi{}i\ifmmode~\check{c}\else \v{c}\fi{}},\ and\ \citenamefont
  {Rosch}}]{lange2018time}%
  \BibitemOpen
  \bibfield  {author} {\bibinfo {author} {\bibfnamefont {F.}~\bibnamefont
  {Lange}}, \bibinfo {author} {\bibfnamefont {Z.}~\bibnamefont {Lenar\ifmmode
  \check{c}\else \v{c}\fi{}i\ifmmode~\check{c}\else \v{c}\fi{}}}, \ and\
  \bibinfo {author} {\bibfnamefont {A.}~\bibnamefont {Rosch}},\ }\href
  {\doibase 10.1103/PhysRevB.97.165138} {\bibfield  {journal} {\bibinfo
  {journal} {Phys. Rev. B}\ }\textbf {\bibinfo {volume} {97}},\ \bibinfo
  {pages} {165138} (\bibinfo {year} {2018})}\BibitemShut {NoStop}%
\bibitem [{\citenamefont {Shirai}\ and\ \citenamefont
  {Mori}(2020)}]{mori2020thermalization}%
  \BibitemOpen
  \bibfield  {author} {\bibinfo {author} {\bibfnamefont {T.}~\bibnamefont
  {Shirai}}\ and\ \bibinfo {author} {\bibfnamefont {T.}~\bibnamefont {Mori}},\
  }\href {\doibase 10.1103/PhysRevE.101.042116} {\bibfield  {journal} {\bibinfo
   {journal} {Phys. Rev. E}\ }\textbf {\bibinfo {volume} {101}},\ \bibinfo
  {pages} {042116} (\bibinfo {year} {2020})}\BibitemShut {NoStop}%
\bibitem [{\citenamefont {Schuckert}\ and\ \citenamefont
  {Knap}(2020)}]{schuckert2020probing}%
  \BibitemOpen
  \bibfield  {author} {\bibinfo {author} {\bibfnamefont {A.}~\bibnamefont
  {Schuckert}}\ and\ \bibinfo {author} {\bibfnamefont {M.}~\bibnamefont
  {Knap}},\ }\href {https://arxiv.org/abs/2007.10347} {\bibfield  {journal}
  {\bibinfo  {journal} {arXiv preprint arXiv:2007.10347}\ } (\bibinfo {year}
  {2020})}\BibitemShut {NoStop}%
\bibitem [{\citenamefont {Rubio-Abadal}\ \emph {et~al.}(2020)\citenamefont
  {Rubio-Abadal}, \citenamefont {Ippoliti}, \citenamefont {Hollerith},
  \citenamefont {Wei}, \citenamefont {Rui}, \citenamefont {Sondhi},
  \citenamefont {Khemani}, \citenamefont {Gross},\ and\ \citenamefont
  {Bloch}}]{abadal2020floquet}%
  \BibitemOpen
  \bibfield  {author} {\bibinfo {author} {\bibfnamefont {A.}~\bibnamefont
  {Rubio-Abadal}}, \bibinfo {author} {\bibfnamefont {M.}~\bibnamefont
  {Ippoliti}}, \bibinfo {author} {\bibfnamefont {S.}~\bibnamefont {Hollerith}},
  \bibinfo {author} {\bibfnamefont {D.}~\bibnamefont {Wei}}, \bibinfo {author}
  {\bibfnamefont {J.}~\bibnamefont {Rui}}, \bibinfo {author} {\bibfnamefont
  {S.~L.}\ \bibnamefont {Sondhi}}, \bibinfo {author} {\bibfnamefont
  {V.}~\bibnamefont {Khemani}}, \bibinfo {author} {\bibfnamefont
  {C.}~\bibnamefont {Gross}}, \ and\ \bibinfo {author} {\bibfnamefont
  {I.}~\bibnamefont {Bloch}},\ }\href {\doibase 10.1103/PhysRevX.10.021044}
  {\bibfield  {journal} {\bibinfo  {journal} {Phys. Rev. X}\ }\textbf {\bibinfo
  {volume} {10}},\ \bibinfo {pages} {021044} (\bibinfo {year}
  {2020})}\BibitemShut {NoStop}%
\bibitem [{\citenamefont {Avdoshkin}\ and\ \citenamefont
  {Dymarsky}(2020)}]{Avdoshkin2020}%
  \BibitemOpen
  \bibfield  {author} {\bibinfo {author} {\bibfnamefont {A.}~\bibnamefont
  {Avdoshkin}}\ and\ \bibinfo {author} {\bibfnamefont {A.}~\bibnamefont
  {Dymarsky}},\ }\href {\doibase 10.1103/PhysRevResearch.2.043234} {\bibfield
  {journal} {\bibinfo  {journal} {Phys. Rev. Research}\ }\textbf {\bibinfo
  {volume} {2}},\ \bibinfo {pages} {043234} (\bibinfo {year}
  {2020})}\BibitemShut {NoStop}%
\bibitem [{\citenamefont {Keser}\ \emph {et~al.}(2016)\citenamefont {Keser},
  \citenamefont {Ganeshan}, \citenamefont {Refael},\ and\ \citenamefont
  {Galitski}}]{Keser2016}%
  \BibitemOpen
  \bibfield  {author} {\bibinfo {author} {\bibfnamefont {A.~C.}\ \bibnamefont
  {Keser}}, \bibinfo {author} {\bibfnamefont {S.}~\bibnamefont {Ganeshan}},
  \bibinfo {author} {\bibfnamefont {G.}~\bibnamefont {Refael}}, \ and\ \bibinfo
  {author} {\bibfnamefont {V.}~\bibnamefont {Galitski}},\ }\href {\doibase
  10.1103/PhysRevB.94.085120} {\bibfield  {journal} {\bibinfo  {journal} {Phys.
  Rev. B}\ }\textbf {\bibinfo {volume} {94}},\ \bibinfo {pages} {085120}
  (\bibinfo {year} {2016})}\BibitemShut {NoStop}%
\bibitem [{\citenamefont {Bartsch}\ and\ \citenamefont
  {Gemmer}(2009)}]{bartsch2009dynamical}%
  \BibitemOpen
  \bibfield  {author} {\bibinfo {author} {\bibfnamefont {C.}~\bibnamefont
  {Bartsch}}\ and\ \bibinfo {author} {\bibfnamefont {J.}~\bibnamefont
  {Gemmer}},\ }\href {\doibase 10.1103/PhysRevLett.102.110403} {\bibfield
  {journal} {\bibinfo  {journal} {Phys. Rev. Lett.}\ }\textbf {\bibinfo
  {volume} {102}},\ \bibinfo {pages} {110403} (\bibinfo {year}
  {2009})}\BibitemShut {NoStop}%
\bibitem [{\citenamefont {Reimann}(2018)}]{reimann2018dynamical}%
  \BibitemOpen
  \bibfield  {author} {\bibinfo {author} {\bibfnamefont {P.}~\bibnamefont
  {Reimann}},\ }\href {\doibase 10.1103/PhysRevE.97.062129} {\bibfield
  {journal} {\bibinfo  {journal} {Phys. Rev. E}\ }\textbf {\bibinfo {volume}
  {97}},\ \bibinfo {pages} {062129} (\bibinfo {year} {2018})}\BibitemShut
  {NoStop}%
\bibitem [{\citenamefont {Reimann}\ and\ \citenamefont
  {Dabelow}(2019)}]{reimann2019typicality}%
  \BibitemOpen
  \bibfield  {author} {\bibinfo {author} {\bibfnamefont {P.}~\bibnamefont
  {Reimann}}\ and\ \bibinfo {author} {\bibfnamefont {L.}~\bibnamefont
  {Dabelow}},\ }\href {\doibase 10.1103/PhysRevLett.122.080603} {\bibfield
  {journal} {\bibinfo  {journal} {Phys. Rev. Lett.}\ }\textbf {\bibinfo
  {volume} {122}},\ \bibinfo {pages} {080603} (\bibinfo {year}
  {2019})}\BibitemShut {NoStop}%
\bibitem [{\citenamefont {Richter}\ and\ \citenamefont
  {Steinigeweg}(2019)}]{richter2019combining}%
  \BibitemOpen
  \bibfield  {author} {\bibinfo {author} {\bibfnamefont {J.}~\bibnamefont
  {Richter}}\ and\ \bibinfo {author} {\bibfnamefont {R.}~\bibnamefont
  {Steinigeweg}},\ }\href {\doibase 10.1103/PhysRevB.99.094419} {\bibfield
  {journal} {\bibinfo  {journal} {Phys. Rev. B}\ }\textbf {\bibinfo {volume}
  {99}},\ \bibinfo {pages} {094419} (\bibinfo {year} {2019})}\BibitemShut
  {NoStop}%
\bibitem [{\citenamefont {Weinberg}(2021)}]{weinberg2021enhanced}%
  \BibitemOpen
  \bibfield  {author} {\bibinfo {author} {\bibfnamefont {P.}~\bibnamefont
  {Weinberg}},\ }\href {https://arxiv.org/abs/2102.02293} {\bibfield  {journal}
  {\bibinfo  {journal} {arXiv preprint arXiv:2102.02293}\ } (\bibinfo {year}
  {2021})}\BibitemShut {NoStop}%
\bibitem [{\citenamefont {Prelovsek}\ and\ \citenamefont
  {Bonca}(2013)}]{prelovsek2013ground}%
  \BibitemOpen
  \bibfield  {author} {\bibinfo {author} {\bibfnamefont {P.}~\bibnamefont
  {Prelovsek}}\ and\ \bibinfo {author} {\bibfnamefont {J.}~\bibnamefont
  {Bonca}},\ }\href {\doibase 10.1007/978-3-642-35106-8} {\bibfield  {journal}
  {\bibinfo  {journal} {Springer Series in Solid-State Sciences}\ } (\bibinfo
  {year} {2013}),\ 10.1007/978-3-642-35106-8}\BibitemShut {NoStop}%
\bibitem [{\citenamefont {Bertini}\ \emph {et~al.}(2015)\citenamefont
  {Bertini}, \citenamefont {Essler}, \citenamefont {Groha},\ and\ \citenamefont
  {Robinson}}]{Bertini2015}%
  \BibitemOpen
  \bibfield  {author} {\bibinfo {author} {\bibfnamefont {B.}~\bibnamefont
  {Bertini}}, \bibinfo {author} {\bibfnamefont {F.~H.~L.}\ \bibnamefont
  {Essler}}, \bibinfo {author} {\bibfnamefont {S.}~\bibnamefont {Groha}}, \
  and\ \bibinfo {author} {\bibfnamefont {N.~J.}\ \bibnamefont {Robinson}},\
  }\href {\doibase 10.1103/PhysRevLett.115.180601} {\bibfield  {journal}
  {\bibinfo  {journal} {Phys. Rev. Lett.}\ }\textbf {\bibinfo {volume} {115}},\
  \bibinfo {pages} {180601} (\bibinfo {year} {2015})}\BibitemShut {NoStop}%
\bibitem [{\citenamefont {Pandey}\ \emph {et~al.}(2020)\citenamefont {Pandey},
  \citenamefont {Claeys}, \citenamefont {Campbell}, \citenamefont
  {Polkovnikov},\ and\ \citenamefont {Sels}}]{pandey2020adiabatic}%
  \BibitemOpen
  \bibfield  {author} {\bibinfo {author} {\bibfnamefont {M.}~\bibnamefont
  {Pandey}}, \bibinfo {author} {\bibfnamefont {P.~W.}\ \bibnamefont {Claeys}},
  \bibinfo {author} {\bibfnamefont {D.~K.}\ \bibnamefont {Campbell}}, \bibinfo
  {author} {\bibfnamefont {A.}~\bibnamefont {Polkovnikov}}, \ and\ \bibinfo
  {author} {\bibfnamefont {D.}~\bibnamefont {Sels}},\ }\href {\doibase
  10.1103/PhysRevX.10.041017} {\bibfield  {journal} {\bibinfo  {journal} {Phys.
  Rev. X}\ }\textbf {\bibinfo {volume} {10}},\ \bibinfo {pages} {041017}
  (\bibinfo {year} {2020})}\BibitemShut {NoStop}%
\bibitem [{\citenamefont {Kollar}\ \emph {et~al.}(2011)\citenamefont {Kollar},
  \citenamefont {Wolf},\ and\ \citenamefont {Eckstein}}]{Kollar2011}%
  \BibitemOpen
  \bibfield  {author} {\bibinfo {author} {\bibfnamefont {M.}~\bibnamefont
  {Kollar}}, \bibinfo {author} {\bibfnamefont {F.~A.}\ \bibnamefont {Wolf}}, \
  and\ \bibinfo {author} {\bibfnamefont {M.}~\bibnamefont {Eckstein}},\ }\href
  {\doibase 10.1103/PhysRevB.84.054304} {\bibfield  {journal} {\bibinfo
  {journal} {Phys. Rev. B}\ }\textbf {\bibinfo {volume} {84}},\ \bibinfo
  {pages} {054304} (\bibinfo {year} {2011})}\BibitemShut {NoStop}%
\bibitem [{\citenamefont {Calabrese}\ \emph {et~al.}(2011)\citenamefont
  {Calabrese}, \citenamefont {Essler},\ and\ \citenamefont
  {Fagotti}}]{Calabrese2011}%
  \BibitemOpen
  \bibfield  {author} {\bibinfo {author} {\bibfnamefont {P.}~\bibnamefont
  {Calabrese}}, \bibinfo {author} {\bibfnamefont {F.~H.~L.}\ \bibnamefont
  {Essler}}, \ and\ \bibinfo {author} {\bibfnamefont {M.}~\bibnamefont
  {Fagotti}},\ }\href {\doibase 10.1103/PhysRevLett.106.227203} {\bibfield
  {journal} {\bibinfo  {journal} {Phys. Rev. Lett.}\ }\textbf {\bibinfo
  {volume} {106}},\ \bibinfo {pages} {227203} (\bibinfo {year}
  {2011})}\BibitemShut {NoStop}%
\bibitem [{\citenamefont {Russomanno}\ \emph {et~al.}(2012)\citenamefont
  {Russomanno}, \citenamefont {Silva},\ and\ \citenamefont
  {Santoro}}]{Russomanno2012}%
  \BibitemOpen
  \bibfield  {author} {\bibinfo {author} {\bibfnamefont {A.}~\bibnamefont
  {Russomanno}}, \bibinfo {author} {\bibfnamefont {A.}~\bibnamefont {Silva}}, \
  and\ \bibinfo {author} {\bibfnamefont {G.~E.}\ \bibnamefont {Santoro}},\
  }\href {\doibase 10.1103/PhysRevLett.109.257201} {\bibfield  {journal}
  {\bibinfo  {journal} {Phys. Rev. Lett.}\ }\textbf {\bibinfo {volume} {109}},\
  \bibinfo {pages} {257201} (\bibinfo {year} {2012})}\BibitemShut {NoStop}%
\bibitem [{\citenamefont {Lezama}\ \emph {et~al.}(2019)\citenamefont {Lezama},
  \citenamefont {Bera},\ and\ \citenamefont {Bardarson}}]{Talia2019}%
  \BibitemOpen
  \bibfield  {author} {\bibinfo {author} {\bibfnamefont {T.~L.~M.}\
  \bibnamefont {Lezama}}, \bibinfo {author} {\bibfnamefont {S.}~\bibnamefont
  {Bera}}, \ and\ \bibinfo {author} {\bibfnamefont {J.~H.}\ \bibnamefont
  {Bardarson}},\ }\href {\doibase 10.1103/PhysRevB.99.161106} {\bibfield
  {journal} {\bibinfo  {journal} {Phys. Rev. B}\ }\textbf {\bibinfo {volume}
  {99}},\ \bibinfo {pages} {161106} (\bibinfo {year} {2019})}\BibitemShut
  {NoStop}%
\bibitem [{\citenamefont {Singh}\ \emph {et~al.}(2019)\citenamefont {Singh},
  \citenamefont {Fujiwara}, \citenamefont {Geiger}, \citenamefont {Simmons},
  \citenamefont {Lipatov}, \citenamefont {Cao}, \citenamefont {Dotti},
  \citenamefont {Rajagopal}, \citenamefont {Senaratne}, \citenamefont
  {Shimasaki}, \citenamefont {Heyl}, \citenamefont {Eckardt},\ and\
  \citenamefont {Weld}}]{singh2019quantifying}%
  \BibitemOpen
  \bibfield  {author} {\bibinfo {author} {\bibfnamefont {K.}~\bibnamefont
  {Singh}}, \bibinfo {author} {\bibfnamefont {C.~J.}\ \bibnamefont {Fujiwara}},
  \bibinfo {author} {\bibfnamefont {Z.~A.}\ \bibnamefont {Geiger}}, \bibinfo
  {author} {\bibfnamefont {E.~Q.}\ \bibnamefont {Simmons}}, \bibinfo {author}
  {\bibfnamefont {M.}~\bibnamefont {Lipatov}}, \bibinfo {author} {\bibfnamefont
  {A.}~\bibnamefont {Cao}}, \bibinfo {author} {\bibfnamefont {P.}~\bibnamefont
  {Dotti}}, \bibinfo {author} {\bibfnamefont {S.~V.}\ \bibnamefont
  {Rajagopal}}, \bibinfo {author} {\bibfnamefont {R.}~\bibnamefont
  {Senaratne}}, \bibinfo {author} {\bibfnamefont {T.}~\bibnamefont
  {Shimasaki}}, \bibinfo {author} {\bibfnamefont {M.}~\bibnamefont {Heyl}},
  \bibinfo {author} {\bibfnamefont {A.}~\bibnamefont {Eckardt}}, \ and\
  \bibinfo {author} {\bibfnamefont {D.~M.}\ \bibnamefont {Weld}},\ }\href
  {\doibase 10.1103/PhysRevX.9.041021} {\bibfield  {journal} {\bibinfo
  {journal} {Phys. Rev. X}\ }\textbf {\bibinfo {volume} {9}},\ \bibinfo {pages}
  {041021} (\bibinfo {year} {2019})}\BibitemShut {NoStop}%
\bibitem [{\citenamefont {Weinberg}\ and\ \citenamefont
  {Bukov}(2017)}]{weinberg2017quspin}%
  \BibitemOpen
  \bibfield  {author} {\bibinfo {author} {\bibfnamefont {P.}~\bibnamefont
  {Weinberg}}\ and\ \bibinfo {author} {\bibfnamefont {M.}~\bibnamefont
  {Bukov}},\ }\href {\doibase 10.21468/SciPostPhys.2.1.003} {\bibfield
  {journal} {\bibinfo  {journal} {SciPost Phys.}\ }\textbf {\bibinfo {volume}
  {2}},\ \bibinfo {pages} {003} (\bibinfo {year} {2017})}\BibitemShut {NoStop}%
\bibitem [{\citenamefont {Weinberg}\ and\ \citenamefont
  {Bukov}(2019)}]{weinberg2019quspin}%
  \BibitemOpen
  \bibfield  {author} {\bibinfo {author} {\bibfnamefont {P.}~\bibnamefont
  {Weinberg}}\ and\ \bibinfo {author} {\bibfnamefont {M.}~\bibnamefont
  {Bukov}},\ }\href {\doibase 10.21468/SciPostPhys.7.2.020} {\bibfield
  {journal} {\bibinfo  {journal} {SciPost Phys.}\ }\textbf {\bibinfo {volume}
  {7}},\ \bibinfo {pages} {20} (\bibinfo {year} {2019})}\BibitemShut {NoStop}%
\end{thebibliography}%

\clearpage

\appendix

\widetext

\begin{figure*}[t!]
	\includegraphics[width=1\textwidth]{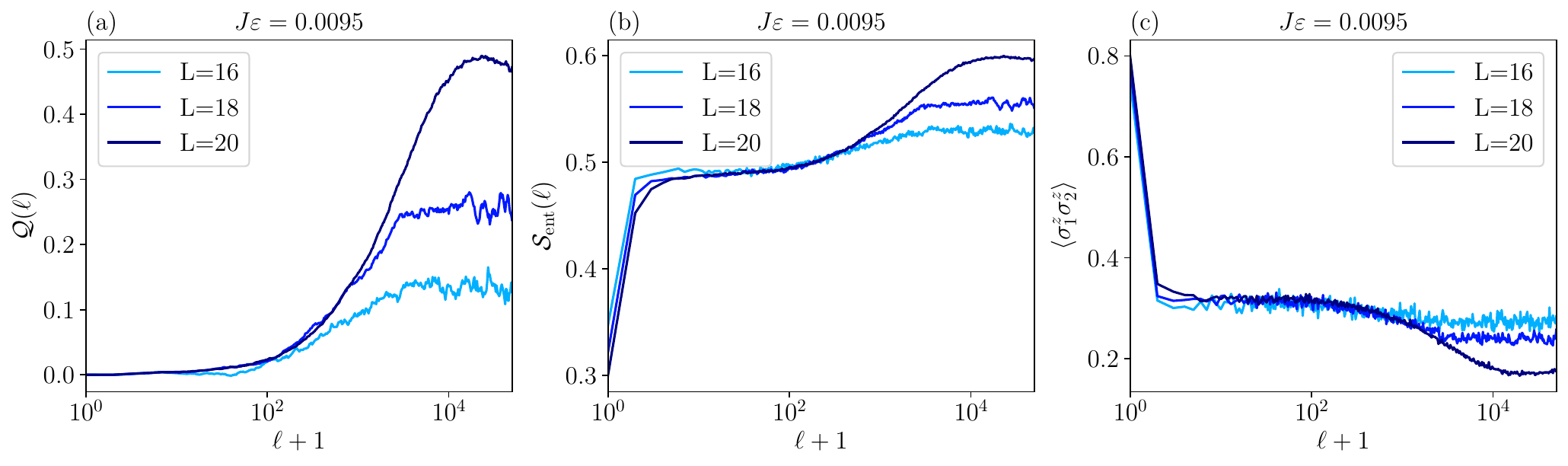}
	\includegraphics[width=1\textwidth]{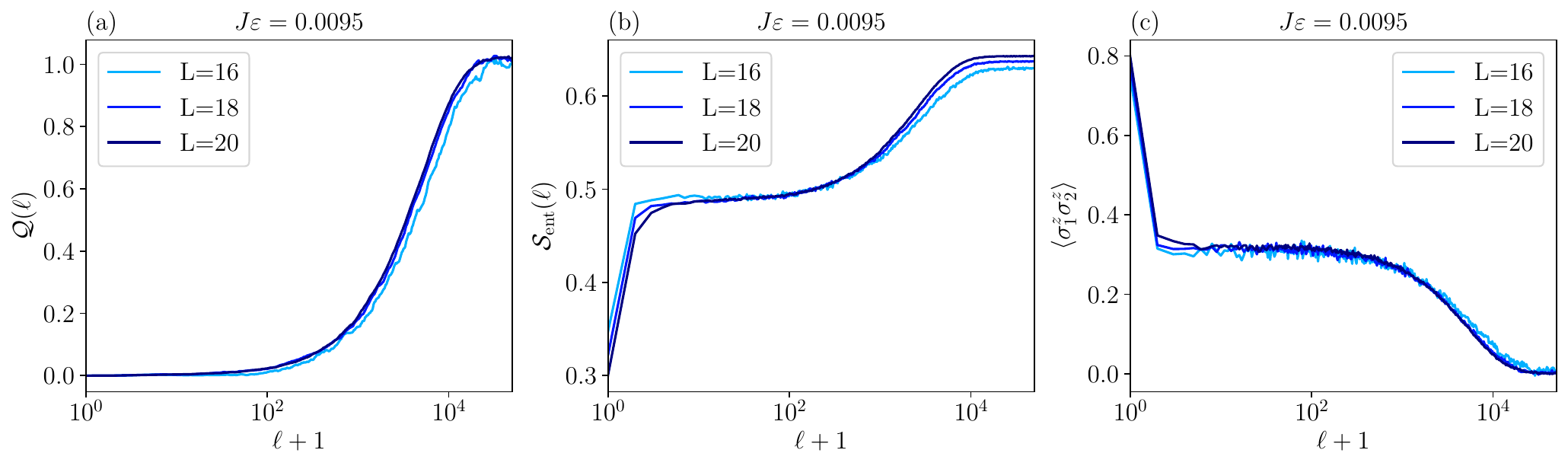}
	\caption{\label{fig:scaling_1}
		Finite size scaling for observables of the mixed-field Ising model drive ($H_1$): 
		\textbf{(a)} relative energy absorption $\mathcal{Q}(\ell)$, 
		\textbf{(b)} entanglement entropy density $\mathcal{S}_\mathrm{ent}$, and 
		\textbf{(c)} a local operator $\langle \sigma^z_1 \sigma^z_2\rangle$. 
		Top row: noise-free case $\delta=0$. Bottom row: $\delta/T=0.005$.
		The parameters are the same as in Fig.~\ref{fig:E_vs_ell_pure} of the main text.
	}
\end{figure*}

\section{\label{app:non-int}Supplementary Data for the Nonintegrable Drive $H_1$}

This appendix contains additional data supporting the simulations performed in Sec.~\ref{sec:nonintegrable}.

\subsection{\label{app:L-dep}Finite-size Dependence}

In this section, we show the finite-size scaling of various quantities discussed in the main text. Figure~\ref{fig:scaling_1} [top row] shows the time evolution of the pure state for three different system sizes. We measure the three quantities (a) $\mathcal{Q}(\ell)$, (b) the entanglement entropy density $\mathcal{S}_\mathrm{ent}(\ell)$ and (c) the local operator $\langle \sigma^z_1\sigma^z_2\rangle$. In all three curves we find the same finite size scaling. In particular, we see two effects: (i) temporal fluctuations die out as $L$ gets increased and (ii) the prethermal physics is, to a good approximation, independent of the system size; however, unconstrained thermalization at later times is affected by the system size: for infinitely large systems, we eventually expect unconstrained thermalization up to infinite temperature. For finite size systems, thermalization might come to a halt as only a portion of the full Hilbert space is active, as we discussed in Sec.~\ref{subsec:noise} of the main text. Most prominently, this is evident from the entanglement entropy density (Fig.~\ref{fig:scaling_1}b), which for ergodic dynamics is directly related to the portion of the Hilbert space that participates in thermalization. Decreasing the system size $L$ leads to plateau values for $\mathcal{S}_\mathrm{ent}(\ell)$ that deviate from the expected Page value given by $\mathrm{log}(2)-1/(2L_A)$ \citep{Page1993}. Interestingly, after adding small noise $\delta> 0$, these finite-size effects disappear almost completely (Fig.~\ref{fig:scaling_1} [bottom row])

\begin{figure*}[t!]
	\includegraphics[width=1\textwidth]{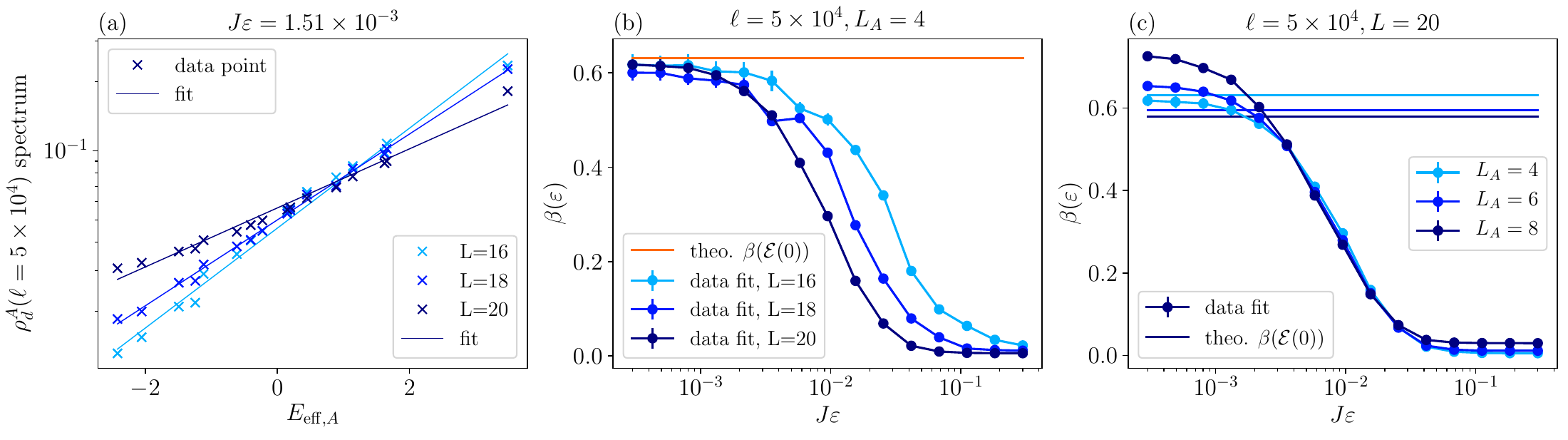}
	\includegraphics[width=1\textwidth]{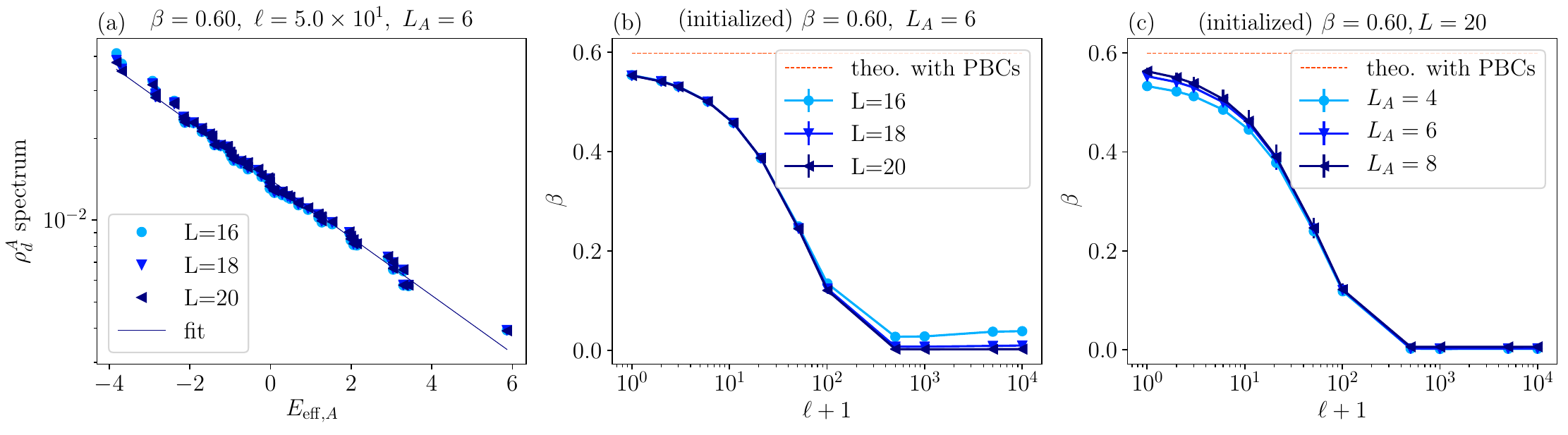}
	\caption{\label{fig:scaling_2}
		Finite size scaling for the mixed-field Ising model drive ($H_1$) of 
		\textbf{(a)} the spectrum of the reduced density matrix, 
		\textbf{(b)} $\beta(\varepsilon)$ for different system size $L$ and 
		\textbf{(c)} for different subsystem size $L_A$.  
		Top tow: pure initial state. Bottom row: Thermal initial state at $\beta(\ell\!=\!0)=0.8$.
		The parameters are the same as in Fig.~\ref{fig:E_vs_ell_pure} of the main text.
	}
\end{figure*}

In Fig.~\ref{fig:scaling_2} [top row], we show (a) the spectrum of the reduced density matrix for the pure state dynamics and (b-c) the scaling of the associated $\beta$ value as a function of $J\varepsilon$ for three different values of the system size $L$, and the subsystem size $L_A$, respectively. Thereby, no significant finite size scaling with respect to $L$ is obtained. A similar behaviour is observed in the finite size scaling with respect to $L_A$. There, increasing $L_A$ leads to worse agreement of the ETH-predicted $\beta$ values with the fitted ones (Fig.~\ref{fig:scaling_2}c [top row]). This is reasonable as ETH is only expected to work properly for sufficiently small ratios $L_A/L$. 

Fig.~\ref{fig:scaling_2} [bottom row] shows the same quantities as Fig.~\ref{fig:scaling_2} [top row], however, this time for the thermal (simulated) state. Here, finite size scaling effects are nearly completely absent. Typicality calculations are thus expected to be rather independent of the system size.

\subsection{\label{app:k-dep}Frequency or ($k$-) Dependence}

\begin{figure*}[t!]
	\includegraphics[width=1\textwidth]{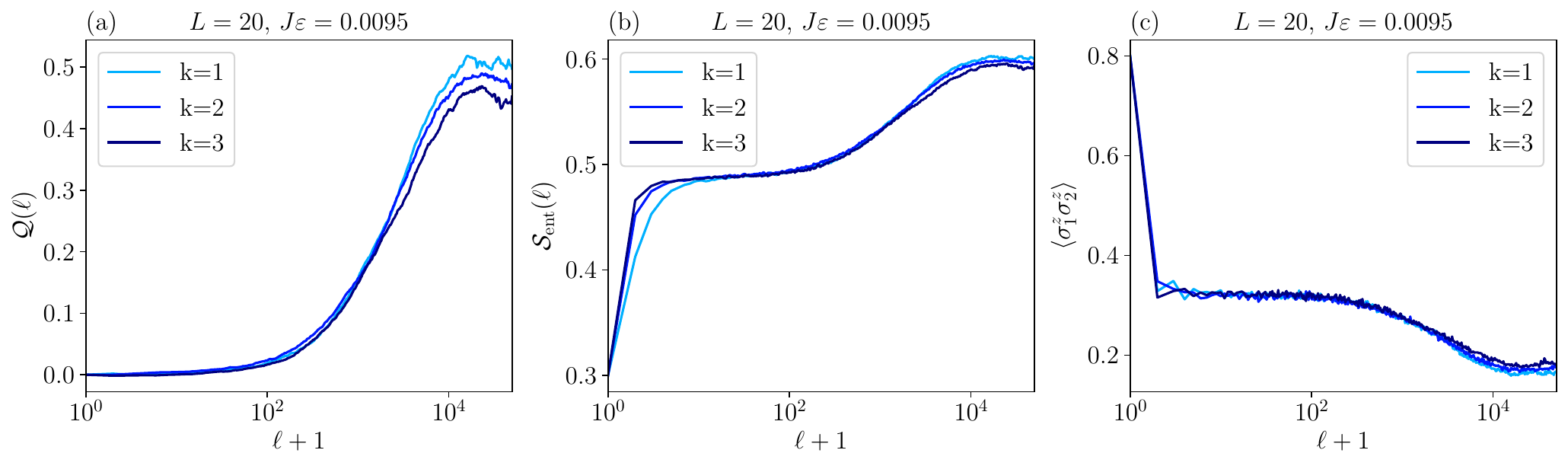}
	\caption{\label{fig:k_scaling}
	Dependence of the dynamics in the mixed-field Ising drive ($H_1$) on the commensurate point $T^\ast_k$: time evolution of the pure state for three different values of $k$ displaying 
	\textbf{(a)} $\mathcal{Q}(\ell)$, 
	\textbf{(b)} entanglement entropy density $\mathcal{S}_{\mathrm{ent}}(\ell)$ and 
	\textbf{(c)} a local operator $\langle \sigma^z_1 \sigma^z_2 \rangle $. 
	The parameters are the same as in Fig.~\ref{fig:E_vs_ell_pure} of the main text.
}
\end{figure*}

The analysis so far provides a strong indication for the qualitatively similar behavior of the $k=0$ and $k>0$ points. Here, we show a direct comparison between different commensurate points $T_k^\ast$.  Unexpectedly, we find that there is no scaling of thermalization behaviour with increasing $k$ (Fig.~\ref{fig:k_scaling}). This is surprising as the frequency drops with increasing $k$ (keeping the duration of the kick, i. e. $\varepsilon$, constant) and the folding window of the Floquet spectrum becomes smaller so more interaction is expected. The lack of a $k$-dependence once more manifests that folding is only a necessary yet insufficient criterion and that the strength of the matrix elements is crucial to determine heating rates.

\section{\label{app:int}Supplementary Data for the transverse-field Ising drive $H_2$}

\begin{figure*}[t!]
	\includegraphics[width=1\textwidth]{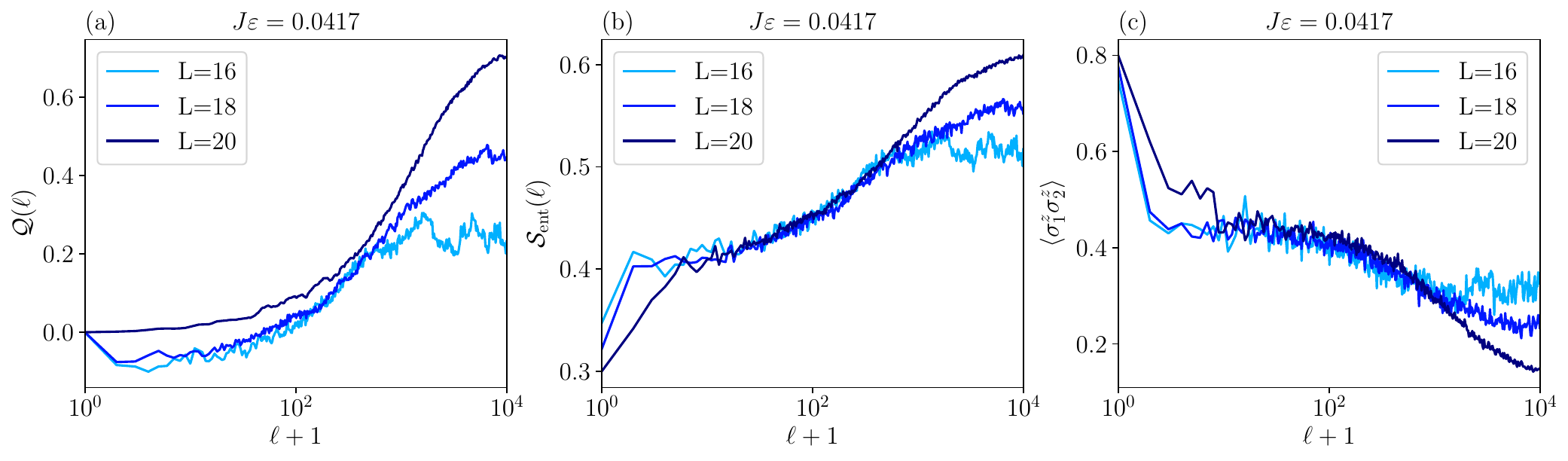}
	\includegraphics[width=1\textwidth]{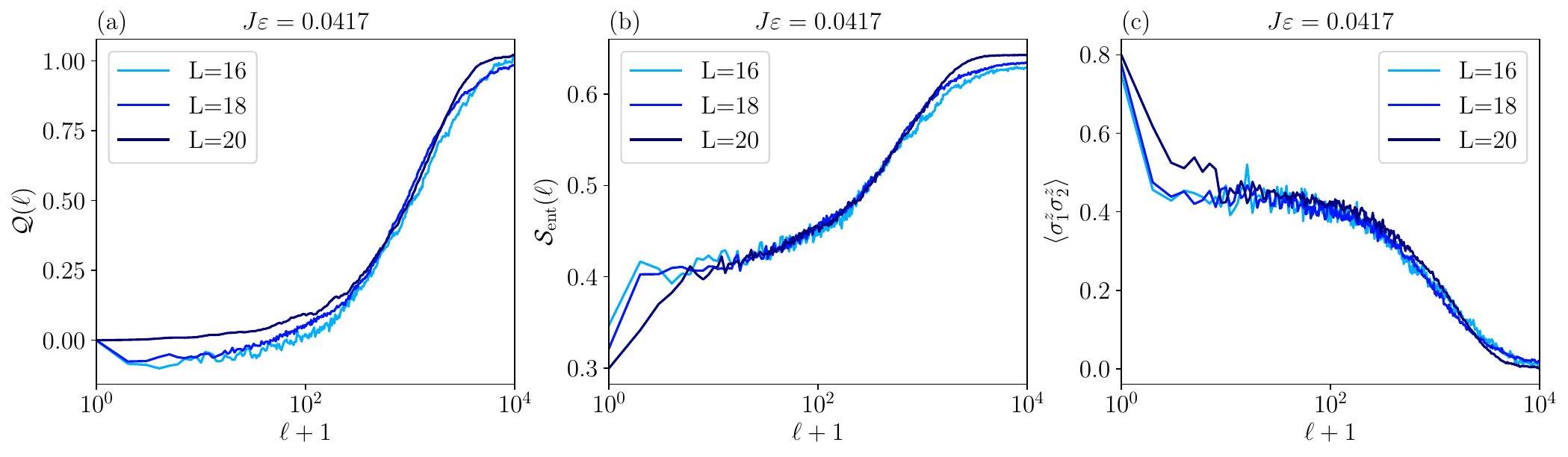}
\caption{\label{fig:scaling_2c_1}
		Finite size scaling for the transverse-field Ising drive ($H_2$): 
		\textbf{(a)} $\mathcal{Q}(\ell)$, 
		\textbf{(b)} entanglement entropy density $\mathcal{S}_\mathrm{ent}(\ell)$, 
		\textbf{(c)} a local operator $\langle \sigma^z_1 \sigma^z_2\rangle$. 
		Top row: noise-free case $\delta=0$. Bottom row: $\delta/T=0.005$.
		The parameters are the same as in Fig.~\ref{fig:E_vs_ell_pure} of the main text.
	}
\end{figure*}

\begin{figure*}[t!]
	\includegraphics[width=1\textwidth]{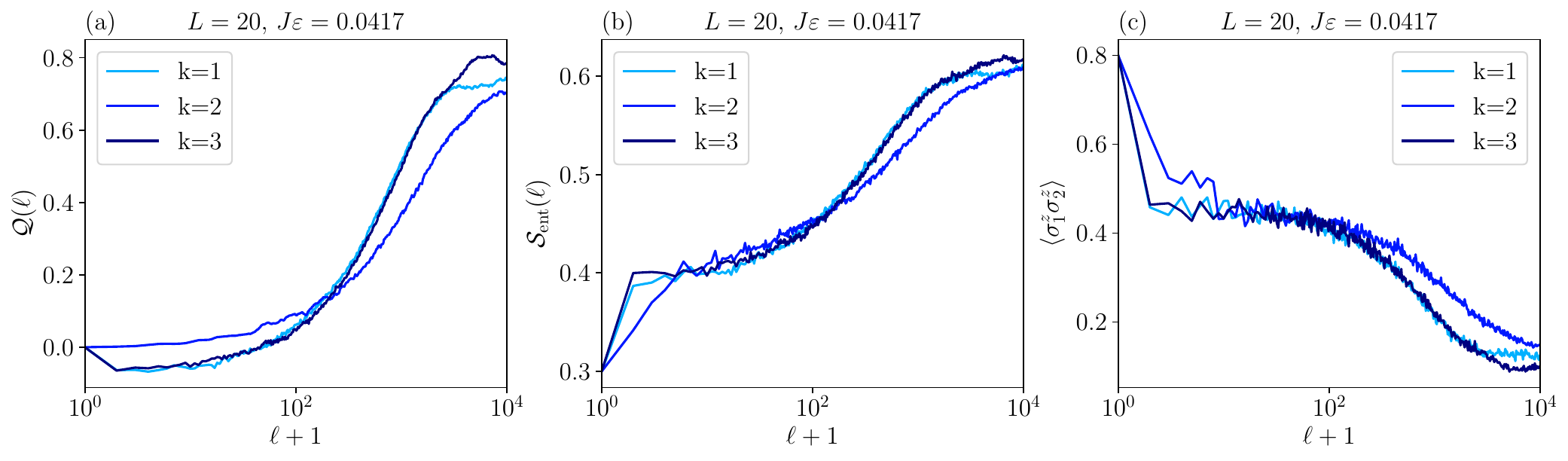}
	\caption{\label{fig:k_scaling_2c}
	Dependence of the dynamics in the transverse-field Ising drive ($H_2$) on the commensurate point $T^\ast_k$: 
	\textbf{(a)} $\mathcal{Q}(\ell)$, 
	\textbf{(b)} entanglement entropy density $\mathcal{S}_{\mathrm{ent}}(\ell)$, and 
	\textbf{(c)} a local operator $\langle \sigma^z_1 \sigma^z_2 \rangle $. 
	The parameters are the same as in Fig.~\ref{fig:E_vs_ell_pure} of the main text.
}
\end{figure*}

\begin{figure*}[h!]
	\includegraphics[width=1\textwidth]{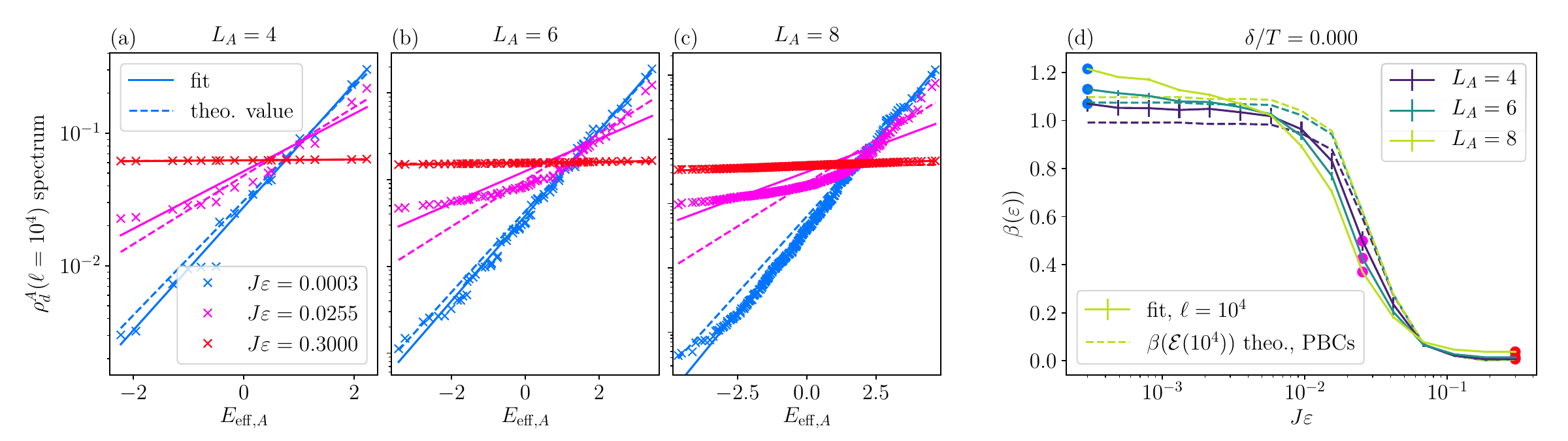}
	\includegraphics[width=1\textwidth]{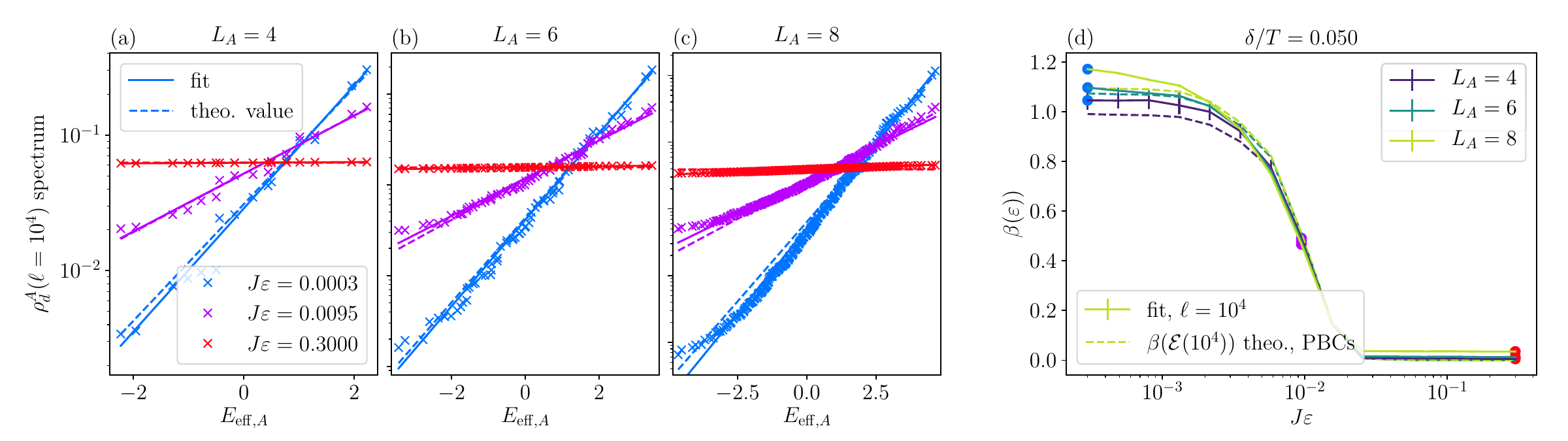}
	\caption{\label{fig:data_TFIM_2_0} Pure state thermalization behaviour of $H_2$: Spectrum of the reduced density matrix for three different values of $\varepsilon$ with $L_A=4$ \textbf{(a)}, $L_A=6$ \textbf{(b)}, and $L_A=8$ \textbf{(c)}. 
	\textbf{(d)} Fitted and ETH-predicted values of the inverse temperature $\beta$ for the different subsystem sizes. 
	In the upper row we display noise-free driving ($\delta=0$), while in the lower row we have added a small noise $\delta/T=0.05$.
}
\end{figure*}

\begin{figure*}[h!]
	\includegraphics[width=1\textwidth]{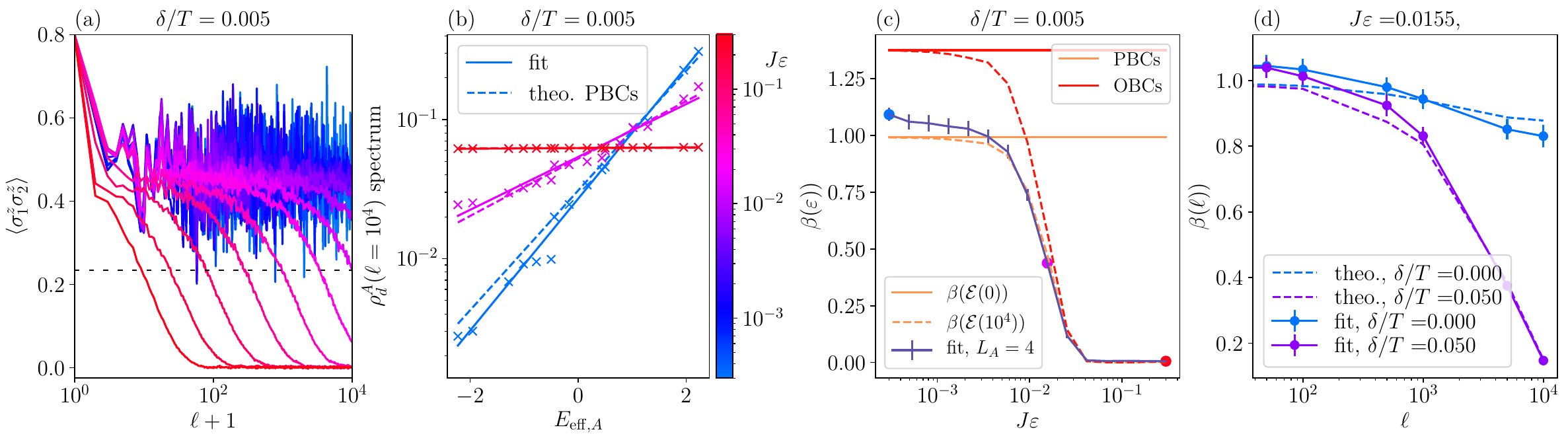}
	\caption{\label{fig:data_TFIM_2}
	Heating in the vicinity of the commensurate points $T^\ast_k$ for the transverse-field Ising drive ($H_2$) at a finite periodicity-breaking noise strength $\delta=0.005$: 
	\textbf{(a)} heating behaviour for different values of $J\varepsilon$ as a function of stroboscopic times , 
	\textbf{(b)} spectrum of the reduced density matrix at $\ell=10^4$ for three different values of $J\varepsilon$, 
	\textbf{(c)} $\beta(\varepsilon)$ obtained from fitting the spectrum of the reduced density matrix (blue curve), and computation with the instantaneous energy at $\ell=10^4$ (orange and red curves). 
	\textbf{(d)} $\beta(\ell)$ for the noise-free (lightblue) and noisy (darkblue) cases. 
	The parameters are the same as in Fig.~\ref{fig:E_vs_ell_pure} of the main text.
}
\end{figure*}

In this Appendix we show the finite size scaling for the transverse-field Ising drive $H_2$ given in Eq.~\eqref{Eq:Ising3}. In Fig.~\ref{fig:scaling_2c_1}, we depict the dependence of the pure state dynamics as a function of the system size $L$ for a noise-free and a noisy drive. In Fig.~\ref{fig:k_scaling_2c}, we show the $k$-dependence of the dynamics. Qualitatively, we find an overall similar scaling behaviour as found in the nonintegrable case in App.~\ref{app:non-int}. 
Finally, in Figs.~\ref{fig:data_TFIM_2_0} and \ref{fig:data_TFIM_2} we display the thermalization behaviour by investigating the spectrum of the reduced density matrix as a function of the subsystem size as well as the associated inverse temperature. We find that the thermal character of the state is lost in the transition regime (i.e.~for $\varepsilon \sim 10^{-2}$) [Fig.~\ref{fig:data_TFIM_2_0} upper row]. Adding small noise $\delta$ to the driving protocol activates the inactive parts of the Hilbert space and in turn reintroduces thermalization even in the transition regime [Fig.~\ref{fig:data_TFIM_2_0} lower panel, Fig.~\ref{fig:data_TFIM_2}].

\section{\label{app:analytical}Replica Expansion for the  analytically tractable drive $H_3$}
In this Appendix we explicitly resum the Replica series for the driving protocol of Sec.~\ref{subsec:clIM}. In contrast to the other parts of this paper, here we use a two-step drive: as we will see below, this has significant computational advantages as compared to the three-step drive. 

We want to resum the product of matrix exponentials $U_F=\exp(-iT/2 H_3)\exp(-i\varepsilon V)$ up to linear order in $\varepsilon$. Clearly, since the expansion parameter $\varepsilon$ only appears in one of the two exponentials, at first sight, this seems to be a formidable task since there are infinitely many terms $\propto \varepsilon T^n$ with $n \in \mathbb{N}$. A neat way to resum these terms and to eventually obtain a closed form expression makes use of the replica expansion~\cite{vajna2018replica}. 

It is easy to verify that the following identity holds
\begin{eqnarray}
TH_{\mathrm{eff}} =i\mathrm{log}\left(U_F\right)=i \lim_{\rho\rightarrow 0}\frac{1}{\rho}\left(U_F^{\rho}-1\right).
\end{eqnarray}
We aim to find an expression resummed in orders of  the small expansion parameter $\varepsilon$. Using a Taylor expansion of $U_F$ in $\varepsilon$ yields the generic expression
\begin{eqnarray}
i \mathrm{log}\left(U_F\right)= i \lim_{\rho \rightarrow 0} \frac{1}{\rho}\left(\sum_{r=0}^{\infty}\frac{1}{r!}\left(\partial_{\varepsilon}^r U_F^{\rho}\right)\bigg\vert_{\varepsilon=0}\varepsilon^r-1\right).
\end{eqnarray}
Interchanging sum and limit yields an algebraic expression for $H_{\mathrm{eff}}$ as a series in $\varepsilon$
\begin{eqnarray}
\label{Eq:Heff_app}
H_{\mathrm{eff}}=\frac{1}{T}\sum_{r=0}^{\infty}\Gamma_r \varepsilon^r,
\end{eqnarray}
where we defined 
\begin{eqnarray}
\Gamma_r=\frac{i}{r!}\lim_{\rho\rightarrow 0}\frac{1}{\rho} \left(\partial_{\varepsilon}^r U_F^{\rho}\right)\bigg \vert_{\varepsilon=0}.
\end{eqnarray}
Next, we can insert the piece-wise constant step drive $U_F=\exp(-iT/2 H_3)\exp(-i\varepsilon V)$. For the case of $r=1$, simple algebraic manipulations lead to
\begin{eqnarray}
\label{Eq:Gamma1}
\Gamma_1=i \lim_{\rho \rightarrow 0} \frac{U_0^{\rho}}{\rho} \left[\sum_{m=0}^{\rho-1} U_0^{-m}V U_0^m \right]
\end{eqnarray}
with $U_0=\exp(-iT/2H_3)$. 

To obtain an expansion with a resummed subseries in $\varepsilon$, we need to evaluate the object
\begin{equation}
\tilde H_m = U_0^{-m} V U_0^m.
\end{equation}
$U_0^m$ consists of two terms: (i) a single particle term, which essentially defines a single particle rotation around the $z$-axis and (ii) a many-body term also along the $z$-axis:
\begin{eqnarray}
U_0^m= U_{0,z}^m U_{0,zz}^m, \qquad U_{0,zz}^m=\exp\left(-imJT/2\sum_j  \sigma^z_{j+1}\sigma^z_j\right), \qquad  U_{0,z}^m=\exp\left(-imh_zT/2\sum_j  \sigma^z_{j}\right).
\end{eqnarray}
Thus, we are allowed to apply each of the two terms separately, where the order does not matter. Let us start with the many-body rotation. Straightforward manipulations yield
\begin{eqnarray}
U_{0,zz}^{-m} V U_{0,zz}^m = -\gamma \sum_j && \frac{1}{2}\sin(2mJT)\left[ \sigma^z_{j-1}\sigma^y_j 
+ \sigma^y_j\sigma^z_{j+1} \right]  + \sin^2(mJT)\sigma^z_{j-1}\sigma^x_{j}\sigma^z_{j+1}
-\cos^2(mJT)\sigma^x_j  .
\end{eqnarray}
Next, we apply the single-particle $z$ rotation and get 
\begin{eqnarray}
\label{Eq:Hm}
\tilde H_m =- \gamma\sum_j && \frac{1}{2}\sin(2mJT)
\bigg[ \sigma^z_{j-1}[\cos(mh_zT)\sigma^y_j + \sin(mh_zT)\sigma^x_j] \nonumber\\
&& + [\cos(mh_zT)\sigma^y_j + \sin(mh_zT)\sigma^x_j]\sigma^z_{j+1} \bigg] \nonumber\\
&& + \sin^2(mJT)\sigma^z_{j-1}[\cos(mh_zT)\sigma^x_{j} - \sin(mh_zT)\sigma^y_j ]\sigma^z_{j+1} \nonumber\\
&& - \cos^2(mJT)[\cos(mh_zT)\sigma^x_{j} - \sin(mh_zT)\sigma^y_j ]  .
\end{eqnarray}
To evaluate the sum over $m$ in the replica resummation, a mode expansion of Eq.~\eqref{Eq:Hm} is required. Collecting terms with different exponents we find
\begin{eqnarray}
\label{Eq:mode_expansion}
\sum_{m=0}^{\rho-1}\tilde{H}_m&=&-\gamma\bigg[\frac{1}{8i}\left(F_{2JT-h_zT}^-(\rho)+F_{2JT+h_zT}^-(\rho)\right)\sum_j (\sigma^z_j\sigma_{j+1}^y+\sigma^y_j\sigma_{j+1}^z)\nonumber\\
&+& \frac{1}{8}\left(F_{2JT-h_zT}^+(\rho)-F_{2JT+h_zT}^+(\rho)\right)\sum_j(\sigma^x_j\sigma_{j+1}^z+\sigma^z_j\sigma_{j+1}^x)\nonumber\\
&-&  \frac{1}{8}\left(F_{2JT-h_zT}^+(\rho)+F_{2JT+h_zT}^+(\rho)-2F_{-h_zT}^+(\rho)\right)\sum_j(\sigma^z_{j-1}\sigma_j^x\sigma_{j+1}^z)\nonumber\\
&+& \frac{i}{8}\left(F_{2JT-h_zT}^-(\rho)-F_{2JT+h_zT}^-(\rho)-2F_{-h_zT}^-(\rho)\right)\sum_j(\sigma^z_{j-1}\sigma_j^y\sigma_{j+1}^z)\nonumber\\
&-& \frac{1}{8}\left(F_{2JT-h_zT}^+(\rho)-F_{2JT+h_zT}^+(\rho)+2F_{-h_zT}^+(\rho)\right)\sum_j(\sigma^x_{j})\nonumber\\
&+& \frac{i}{8}\left(F_{2JT-h_zT}^-(\rho)-F_{2JT+h_zT}^-(\rho)+2F_{-h_zT}^-(\rho)\right)\sum_j(\sigma^y_{j})\bigg]
\end{eqnarray}
with 
\begin{eqnarray}
\label{Eq:Falpha1}
F_{\chi}^-(\rho)&=&-i\frac{\cos(\chi \rho -  \chi/2)-\cos(\chi/2)}{\sin(\chi/2)},\\
\label{Eq:Falpha2}
F_{\chi}^+(\rho)&=&\frac{\sin(\chi \rho-\chi/2)+\sin(\chi/2)}{\sin(\chi/2)}.
\end{eqnarray}
Taking the limit in Eq.~\eqref{Eq:Gamma1} and using Eqs.~\eqref{Eq:Falpha1} and \eqref{Eq:Falpha2}, we obtain
\begin{eqnarray}
\lim_{\rho\rightarrow 0}\frac{U^{\rho}_0}{\rho}F_{\chi}^+(\rho)&=&\chi\cot (\chi/2),\\
\lim_{\rho\rightarrow 0}\frac{U^{\rho}_0}{\rho}F_{\chi}^-(\rho)&=&-i\chi .
\end{eqnarray}
This eventually leads to
\begin{eqnarray}
\label{Eq:Gamma1_final}
\Gamma_1&=&\gamma\bigg[JT/2\sum_j (\sigma^z_j\sigma_{j+1}^y+\sigma^y_j\sigma_{j+1}^z)\nonumber\\
&+& \frac{1}{8}\left[(2JT+h_zT)\cot(JT+h_zT/2)-(2JT-h_zT)\cot(JT-h_zT/2)\right]\sum_j(\sigma^x_j\sigma_{j+1}^z+\sigma^z_j\sigma_{j+1}^x)\nonumber\\
&+&  \frac{1}{8}\left[(2JT-h_zT)\cot(JT-h_zT/2)+(2JT+h_zT)\cot(JT+h_zT/2)-2h_zT\cot(h_zT/2)\right]\sum_j(\sigma^z_{j-1}\sigma_j^x\sigma_{j+1}^z)\nonumber\\
&+& \frac{1}{8}\left[(2JT-h_zT)\cot(JT-h_zT/2)+(2JT+h_zT)\cot(JT+h_zT/2)+2h_zT\cot(h_zT/2)\right]\sum_j(\sigma^x_{j})\nonumber\\
&+& h_zT/2\sum_j(\sigma^y_{j})\bigg].
\end{eqnarray} 
Using the above expression for $\Gamma_1$ in Eq.~\eqref{Eq:Heff_app}, this yields a resummed expression for $H_{\mathrm{eff}}$ up to linear order in $\varepsilon$
\begin{eqnarray}
H_{\mathrm{eff}}&=&\frac{1}{2} H_3+\frac{\varepsilon \gamma}{2}\bigg[J\sum_j (\sigma^z_j\sigma_{j+1}^y+\sigma^y_j\sigma_{j+1}^z)\nonumber\\
&+& \frac{1}{4}\left[(2J+h_z)\cot(JT+h_zT/2)-(2J-h_z)\cot(JT-h_zT/2)\right]\sum_j(\sigma^x_j\sigma_{j+1}^z+\sigma^z_j\sigma_{j+1}^x)\nonumber\\
&+&  \frac{1}{4}\left[(2J-h_z)\cot(JT-h_zT/2)+(2J+h_z)\cot(JT+h_zT/2)-2h_z\cot(h_zT/2)\right]\sum_j(\sigma^z_{j-1}\sigma_j^x\sigma_{j+1}^z)\nonumber\\
&+& \frac{1}{4}\left[(2J-h_z)\cot(JT-h_zT/2)+(2J+h_z)\cot(JT+h_zT/2)+2h_z\cot(h_zT/2)\right]\sum_j \sigma^x_{j}\nonumber\\
&+& h_z\sum_j \sigma^y_{j}\bigg]+\mathcal{O}(\epsilon^2).
\end{eqnarray}
It is possible, though tedious, to evaluate the leading higher-order terms numerically~\cite{vajna2018replica}.

\begin{figure*}[t!]
	\includegraphics[width=0.45\textwidth]{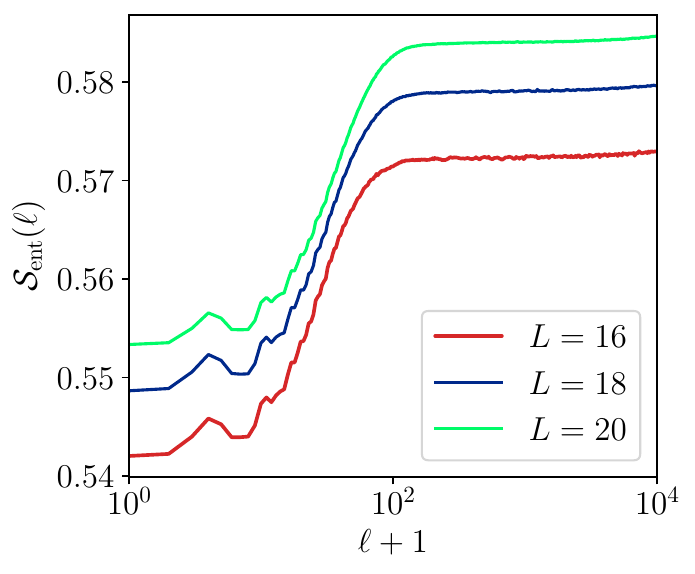}
	\caption{\label{fig:finite_size_case_2a} 
		Finite size scaling for the dynamics generated by $H_3$: Same as Fig.~\ref{fig:scaling_case_2a_3} (b) for two different system sizes and fixed $J\varepsilon=0.1629$ ($J=0.6$). The small overall shift is caused by the sightly different energy densities at the initial temperature $\beta(\ell\!=\! 0)=0.8$.
	}
\end{figure*}

\section{\label{app:case_2a}Supplementary Data for the  analytically tractable drive $H_3$}

In this Appendix we provide more detailed data regarding the thermalization behaviour of analytically tractable drives, using dynamics generated by $H_3$. 

We begin by investigating the finite size scaling of the dynamics generated by $H_3$. In the main text, we observed that intermediate values of $\varepsilon$ can lead to a late-time saturation in the time-evolution curve of given observables away from their infinite-temperature value (see Sec.~\ref{subsec:noise}). Yet, a careful investigation of the dynamics driven by $H_1$ leads to the conclusion that this constitutes a finite size effect, which might be removed by the addition of noise in the drive protocol at finite system sizes [cf.~App.~\ref{app:non-int}]. To rule out the possibility that the observed late-time plateau in the dynamics generated by $H_3$ originates from a similar effect, in Fig.~\ref{fig:finite_size_case_2a} we compare the time-evolution curves of the entanglement entropy density for different system sizes. Notice that the slight shift in the curves by approximately a constant with increasing $L$, is caused by the slight change in the initial energy density, corresponding to the fixed initial temperature, and the observed shift of the prethermal plateau matches the shift of the initial energy density. Thus, it is caused by a systematic mismatch in the initial energy densities and is not a finite-size effect -- a fact corroborated also by the scale on the $y$-axis.

\begin{figure*}[t!]
	\includegraphics[width=1\textwidth]{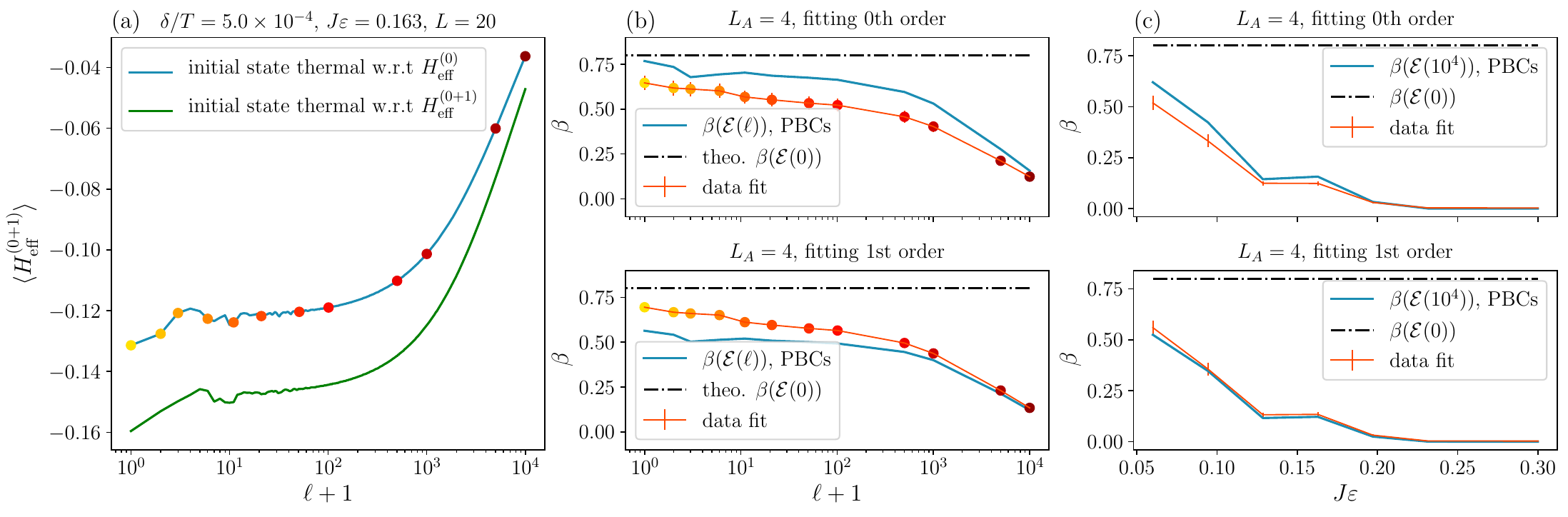}
	\caption{\label{fig:thermal_case_2a_1} 
		Same as Fig.~\ref{fig:thermal_case_2a}, however, inverse temperatures are extracted from instantaneous energies based on the time evolution of an initial state thermal w.r.t.~$H_{\mathrm{eff}}^{(0)}$.
}
\end{figure*}

Although the dynamics of the system at late times is not expected to strongly depend on the initial state, one may want to reason that the observed agreement in the lower panels of Fig.~\ref{fig:thermal_case_2a}(b-c), as well as the corresponding disagreement in the upper panels, arise from the initial state being already thermal w.r.t.~$H_\mathrm{eff}^{(0+1)}$ (as opposed to thermal w.r.t.~$H_{\mathrm{eff}}^{(0)}$). To rule out this possibility, we perform the steps of the above analysis using an initial state which is now thermal w.r.t.~$H_\mathrm{eff}^{(0)}$ [Fig.~\ref{fig:thermal_case_2a_1}]. In this setup, we do not find a good agreement of the fitted and ETH-predicted inverse temperatures, using either of the two effective Hamiltonians at small and intermediate driving times [Fig.~\ref{fig:thermal_case_2a_1}(b-c)]; only at long driving times is the agreement restored when using $H_\mathrm{eff}^{(0+1)}$, since the infinite-temperature state is a universal long-time attractor. 
Thus, using the zeroth-order term, $H_\mathrm{eff}^{(0)}$, an agreement of the numerically-fitted and the ETH-predicted temperatures is only reached close to infinite temperature. 
 
\begin{figure*}
	\includegraphics[width=1\textwidth]{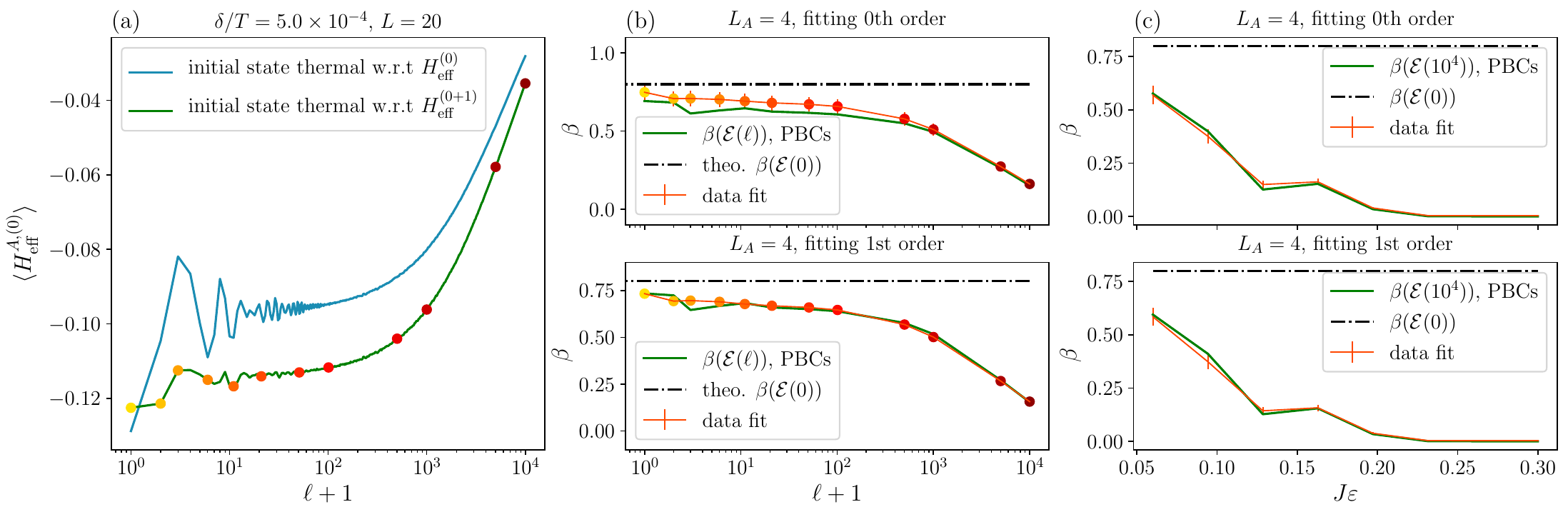}
	\caption{\label{fig:thermal_case_2a_app} 
	Same as Fig.~\ref{fig:thermal_case_2a}, where we changed the observable from $H_{\mathrm{eff}}^{(0+1)}$ to $H_{\mathrm{eff}}^{A,(0)}$. The superscript $A$ here indicates that we only measure $H_{\mathrm{eff}}^{(0)}$ on the subsystem.
}
\end{figure*}

Finally, we would also like to emphasize that the improvement brought by higher-order corrections to the effective Hamiltonian depends on the observable of interest, as  suggested by Fig.~\ref{fig:thermal_case_2a_app}. Indeed, Fig.~\ref{fig:thermal_case_2a_app} is equivalent to Fig.~\ref{fig:thermal_case_2a} of the main text, however, this time using $H_\mathrm{eff}^{A,(0)}$ (i.e.~the effective Hamiltonian to leading order on the subsystem) as observable. The observed deviations are small already on the level of the zeroth order effective Hamiltonian. We emphasize that, Fig.~\ref{fig:thermal_case_2a_app} contains a feature that indirectly proves that $H_\mathrm{eff}^{(0+1)}$ provides a better approximation to the effective Hamiltonian as compared to $H_\mathrm{eff}^{(0)}$: When the system is initialized in a thermal state w.r.t $H_\mathrm{eff}^{(0+1)}$, constrained thermalization is nearly almost absent (as opposed to the initial state being thermal w.r.t. $H_\mathrm{eff}^{(0)}$) [Fig.~\ref{fig:thermal_case_2a_app}a, oscillatory blue line]. This implies that the system is already initially in the correct thermal state  w.r.t.~the generator of dynamics so that no drive-induced initial quench dynamics occurs.

\section{\label{app:errors}Deviations from Pure Thermal States in Floquet Subsystem Thermalization}

In this Appendix we discuss the observed deviations between subsystem density matrices, obtained from numerical simulations, and exact thermal states, for all models studied in this work. 

As mentioned in Sec.~\ref{subsec:subsys}, deviations from a perfect thermal state are expected to occur, due to (i) finite-size effects (i.e., the finite ratio $L_A/L$ of the subsystem to system sizes), (ii) the finitely many states used to construct the diagonal ensemble (in the case of pure initial states) or the thermal ensemble approximated using typicality, and (iii) the approximate character of the effective Hamiltonian $H_\mathrm{eff}$ (as opposed to the exact Floquet Hamiltonian), that the system thermalizes with respect to. 
Throughout the main text, we did a least square fit of the eigenvalues of the reduced density matrix obtained from the numerical simulations, and the thermal density matrix according to ETH predictions. The mismatch between the two is quantified by the uncertainty of the least square fits, and is displayed in form of error bars. This comparison, although natural, does not take into account the deviation between the eigenstates of the numerical and thermal density matrices. 
Therefore, here, we would like to complete this discussion, by investigating additional quantities that directly quantify the difference between the two density matrices. As a measure for the deviation, we focus on (i) the quantum Kullback-Leibler (KL) divergence and (ii) the Uhlmann fidelity. 

The KL divergence, sometimes referred to as cross-entropy or relative entropy, is commonly used to give a measure for the similarity of two classical probability distributions. Similar to the definition of the von Neuman entropy, it can be defined for density matrices. Thus, the KL divergence of a density matrx $\rho$ with respect to the density matrix $\sigma$ is defined as
\begin{eqnarray}
\mathrm{K\! L}(\rho \vert \vert \sigma) = \mathrm{tr}\left(\rho \mathrm{log}(\rho) - \rho \mathrm{log}(\sigma)\right).
\end{eqnarray}
The KL divergence is non-negative and, in general, equal to zero, if and only if, $\rho=\sigma$. Hence, any deviations from zero of the KL divergence quantify the difference between $\rho$ and $\sigma$. 

Likewise, the Uhlmann fidelity between two (mixed) quantum states provides an alternative measure. For any two density matrices $\rho$ and $\sigma$, the fidelity is defined as
\begin{eqnarray}
f(\rho, \sigma) = \mathrm{tr}\left(\sqrt{\sqrt{\rho}\sigma\sqrt{\rho}}\right)^2.
\end{eqnarray}
Unlike the KL divergence, the Uhlmann fidelity has the advantage of being symmetric, i.e., $f(\rho,\sigma) = f(\sigma,\rho)$. Moreover, the fidelity is bounded, $0 \leq f(\rho, \sigma) \leq 1$, with $f(\rho,\sigma)=1$, if and only if $\rho=\sigma$. This complicates a direct comparison of the two measures and motivates to instead investigate $1-f(\rho,\sigma)$ as well as a symmetrized version of the KL divergence, known as the Jensen-Shanon (JS) divergence defined by
\begin{eqnarray}
\mathrm{JSD}(\rho \vert \sigma) =\frac{1}{2} \mathrm{K\! L}(\rho \vert \kappa)  + \frac{1}{2} \mathrm{K\! L}(\sigma\vert \kappa),  
\end{eqnarray}
where $\kappa = 1/2 (\rho+\sigma)$.

\begin{figure}
\centering
\includegraphics[scale=0.5]{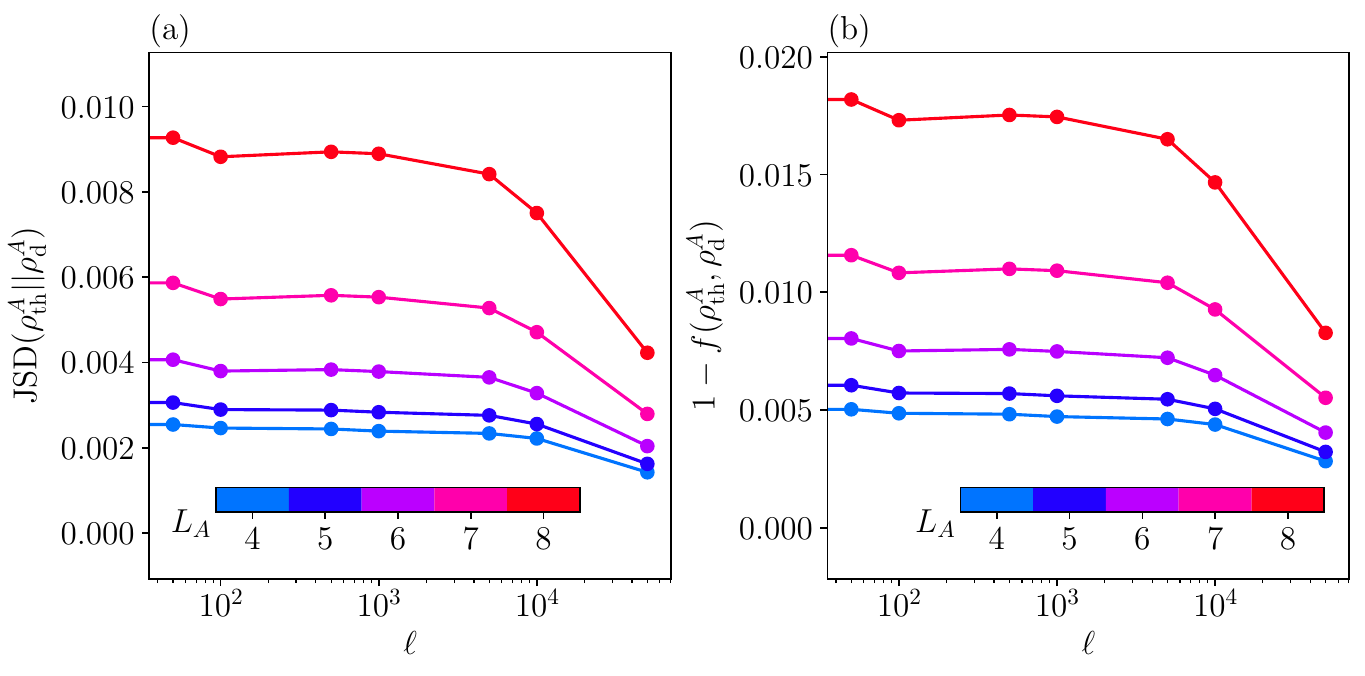}
\caption{KL divergence per spin (a) and fidelity (b) for the dynamics generated by $H_1$: We compare the thermal density matrix $\rho_{\mathrm{th}}^A$ (cf.~Eq.~\eqref{eq:beta_of_E}) on a subsystem of size $L_A$ given the effective Hamiltonian $H_{\mathrm{eff}}^A$ (with OBC) with the reduced density matrix of the empirical diagonal ensemble, constructed from exact time evolution, as a function of the evolution cycle $\ell$. We apply a small noise of $\delta/T=0.005$ to the driving protocol (compare Eq.~\ref{eq:noisy_drive}). The parameters are the same as in Fig.~\ref{fig:E_vs_ell_pure}.}
\label{fig:error_case_3}
\end{figure}

\begin{figure}
\centering
\includegraphics[scale=0.5]{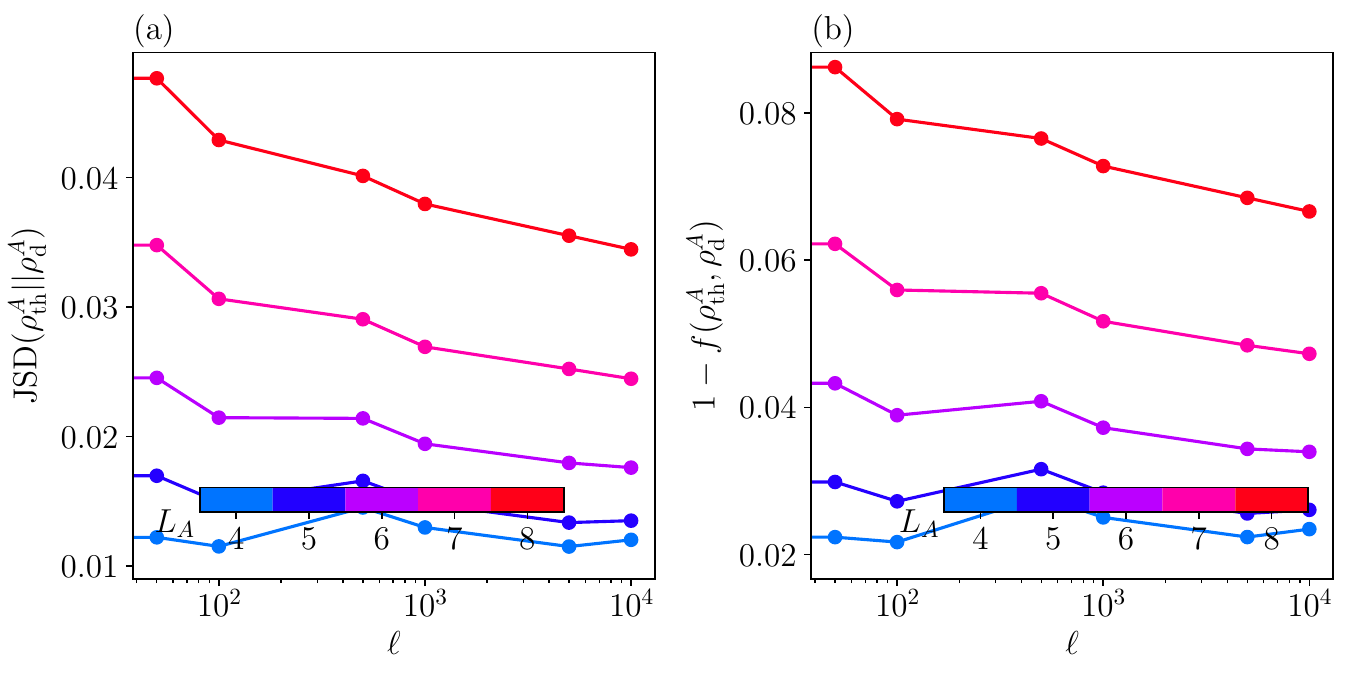}
\caption{Same as Fig.~\ref{fig:error_case_3}, for the dynamics generated by $H_2$.}
\label{fig:error_case_2c}
\end{figure}

\begin{figure}
\centering
\includegraphics[scale=0.5]{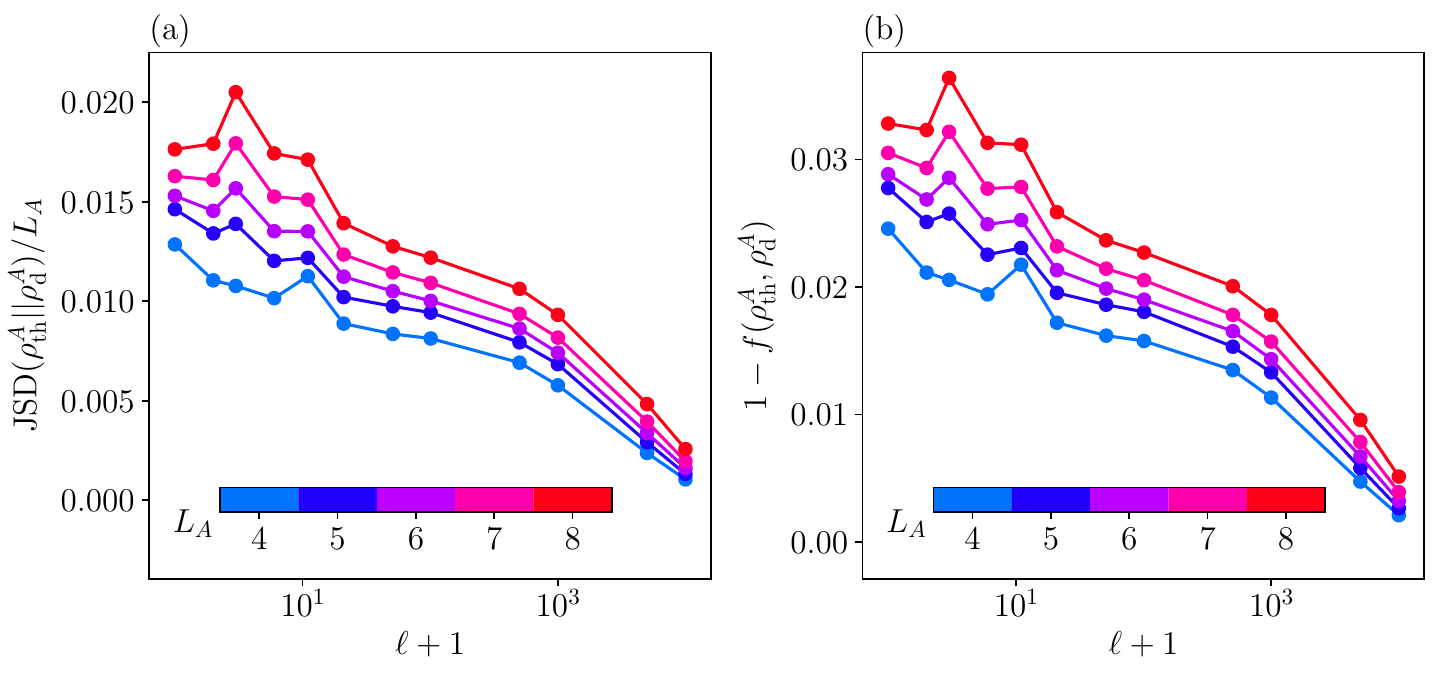}
\caption{Same as Fig.~\ref{fig:error_case_3}, for the dynamics generated by $H_3$. In analogy to the main text, for this model we use PBC to the subsystem effective Hamiltonian. The parameters are the same as in Fig.~\ref{fig:thermal_case_2a}.}
\label{fig:error_case_2a}
\end{figure}

In this Appendix, we compare (1) the exact thermal density matrix $\rho_{\mathrm{th}}^A\propto\exp(-\beta H_\mathrm{eff}^A)$, constructed with the help of the corresponding approximate effective Hamiltonians and temperature $\beta$ set by the initial energy density (see Eq.~\eqref{eq:beta_of_E} and the associated discussion), with (2) the subsystem density matrix obtained numerically using the reduced approximate diagonal ensemble $\rho_d^A$. Comparisons are shown in Figs.~\ref{fig:error_case_3}, \ref{fig:error_case_2c} and \ref{fig:error_case_2a}, corresponding to the drives generated by $H_1$, $H_2$ and $H_3$, respectively. 
All three models show reasonably small JS divergence as well as a fidelity close to unity. Yet, the drive generated by $H_2$ shows visibly increased (decreased) values of the KL divergence (fidelity). This agrees with the observed behavior of the corresponding subsystem density matrix eigenvalues (cf.~Fig.~\ref{fig:data_TFIM_2_0}), which shows increased deviations from the thermal state, especially for larger subsystems. Moreover, the KL divergence and fidelity also capture the finite size sensitivity of the subsystem thermalization seen for drives with pure initial states ($H_1$ and $H_2$). On the other hand, thermal initial states do not show this sensitivity, which is also consistent with Fig.~\ref{fig:error_case_2a}. In this case, as the system is thermal right away, it is also meaningful to investigate both quantities at $\ell =0$. This provides us with a measure for the quality of the approximate thermal initial state. 


\end{document}